\RequirePackage{fix-cm}
\documentclass[smallextended]{svjour3}       % onecolumn (second format)
\smartqed  % flush right qed marks, e.g. at end of proof
\usepackage{graphicx}
\usepackage{amssymb}
\usepackage{color}
\def\Tc{{\mathcal T}}

\def\Fc{{\mathcal F}}

\def\Dc{{\mathcal D}}
\def\Pc{{\mathcal P}}
\def\Qc{{\mathcal Q}}

\def\Xc{{\mathcal X}}

\def \b{\beta}

\def \d{\delta}

\def \l{\lambda}

\def \p{\pi}
\def \r{\rho}
\def \s{\sigma}
\def \t{\tau}

\def \f{\varphi}
\def \o{\omega}

\def \e{\varepsilon}
\def \one{\hbox{I\hskip-.60em 1}}
\def \[{I\!\![}
\def \[{\rrbracket}
\def \]{I\!\!]}
\def \]{\llbracket}

\spnewtheorem{assumption}{Assumption}{\bf}{\rm}
\usepackage{latexsym}
\newcommand{\eqref}[1]{(\ref{#1})}

\newcommand{\E}    {\mathbb{E}}

\newcommand{\N}    {\mathbb{N}}
\renewcommand{\P}  {\mathbb{P}}
\newcommand{\Q}    {\mathbb{Q}}
\newcommand{\R}    {\mathbb{R}}

\newcommand{\la}{\langle}
\newcommand{\ra}{\rangle}
\begin{document}

\title{No free lunch for markets with multiple num\'{e}raires%\thanks{Grants or other notes
}


\author{Laurence Carassus}


\institute{\at
L\'{e}onard de Vinci P\^ole Universitaire, Research Center, 92 916 Paris La D\'{e}fense, France \\
and LMR, UMR 9008, Universit\'e Reims Champagne-Ardenne.\\
\email{laurence.carassus@devinci.fr}
}
\date{\today}

\maketitle

\begin{abstract}
We consider a new framework, that of a global market with a finite number of submarkets, where there is a tradable num\'{e}raire for each submarket, but no tradable num\'{e}raire for the global market.  Under a global no arbitrage condition, we show the existence of a common density from which equivalent (local) martingale measures are constructed for each submarket. We also introduce several superreplication prices, depending on the  chosen type of hedging: on the global market, on a given submarket or on all submarkets separably. We prove duality results on these prices that allow 
 to assess differences in characteristics between the submarkets, such as liquidity or credit quality. The results are applied in concrete situations, in particular in a Brownian  setup. 
\end{abstract}

\section{Introduction}
\label{intro}
The concept of no arbitrage (NA) is fundamental in the modern theory of mathematical finance. Let us consider a one period market with $d+1$ assets with price process 
$(\hat S_{t\in\{0,1\}})$. Let $\hat \Phi \in \R^{d+1}$ be the number of shares  purchased at time zero. It is an arbitrage if $\hat \Phi .  \hat S_0 \leq 0$, $\hat \Phi .  \hat S_1 \geq 0$ $\P$-a.s. and 
$\P(\hat \Phi .  \hat S_1>0) >0,$ where $\P$ is the probability measure prevailing on the state space. This condition has a good mathematical characterization in terms of martingale measures, which is called the fundamental theorem of asset pricing (FTAP). To state the FTAP, one has to consider a specific financial asset called num\'{e}raire, which allows  to deflate the other assets. They are then denominated in units of the num\'{e}raire as units of account. 
It is important to note that the NA condition does not depend on the num\'{e}raire, but  the martingale measures do. To see this, 
let $\hat S=(S^0,S)$ where  $S^0$ is the num\'{e}raire and $S$ is the price process of the $d$ other assets. 
If $S^0$ is a tradable asset, and only in this case, NA is equivalent to the following property : any investment $\Phi$   in the other $d$ traded assets  which yields with positive probability  a better result than investing the same amount (divided by the initial value of the num\'{e}raire)  in the num\'{e}raire must be exposed to some downside risk, i.e., setting $\tilde{S}=S/S^0$ there is not $\Phi \in \R^d$ such that 
%$\Phi. S_1 \geq S_1^0\frac{ \Phi. S_0}{S_0^0}$ $P$-a.s. and  $P(\Phi. S_1 > S_1^0\frac{ \Phi. S_0}{S_0^0}) >0$ or setting $\tilde S=\frac{ S}{S^0}$ 
$\Phi. (\tilde S_1- \tilde S_0)\geq0 $ $\P$-a.s. and 
$\P(\Phi.  (\tilde S_1- \tilde S_0)>0) >0.$ 
 Then, the FTAP asserts that NA is equivalent to the existence of a probability measure $\Q$ equivalent to $\P$ and such that $\tilde S$ is martingale under $\Q$. 
In general market, under integrability conditions, the FTAP asserts that an appropriate notion of no arbitrage is equivalent to the existence of equivalent local martingale (or risk-neutral) probability measures for the  price process of the assets deflated by the tradable num\'{e}raire.
%The important assumption is that the num\'{e}raire needs to be a strictly positive tradable asset. 
One usually takes the  num\'{e}raire to be a savings account (or zero-coupon bond) expressed in the currency of the country. 
%%EDITOR'S NOTE: Please ensure that the intended meaning has been maintained in this edit.
The FTAP was initially formalised in \cite{HK79,HP81},
while \cite{dmw} established it in a general discrete-time setting, and \cite{DS94} did so in continuous-time models. The literature on the subject is vast, and we refer to \cite{DelSch05,KaS} for a general overview.
The FTAP is essential for pricing issues, namely, for computing the superreplication price, which is, for a given claim, the minimum selling price needed to superreplicate it by trading in the market. This is the hedging price with no risk, and to the best of our knowledge, it was first introduced in \cite{Ben91} in the context of transaction costs. In complete markets, the superreplication cost is just the cash flow expectation computed under the unique risk-neutral measure. However, in incomplete markets, the superreplication cost is equal to the supremum of those expectations computed under the different risk-neutral probability measures. This is the so-called dual formulation of the superreplication price or superhedging theorem (see, for instance, \cite{ElQu91} or \cite{CvKa92}).

%The absence of arbitrage condition is independent of the choice, and even the existence, of a num\'{e}raire but the set of equivalent martingale measures in the FTAP does depend on 
%the  num\'{e}raire. Moreover, the FTAP requires the existence of  a tradable num\'{e}raire 
%%i.e. an asset (or a synthetic one as a sum of assets)
%in units of which all other assets are expressed.  Here,  this is no longer the case. 
In this paper, we assume that 
 there is a global market made of several submarkets and {we suppose that one cannot trade between submarkets}. 
%on ne peut tariter entre les marché car l\rq{}autofi est marché par marché et chauqe richesse initiale est \geq 0.  
 So, 
there is not one asset that is traded in all the submarkets and that could be used  as a  num\'{e}raire in the FTAP. 
Still we consider the absence of arbitrage at the level of the global market and not only in each submarket.  
 Here, we want to capture the situation where each submarket is arbitrage free but where, without further assumption, there may be arbitrage in the global market. 

One may think of a big company with several Business Units (BU). Each BU acts as a separate part of the company and has some form of autonomy in its operations. Each BU  is implemented around a single area of activity with its own objectives and resources. 
%So, some BU may have good performance while other may perform badly. 
Nevertheless, the total wealth of the company is the sum of the wealth of each BU and an arbitrage opportunity may occur at the level of the company even if every BU is arbitrage free.  One may also think of a financial institution with several trading desks specialized by financial products or market segments: stocks, currencies, commodities, bonds, etc. 
As argued by \cite{HHP}, cognitive restrictions require traders to restrict attention to a given subset  of assets. 
Each desk is independent but the  P\&L of the financial institution  is the sum of the P\&L of each desks. 

So, we suppose that the economic agent (the company or the financial institution) starts with an endowment in each submarket (BU or desks), trades separately in each submarket and that her total wealth is the sum of her wealth in each submarket. Note that we do not allow borrowing on a submarket against others. 
This is justified by the fact that a company does not allocate its wealth by betting on one BU against another or even that a bank has risk limits and cannot invest massively on a desk. 
Going back to the example of a one period model, we suppose that there are  two submarkets : the submarket $\tau_1$  with $d_1+1$ assets with price process 
$(\hat S^{\t_1}_{t\in\{0,1\}})$ and the submarket $\tau_2$  with $d_2+1$ assets with price process 
$(\hat S^{\t_2}_{t\in\{0,1\}})$. Then,  $(\hat \Phi^{\t_1}, \hat \Phi^{\t_2}) \in \R^{d_1+d_2+2}$  is an arbitrage in the global market if $\hat \Phi^{\t_1} .  \hat S^{\t_1}_0 \leq 0$ and $\hat \Phi^{\t_2} .  \hat S^{\t_2}_0 \leq 0$
and $\hat  \Phi^{\t_1} .  \hat S^{\t_1}_1 +\hat \Phi^{\t_2} .  \hat S^{\t_2}_1\geq 0$ $\P$-a.s. and 
$\P(\hat \Phi^{\t_1} .  \hat S^{\t_1}_1 +\hat \Phi^{\t_2} .  \hat S^{\t_2}_1>0) >0.$

Our purpose is to extend the FTAP and the Superhedging Theorem when no  tradable num\'{e}raire exists for the global market but when there is a tradable num\'{e}raire per submarket. 
Under the assumption that there is no arbitrage in the global market, there exist risk-neutral measures (or state price deflators) that depend on the submarket but are constructed from a common random variable (see Theorem \ref{main1}).
If we apply the classical FTAP separately to each submarkets (which are arbitrage free),  we obtain the existence of submarket-dependent risk-neutral measures, but we are not able to find some common factor between them. 
%, which  should appear because the submarkets are part of a global market. 
We find that in general, it is not possible to identify a common risk-neutral measure. Nevertheless, we prove in Section \ref{ecoillus} that this is true when the  spreads between the different submarkets are deterministic (and of course when there is a common num\'eraire). In this section, we also show by examples and by counter-examples that there is no relationship between  the completeness of all submarkets and that of the global market.

% Mathematically, this means that we can not use  a common  num\'{e}raire and the associated self-financing condition  in order to reduce the dimension of the problem.   

The initial motivation for this modelisation was the multicurve setting of the post-2007 interest rate market. 
Since 2007,  significant spreads are observed  between the zero-coupon (ZC) curves associated to different frequencies (or tenors) and the market practice for interest rate swap valuation has evolved and consider several ZC curves which are tenor based. This means  that each curve is build using instruments, like Forward Rate Agreement (FRA), with the same tenor. 
%Note that we are not here in the well studied case of a multicurrency markets  where we have to choose one currency and change others to this currency and then use the associated  num\'{e}raire to discount: here the different num\'{e}raires are used together to express the value of the portfolio.
In Section \ref{remtaux}, we show by a text-book no-arbitrage argument  that if 
one  assume the existence of ZC bonds traded in the whole market for all maturities, then there should not co-exist the FRA of different tenors. 
This justifies why one should not assume that there is a tradable num\'{e}raire available for the entire interest rate multicurve market.

Our model could also apply to other situations.  One may incorporate the risks implicit in interbank transactions like for example  liquidity risk or credit risk. 
 %as we can model the case where  several submarkets do not have the same liquidity or the same credit risk. 
 For that we assume that the different submarkets correspond to different classes of investments with the same level of liquidity or of credit spread.  The credit spread may be defined as the market unit that remunerates investors for the risk of default inherent in any debt instrument not considered to be risk free.  
Consider as a final example   the case of multi-currency markets. Usually, one specific currency is chosen, and the other currencies are expressed using this specific currency and then deflated using the associated num\'{e}raire.  
Our model is different since the num\'{e}raires associated with each currency are used jointly to express the value of the portfolio. Taking the example of the European Union, the German and French markets use their own num\'{e}raires, and there is a significant spread between their sovereign rates (OAT in France and Bund in Germany). A European financial institution has positions in both countries, each of which will be discounted using its own num\'{e}raire, and the institution will at the end of the day compute its wealth by summing its positions in both countries.

A natural question is how to price a contract $H$. The payoff of this contract may be associated to the activities of one or of several BUs or of financial products treated by several desks.   We define different superreplication prices that depend on the way the superhedging is performed. If one uses only the submarket $\tau,$ the price $\hat \pi^{\tau}(H)$ is the classical superreplication price in the market $\tau$. If one invests the initial wealth in the submarket $\tau$ and then uses all the other submarkets to hedge, the price is called $\pi^{\tau}(H)$. Now, if the initial investment and the hedging strategy use all submarkets, 
 the price is called $\pi(H)$.  We also introduce the minimum cost  $\underline{\pi}(H)$ (resp. $\overline{\pi}(H)$) for which it is possible to find one specific submarket where $H$ can be superreplicated using only assets from this submarket (resp. to superreplicate $H$ in all submarkets). 
 
This means that the same claim can have various superhedging prices depending on how it is hedged. 
So,  considering submarkets without the possibility to neither trade between them nor borrow on one again the others allows to  model assets having the same payoff but  different initial prices. 
In the example of a one period model with two arbitrage-free submarkets and $d_1=d_2=1,$  assume that $S_1^{\tau_1}=S_1^{\tau_2}.$ 
Then, one may have $S_0^{\tau_2}>S_0^{\tau_1}$. Clearly, considering that both assets can be traded together, there is an arbitrage opportunity. But if it is not possible to short in the market $\tau_2$ and be long in the market $\tau_1,$ there is no arbitrage in the global market. 
%This kind of feature is important in the multicurve model, see Remark \ref{remtaux}.  
Moreover, $S_0^{\tau_2}-S_0^{\tau_1}$ is a measure of the difference of features between submarkets $\tau_1$ and $\tau_2,$  for example a difference of liquidity.  
%For example, it may measure the difference of liquidity. Indeed, assume in a multiperiod model that the market $\tau_2$ is more liquid than $\tau_1.$ One may want to hedge the claim of the market $\tau_1$ with payoff $S_T^{\tau_1}$ in the market $\tau_2$. Let $S_0^{\tau_2}$ be the initial cost of the hedge of $S_T^{\tau_1}$ in the market $\tau_2$. Then, the cost of illiquidity, i.e., not being able  to trade $S^{\tau_1}$ in the market $\tau_2$ is measured by $S_0^{\tau_1}-S_0^{\tau_2}$. 

%{ \red This does not imply that there is some arbitrage AVOIR? and thus the risk that is accepted. These superreplication prices can also be used to price some asset of a one submarket in another one and  Recall that we are in a situation where there are tenor-specific risks, which implies that }
%Indeed depending considering a superhedge using all submarkets whether the different submarkets are used together (or not) to super hedge the contingent claim.
In Theorem \ref{main2}, we provide some inequalities and equalities between the different superreplication prices. We give lower and upper bounds for ${\pi}(H) $, but in the general case, we do not get an exact duality formulation. 
We show  that $\overline{\pi}(H) \geq \underline{\pi}(H) \geq \min_{\t \in \Tc} {\pi}^{\t}(H)\geq \pi(H)$. 
This proves that using all submarkets together to superreplicate $H$ yields the lowest price when requiring that superreplication holds true in all the submarkets separably yields the highest price. Taking the example of the credit risk markets, the price $\pi(H)$ corresponds to a hedge through markets with different credit risk levels as D assets while $\overline{\pi}(H)$ ensures a hedge in all markets and in particular with AAA  assets. 
%We show that the cheapest superreplication price for $H$ is $\pi(H)$ the one where the initial investment and the hedging strategy use all the submarkets. 
We propose in  Proposition \ref{propmax} and Section \ref{Examples} particular types of markets, where the duality formulation for $\pi(H)$ is exact. This is the case when the spreads between the different submarkets are deterministic and we find that $\pi(H)={\pi}^{\t_{max}}(H),$ where $\t_{max}$ is the submarket where the spread is maximum. 

When there are two submarkets $\tau_1$ and $\tau_2$ with one risky asset per submarket, we are able to fully compute in Proposition \ref{propdeux} the prices of  $S_T^{\tau_1}$ and 
$S_T^{\tau_2}$ and also of $S_T^{\tau_1}-S_T^{\tau_2},$ the instrument which allows to be short of the  asset of the submarket $\tau_2$  and to be long of the  asset of the submarket $\tau_1$. This instrument is like a Basis swap in the interest rate market. 
We find that $\pi(S_T^{\tau_1})$ is the infimum between  $\pi^{\tau_1}(S_T^{\tau_1})$ and $\pi^{\tau_2}(S_T^{\tau_1}).$ We also show that our measure of difference of features between submarkets $\tau_1$ and $\tau_2$ is indeed relevant as the initial investment(s) is (are) made in the submarket(s) where this measure equals zero.

Then, we  propose several economic illustrations. First, we show the appropriateness of our model by detailing the case of the multi-curve market. Then, we  prove that in the case of constant  num\'eraire spreads, a common martingale measure exists for the whole market. We also discuss completeness issues.  We finish with a Brownian illustration, where under the assumption of a time dependent Vasicek model for the num\'eraires spread, we provide a characterisation of the sets of martingale measures. 
When there are two sub-markets, each with only one risky asset, we fully calculate the different superreplication prices.

%This allows to have some insight on the existence of a martingale measure in the case of multi-curve term structure. Indeed in this case it is usually assumed (see \cite{GrbacRunggaldier16} and \cite{CuchieroFontanaGnoatto16} and the references therein)  that there exists a (unique) martingale measure, but this assumption is not justified by some adapted FTAP. In Example \ref{exple}we prove  that it is justified in the case of deterministic spreads.

The paper is structured as follows: Section \ref{s:notions} presents the framework and notations of the paper. In Section \ref{s:no-free-lunch}, we present the FTAP, while in Section \ref{s:surepli}, we study various superreplication prices depending on the choice of the submarket(s) allowed in the superreplicating portfolio. Section \ref{ecoillus} proposes some economic illustrations and we conclude in Section \ref{Conclu}. 
Finally, Section \ref{appendix} collects the proofs.

\section{The model}
\label{s:notions}
Let $\Tc$ be the finite set of all submarkets for a given global market. 
%By analogy to the multi-curve term structure model, we will call $\tau \in \Tc$ a submarket of tenor $\tau$. 
For all $\t \in \Tc$, the stochastic processes $S^{\t}=(S^{\t}_t)_{t \geq 0}$ and $S^{\t,0}=(S^{\t,0}_t)_{t \geq 0}$ are $\R^{d_{\t}}$ (for some given $(d_{\t})_{\t \in \Tc} \subset \N$) and $(0,\infty)$-valued, respectively,  have c\`ad-l\`ag trajectories and are adapted to a filtered probability space $(\Omega,\Fc,(\Fc_t)_{t \geq 0})$.
For all $\t \in \Tc$, we denote by $\tilde{S}^{\t}={S^{\t}}/{S^{\t,0}}$ the $\R^{d_{\t}}$-valued adapted price process of the risky assets of the submarket $\t$ deflated (or discounted) by the num\'eraire $S^{\t,0}$, which is  tradable only in the submarket $\tau$.
We note that $(S,S^0)=(S^{\t},S^{\t,0})_{\t \in \Tc}$ and $\tilde{S}=(\tilde{S}^{\t})_{\t \in \Tc}.$  We assume that the filtration $(\Fc_t)_{t \geq 0}$ satisfies the usual conditions (right continuous, $\Fc_0$ contains all null sets of $\Fc_{\infty}=\Fc$). 

We denote by $L^{1}=L^{1}(\Omega,\Fc,\P)$ the set of integrable, $\R$-valued and $\Fc$-adapted random variables, by $L_+^{1}=L_+^{1}(\Omega,\Fc,\P)$ the set of non-negative elements of $L^{1}$ and by $L_{>0}^{1}=L_{>0}^{1}(\Omega,\Fc,\P)$ the set of $X\in L^{1}$ such that $\P(X>0)=1$. The same notations apply for $L^{\infty}=L^{\infty}(\Omega,\Fc,\P)$: the set of essentially bounded, $\R$-valued and $\Fc$-adapted random variables. Moreover, $x\wedge y=\min(x,y).$ 

In order to prove our main results, we need to ensure that the wealth process belongs to $L^{\infty}$. So, we first make a technical assumption.
\begin{assumption}
\label{locbound}
For all $\t \in \Tc,$ the process $\tilde{S}^{\t}$ is locally bounded. 
\end{assumption}
Note that if  $\tilde{S}^{\t}$ is an adapted c\`ad-l\`ag process with uniformly bounded jumps, then it is locally bounded.

In the setup of the paper, the wealth process   can not be discounted as there is no common tradable num\'eraire.  As we will see in \eqref{stratricht} below, $S_t^{\t,0}$ appears in the expression of the wealth process at time $t,$ which needs to be bounded. So, we have to guarantee that $S^{\t,0}$ remains bounded by some real number $M.$  
We fix some real number $T_0>0$ which will be the time horizon of the market if all num\'eraires $S^{\t,0}$ are  essentially bounded. In this case, we define $M$ as the maximum of their norms. 
Otherwise, we will choose as time horizon for the market a finite stopping time $T$ guaranteeing that all the  $S^{\t,0}$ are bounded. 
Let $M$ be any fixed positive real number, $T^{\t}_M=\inf\{t>0|\, S^{\t,0}_t>M\}$ and $T= T_0 \wedge \inf\{T^{\t}_M|\t \in \Tc \}.$ 
Then, $T^{\t}_M$ and $T$ are stopping times. It is clear that $\P(\{T<\infty\})=1.$ We also remark that $S^{\t,0}_t \leq M$ on $\{t \leq T\}$ for all  $\t \in \Tc.$ 
%As $T\leq T_0$ and $\{T\leq T_0\} \subset \{T <\infty\}$, $\P(\{T<\infty\})=1.$ Let\rq{}s show that $S_T \leq M.$ \\
Indeed, fix some $\t \in \Tc.$ If $\o \in \{T^{\t}_M=\infty\},$ then $S^{\t,0}_t (\o)\leq M$ for all $t>0$ and, in particular,   if $t \leq T(\o)$. If now 
 $\o \in \{T^{\t}_M<\infty\},$ then $S^{\t,0}_{t} (\o) \leq M$ if $t \leq T(\o)\leq T^{\t}_M(\o).$  
 Then, for all  $\t \in \Tc,$ we assume  that ${S}^{\t}_t={S}^{\t}_T$ and  ${S}^{\t,0}_t={S}^{\t,0}_T$ on $\{t>T\}.$

In the sequel, in order to avoid to many fractions we also write $\bar S_T^{\t,0}=  S_T^{\t,0}/ S_0^{\t,0}$. Indeed, as there are several num\'{e}raires, there is no reason to suppose that  $S_0^{\t,0}=1$ for all $\t \in \Tc$.

%\begin{remark}
%The finite stopping time $T$ is the time horizon of the market.
%% and  play a central role, see Remark \ref{r:finite-stop}.
%We implicitly assume that ${S}^{\t}_t=0$ on $\{t>T\}.$
%If $\tilde{S}^{\t}$ and $S^{\t,0}$ are locally bounded, then Assumption \ref{locbound} holds true. In particular, Assumption \ref{locbound} holds true if $\tilde{S}^{\t}$ and $S^{\t,0}$ are adapted c\`ad-l\`ag processes with uniformly bounded jumps.
%\end{remark}
We will use simple trading strategies to define the no free lunch condition as in \cite{HK79}, \cite{K81} and \cite{HP81}. Using this kind of strategy, it is not possible to construct doubling strategies. 
%We denote by $\Sc_{T}$ the set of stopping times $\d$ such that $0\leq \d \leq T$.

\begin{definition}
\label{defstrat}
For all $\t \in \Tc$, a (simple) strategy in the submarket $\t$ is given by $\R^{d_{\t}}\times \R$-valued processes $(\Phi^{\t},\Phi^{\t,0})=(\Phi^{\t}_t,\Phi^{\t,0}_t)_{t \geq 0}$ of the form
\begin{eqnarray*}
%\label{strat}
\Phi^{\t}=\sum_{j=1}^{n_{\t}} \f_j^{\t} \one_{(\d_{j-1}^{\t},\d_j^{\t}]} & & \Phi^{\t,0}=\sum_{j=1}^{n_{\t}} \f_j^{\t,0} \one_{(\d_{j-1}^{\t},\d_j^{\t}]},
\end{eqnarray*}
where $n_{\t}\geq 1$ and $0=\d_0^{\t} \leq \d_1^{\t} \leq \ldots \leq \d_{n_{\t}}^{\t}<\infty$ are $n_{\t}+1$ finite stopping times  and $\f_{j}^{\t}$ (resp. $\f_{j}^{\t,0}$) are $\R^{d_{\t}}$ (resp. $\R$)-valued, $\Fc_{\d_{j-1}^{\t}}$-measurable random variables for all $j\in \{1,\ldots,n_{\t}\}$. We denote the global strategy by $(\Phi,\Phi^0)= (\Phi^{\t},\Phi^{\t,0})_{\t \in \Tc}$.
\end{definition}

\begin{definition}
\label{defautofi}
A strategy $(\Phi, \Phi^{0})$ is self-financing if for all $\t \in \Tc$ and all $j \in \{1,\ldots,n_{\t}\}$
\begin{eqnarray}
\label{autofo}
\Phi^{\t}_{\d_{j-1}^{\t}} S^{\t}_{\d_{j-1}^{\t}} + \Phi^{\t,0}_{\d_{j-1}^{\t}} S^{\t,0}_{\d_{j-1}^{\t}} =\Phi^{\t}_{\d_j^{\t}} S^{\t}_{\d_{j-1}^{\t}} + \Phi^{\t,0}_{\d_j^{\t}} S^{\t,0}_{\d_{j-1}^{\t}}.
\end{eqnarray}
\end{definition}
The economic agent trades in each submarket  separately. This is why the self-financing condition is given submarket by submarket and we do not define a global self-financing condition (i.e., summing on $\tau$ on the left- and right-hand sides of \eqref{autofo}). 
Nevertheless, the total wealth of the agent is the sum of her wealth in each submarket. \\
%as the agent has access to the global market, her wealth is the aggregated value of her wealth in each submarket. 
%Indeed, thinking of the multi-curve model, she trades separately in the 3-month submarket and in the 6-month submarket with specific num\'{e}raires and self-financed trading strategies, but at the end of the day, her global wealth is the sum of her wealth in the 3-month and the 6-month submarkets because she has, for example, to compute some global risk measures.
Let $(\Phi,\Phi^0)$ be a self-financing strategy. We denote by $V^{\Phi,\Phi^0}$ the $\R$-valued adapted wealth process obtained by summing the wealth of each submarket, i.e., on $\{t \leq  \d_{n_{\t}}^{\t}\}$
\begin{eqnarray}
\label{wealth-proc}
V_t^{\Phi,\Phi^0}= \sum_{\t \in \Tc} \left(\Phi^{\t}_t S^{\t}_t +\Phi^{\t,0}_t S^{\t,0}_t\right).
\end{eqnarray}
Note that on $\{t \geq T \},$ $V_t^{\Phi,\Phi^0}=V_T^{\Phi,\Phi^0}$. 
Using the self-financing condition, it is easy to see that on $\{t \leq  \d_{n_{\t}}^{\t}\}$
\begin{eqnarray}
\nonumber
V_t^{\Phi,\Phi^0} 
%& = & \sum_{\t \in \Tc}S^{\t,0}_t\left(\Phi^{\t}_t \tilde{S}^{\t}_t +\Phi^{\t,0}_t \right) \\
%\label{stratautofi}
& = & \sum_{\t \in \Tc}S^{\t,0}_{t }\left(\f^{\t}_0 \tilde{S}^{\t}_0 +\f^{\t,0}_0  + \sum_{j=1}^{n_{\t}}
\f^{\t}_j\left(\tilde{S}^{\t}_{\d_j^{\t} \wedge t}-\tilde{S}^{\t}_{\d_{j-1}^{\t} \wedge t} \right)\right)\\
\label{stratricht}
& = & \sum_{\t \in \Tc} x^{\tau} \bar S^{\t,0}_t   +  \sum_{\t \in \Tc} S^{\t,0}_{t }\sum_{j=1}^{n_{\t}}
\f^{\t}_j\left(\tilde{S}^{\t}_{\d_j^{\t} \wedge t}-\tilde{S}^{\t}_{\d_{j-1}^{\t} \wedge t} \right) = V_t^{x,\Phi},
\end{eqnarray}
where $x^{\tau}=\f^{\t}_0 {S}^{\t}_0 +\f^{\t,0}_0{S}^{\t,0}_0$ is the initial wealth in the submarket $\tau$ and $x=(x^{\tau})_{\t \in \Tc}$  the associated vector.  
%  $x=(x^{\tau})_{\t \in \Tc}$ is the vector of initial wealth. 
%Therefore, if we assume that $(\Phi,\Phi^0)$ are given and the self-financing condition holds true, setting $x^{\tau}=\f^{\t}_0 {S}^{\t}_0 +\f^{\t,0}_0{S}^{\t,0}_0$, we obtain that$V_t^{\Phi,\Phi^0} =V_t^{x,\Phi}.$ 
Conversely, starting from some vector of initial wealth $x=(x^{\tau})_{\t \in \Tc}$ and some strategy in the risky assets $\Phi$, it is possible, submarket by submarket using the self-financing condition, to construct some $\Phi^{\t,0}$ such that  on $\{t \leq  \d_{n_{\t}}^{\t}\}$
\begin{eqnarray}
 \label{autofirichini}
 S^{\t,0}_t\left(\frac{x^{\tau}}{{S}^{\t,0}_0}  + \sum_{j=1}^{n_{\t}}
\f^{\t}_j\left(\tilde{S}^{\t}_{\d_j^{\t}\wedge t}-\tilde{S}^{\t}_{\d_{j-1}^{\t}\wedge t } \right)\right)=\Phi^{\t}_t {S}^{\t}_t +\Phi^{\t,0}_t S^{\t,0}_t
\end{eqnarray}
and $x^{\tau}=\f^{\t}_0 {S}^{\t}_0 +\f^{\t,0}_0{S}^{\t,0}_0$. This implies that
$V_t^{x,\Phi}= V_t^{\Phi,\Phi^0}.$ We see in this argument that the assumption that all the  $S^{\t,0}$ are tradable is crucial. 
%\begin{remark}
%\label{r:finite-stop}
%In the usual set-up, i.e., when $\Tc=\{\hat \tau \}$, using the discounted wealth
%$V_T^{x,\Phi} /S_T^{\hat \tau,0}$ given by \eqref{stratrich} allows us to reduce the dimension of the problem, as the strategy $\Phi^{\hat \tau,0}$ in the num\'{e}raire no longer appears. This is one ingredient of the proof of the FTAP and of the superhedging theorem. Here, we do not assume the existence of a num\'{e}raire $B$ for the global market to discount $V^{\Phi,\Phi^0}$ in \eqref{wealth-proc}. This is why we cannot use a discounted wealth process to eliminate the strategies $\Phi^{\t,0}$ in the  num\'{e}raires and instead have to use the expression given in  \eqref{stratrich}. %%EDITOR'S NOTE: Please ensure that the intended meaning has been maintained in this edit. Alternatively, consider "wealth process to identify the strategies" or "replace the strategies".
%
%%As usual we use the self-financing condition \eqref{autofo} in order to reduce the dimension of the problem and to express the wealth process without using the strategies in the num\'eraire assets. Due to simultaneous presence of multiple num\'eraire, we cannot use a discounted wealth process. That is why we multiply by $S_T^{\tau, 0}$, for all $\tau \in \mathcal{T}$ and need a fixed finite stopping time $T$ for trading horizon.
%\end{remark}
Note that on $\{T >\d_{n_{\t}}^{\t}\},$ we add one step to the strategy ending at the stopping time $\d_{n_{\t}}^{\t}:$ $\Phi_T^{\t}=0$ and using the self-financing condition, we have that 
\begin{eqnarray*} \Phi_T^{\t,0} & = & \Phi_{\d_{n_{\t}}^{\t}}^{\t,0} + \left(\Phi_{\d_{n_{\t}}^{\t}}^{\t}-\Phi_T^{\t} \right)\frac{S_{\d_{n_{\t}}^{\t}}^{\t}}{S_{\d_{n_{\t}}^{\t}}^{\t,0}}=
\frac{\Phi_{\d_{n_{\t}}^{\t}}^{\t,0}S_{\d_{n_{\t}}^{\t}}^{\t,0}+ \Phi_{\d_{n_{\t}}^{\t}}^{\t}S_{\d_{n_{\t}}^{\t}}^{\t}}{S_{\d_{n_{\t}}^{\t}}^{\t,0}} \\
 & =& \frac{x^{\t} \bar S_{\d_{n_{\t}}^{\t}}^{\t,0}  +  S_{\d_{n_{\t}}^{\t}}^{\t,0}\sum_{j=1}^{n_{\t}}
\f^{\t}_j\left(\tilde{S}^{\t}_{\d_j^{\t} }-\tilde{S}^{\t}_{\d_{j-1}^{\t} }\right)}{S_{\d_{n_{\t}}^{\t}}^{\t,0}}.
\end{eqnarray*}
Thus, we obtain that
\begin{eqnarray}
\label{stratrich}
V_T^{x,\Phi} & = & \sum_{\t \in \Tc}(\Phi_T^{\t,0} S^{\t,0}_T + \Phi_T^{\t} S^{\t}_{T } ) 
=  \sum_{\t \in \Tc} x^{\tau} \bar S^{\t,0}_T  + \sum_{\t \in \Tc}S^{\t,0}_{T } \sum_{j=1}^{n_{\t}}
\f^{\t}_j\left(\tilde{S}^{\t}_{\d_j^{\t} }-\tilde{S}^{\t}_{\d_{j-1}^{\t} } \right). 
\end{eqnarray}
The same formula holds true on $\{T \leq \d_{n_{\t}}^{\t}\}$  using \eqref{stratricht}, because  $\tilde{S}^{\t}_{\d_j^{\t} \wedge T}=\tilde{S}^{\t}_{\d_j^{\t} }$ on $\{\d_j^{\t}<T\}$ and $\tilde{S}^{\t}_{\d_j^{\t} \wedge T}=\tilde{S}^{\t}_{T }=\tilde{S}^{\t}_{\d_j^{\t} }$ on $\{\d_j^{\t}\geq T\}.$ \\
\begin{definition}
\label{defadmi}
A strategy $(\Phi,\Phi^0)$ as in Definition \ref{defstrat} is called admissible if  the processes $ (\tilde{S}^{\t}_{t\wedge\d_{n_{\t}}^{\t}})_{t>0}$ and the random variables $(\f_{j}^{\t},\f_{j}^{\t,0})_{j \in \{1,\ldots,n_{\t}\}}$ are uniformly  bounded for all
$\t \in \Tc$ and if the initial wealth in each submarket is non negative, i.e., $x^{\tau}=\f^{\t}_0 {S}^{\t}_0 +\f^{\t,0}_0{S}^{\t,0}_0\geq 0$ for all
$\t \in \Tc$. 
\end{definition}
The assumption that $x^{\tau}$ is non-negative  is made because it is not permitted to borrow on a submarket against the others. 

We introduce now the set of contingent claims available at the stopping time $T$ at cost zero (i.e., $x^{\tau}=\f^{\t}_0 {S}^{\t}_0 +\f^{\t,0}_0 {S}^{\t,0}_0=0$) via an admissible self-financing simple strategy using only the submarket $\t \in \Tc$
\begin{eqnarray*}
%\nonumber
K^{\t} & = &
\left\{ S^{\t,0}_T \left(\sum_{j=1}^{n_{\t}}
\f^{\t}_j \left(\tilde{S}^{\t}_{\d_j^{\t}}-\tilde{S}^{\t}_{\d_{j-1}^{\t}} \right) \right)| \; 
n_{\t} \geq1, \;0=\d_0^{\t} \leq \ldots \leq \d_{n_{\t}}^{\t} <\infty,  \right.\\
& & \left.  (\tilde{S}^{\t}_{t\wedge\d_{n_{\t}}^{\t}})_{t>0}  \mbox{uniformly bounded,} \;(\f_{j}^{\t})_{j \in \{1,\ldots,n_{\t}\}}
%\label{defKtau}   \\ & & \R^{d_{\t}}\mbox{-valued, } \left.
 \Fc_{\d_{j-1}^{\t}} \mbox{-adapted and bounded}
\right\}
\end{eqnarray*}
which belongs to  $L^{\infty}$ using Assumption \ref{locbound} and the definition of $T$.

We denote by $K$ the set of contingent claims available at the stopping time $T$ at total cost zero via an admissible self-financing simple strategy 
%satisfying $\f^{\t}_0 {S}^{\t}_0 +\f^{\t,0}_0{S}^{\t,0}_0$ non-negative and thus equal to zero for each $\tau$
\begin{eqnarray}
  \label{defK}
K & = & \left\{V_T^{\Phi,\Phi^0}|\,  \Phi, \Phi^0, \; \f^{\t}_0 {S}^{\t}_0 +\f^{\t,0}_0{S}^{\t,0}_0 = 0\; \forall \tau \right\} =  
\left\{
\sum_{\t \in \Tc} W^{\t}| \, W^{\t} \in K^{\t}  \right\}
\end{eqnarray}
which belongs to  $L^{\infty}$ using  again Assumption \ref{locbound} and the definition of $T$. Note that it is equivalent to assume that  $\f^{\t}_0 {S}^{\t}_0 +\f^{\t,0}_0{S}^{\t,0}_0=0$  for all $\t \in \Tc$ or that  $\sum_{\t \in \Tc} (\f^{\t}_0 {S}^{\t}_0 +\f^{\t,0}_0{S}^{\t,0}_0)=0$ because the admissibility condition requires 
%that the investor is not allowed to borrow on a specific submarket against the others and thus 
$\f^{\t}_0 {S}^{\t}_0 +\f^{\t,0}_0{S}^{\t,0}_0 \geq 0$ for all $\t \in \Tc$. 
%We are in the situation where the agent separately trades in the different submarkets and begins with a vector $(x^{\tau})_{\tau \in \Tc}$ of initial wealth. We assume that the investor is not allowed to borrow on some specific tenors against the others and  that each $x^{\tau}=\f^{\t}_0 {S}^{\t}_0 +\f^{\t,0}_0{S}^{\t,0}_0$ is non-negative. 
%Therefore, this is equivalent to assuming that $\sum_{\t \in \Tc} x^{\tau}=0$ or that $x^{\tau}=0$ for all $\tau.$ 
We finally define
$$C=K -L^{\infty}_+=\{V-Z| \; V \in K, Z  \in L_+^{\infty}\}.$$

\section{Fundamental theorem of asset pricing}

\label{s:no-free-lunch}
The classical no arbitrage (NA) condition for simple strategies stipulates that $K \cap   L^{\infty}_+ =\{0\}$, and this is equivalent to
$C \cap   L^{\infty}_+ =\{0\}$. It is well known that in continuous-time models, stipulating the (NA) condition is insufficient to obtain an equivalent local martingale measure (see, for example, Proposition 5.1.7 in \cite{DelSch05}). Therefore, following \cite{K81}, we use the no free lunch (NFL) condition.
\begin{definition}
\label{defNFL}
The process $(S,S^0)=(S^{\t},S^{\t,0})_{\t \in \Tc}$ satisfies the no free lunch (NFL) condition if
$$\bar{C} \cap L^{\infty}_+=\{0\},$$
where $\bar{C}$ is the closure of $C$ taken with respect to the weak-star topology of $L^{\infty}$.
\end{definition}
Definition \ref{defNFL} ensures the NFL condition for the global market. %%EDITOR'S NOTE: Please ensure that the intended meaning has been maintained in this edit. Alternatively, consider "We assume no free lunch", "We seek to ensure that there is no free lunch", or "Our objective is to ensure the NFL condition for the global market".
It is clear from the definition that this implies the NFL in each submarket $\tau$, i.e.,
$$\bar{C}^{\tau} \cap L^{\infty}_+=\{0\} \mbox{  with } C^{\tau}=K^{\tau} -L^{\infty}_+.$$
The reverse implication does not hold true~: there may be a free lunch in the global market while the different submarkets satisfy the NFL condition. 
%the different submarkets   can be arbitrage free, and the market considered as global  market may display a free lunch. For example in the multi-curve setup,  the 3-month and the 6-month markets considered separately are arbitrage free, but there may be some arbitrage when considering both markets as a unique market without any constraints, see the arbitrage between Zero-Coupon bond in the 3-month tenor with maturity  6 months and the one with the same maturity in the 6-month tenor in Remark \ref{remtaux}. 
Consider a one period market with finite $\Omega = \{\o_1,\o_2,\o_3\}$ and two submarkets $\tau_1$ and $\tau_2$ each with one risky asset such that  
$ S^{\tau_1}_0= S^{\tau_2}_0=0,$  $ S^{\tau_1}_1=-\one_{\o_2}+\one_{\o_3}$ and $ S^{\tau_2}_1=\one_{\o_1}+\one_{\o_2}-\one_{\o_3},$ where  
$\one_{\o_i}(\o)=1$ if $\o=\o_i$ and zero else. 
The num\'eraires are such that $S^{\t_1,0}_0=S^{\t_2,0}_0=1$ and $S^{\t_1,0}_1=\frac32$ and $S^{\t_2,0}_1=1.$ 
Then, both submarkets satisfy the NFL condition. However, the global market admits an arbitrage : if we buy one unit of asset $ S^{\tau_1}$ and 
 of asset $ S^{\tau_2}$, the initial cost is $x=0$ and 
$$V_1=S^{\t_1,0}_1 \left(\frac{S^{\tau_1}_1}{S^{\t_1,0}_1}-0\right)+ S^{\t_2,0}_1  \left(\frac{S^{\tau_2}_1}{S^{\t_2,0}_1}-0\right)=\one_{\o_1}.$$

\begin{definition}
%Let $\Xc^*$ be the set of $X^* \in L_{>0}^1$ such that for all  $\t \in \Tc$, all stopping times $ \b_1 \leq \b_2 \leq T$ and all $\R^{d_{\t}}$-valued, $\Fc_{\b_1}$-measurable random variables $\f,$
%\begin{eqnarray}
%\label{condmart}
%\E\left(X^*S^{\t,0}_T \f \left(\tilde{S}^{\t}_{\b_2}-\tilde{S}^{\t}_{\b_1} \right)\right)=0.
%\end{eqnarray}
We denote by $\Xc^{\t}$ the set of $X^* \in L_{>0}^1$ such that for all stopping times $ \b_1 \leq \b_2 \leq T$   and all $\R^{d_{\t}}$-valued, $\Fc_{\b_1}$-measurable random variables $\f$,
\begin{eqnarray}
\label{condmart}
%\label{condmarttau}
\E\left(X^*S^{\t,0}_T \f \left(\tilde{S}^{\t}_{\b_2}-\tilde{S}^{\t}_{\b_1} \right)\right)=0.
\end{eqnarray}
\end{definition}
We set $\Xc^* =\cap_{\t \in \Tc} \Xc^{\t}$.
\begin{theorem}
\label{main1}
Under Assumption \ref{locbound}, $(S,S^0)$ satisfies the no free lunch (NFL) condition if and only if $\Xc^* \neq \emptyset$.
\end{theorem}
\begin{proof} See Section \ref{preuve1}. $\Box$\end{proof}

As NFL condition implies NFL in each submarket $\tau$ for all
$\t \in \Tc$, the classical Kreps-Yan theorem (see \cite{K81}, \cite{Y80}) shows that $\Xc^{\t} \neq \emptyset$. Theorem \ref{main1} also proves that this holds true since $\Xc^*  \subset \Xc^{\t}$ for all
$\t \in \Tc$. However, starting from the fact that  $\Xc^{\t} \neq \emptyset$ for all
$\t \in \Tc$, Kreps-Yan theorem does not show that the global no free lunch holds true or that  there exists some $X^*$ belonging to all $\Xc^{\t}$.\\
%From Theorem \ref{main1} it is clear that if $(S,S^0)$ satisfies the  no free lunch (NFL) condition then

One may wonder whether the fact that each submarket $\tau$  is complete implies that the global market is complete and vice versa. In  Section \ref{complet}, we will precise 
the notion of completeness for the global market and give examples and counter-examples for both implications.
% must be clarified because $\Xc^* $ cannot be a singleton (if $X^* \in \Xc^*$, then $\l X^* \in \Xc^*$ for all $\l>0$). We will discuss this point in Section \ref{complet}. 

\section{Martingale measures}
The condition $\Xc^* \neq \emptyset$ is equivalent to the existence for all  $\t \in \Tc$ of an equivalent local martingale 
\footnote{We obtain local martingale instead of true martingale because the $\tilde{S}^{\t}$ are locally bounded.}
% and not continuous processes.} 
(or risk-neutral) measure $\Q^{\t}$ for the process $\tilde{S}^{\t}$ (i.e., $\tilde{S}^{\t}$ is a local martingale under $\Q^{\t}$ and $\Q^{\t} \sim \P$). Let 
$X^* \in \Xc^*.$ Those  martingale measures are defined by
\begin{eqnarray}
\label{Qtau}
\frac{d\Q^{\t}}{d\P}\Big|_{\Fc}=\frac{X^*S^{\t,0}_T}{\E(X^*S^{\t,0}_T)}.
\end{eqnarray}
In the financial literature, a state price deflator is a process $(D_t)_{t>0}$ such that $(D_t {S}_t^{\t})_{t>0}$ is a local martingale under the initial probability measure $\P$. Here, $(D^{\t}_t)_{t>0}$ is a state price deflator, where
$$D^{\t}_t=
\frac{\E\left(X^* S^{\t,0}_T| \Fc_t
\right)}{S^{\t,0}_t}. 
$$
Therefore, Theorem \ref{main1} shows that the martingale measures or the state price deflators are  dependent of the submarket but constructed from a common random variable $X^*$.

We now study the problem called \lq{}\lq{}change of num\'{e}raire\rq\rq{}. We use \lq{}\lq{} \rq\rq{} to emphasize that   the word \lq{}\lq{}num\'{e}raire\rq\rq{}  in this paragraph has not the same meaning as in the rest of the paper. Indeed, a \lq{}\lq{}num\'{e}raire\rq\rq{} $Z$ does not need to be tradable as $S^{\tau,0}$ is assumed to be. 
%In the context of a multi-currency framework, one usually takes for each submarket (or each country) the num\'{e}raire to be a savings account (or zero-coupon bond) expressed in the currency of the country. %%EDITOR'S NOTE: Please ensure that the intended meaning has been maintained in this edit.
%One might change the num\'{e}raire by changing to the currency of another country. 
It is sometimes convenient to change the \lq{}\lq{}num\'{e}raire\rq\rq{} because of modelling considerations. Indeed, when we change the \lq{}\lq{}num\'{e}raire\rq\rq{}, we change the risk-neutral measures, which may simplify the model, facilitating calculations of prices. 
%This is in particular the case using the so-called zero-coupon bond as a num\'{e}raire with the associated forward measures since forward prices are martingales under forward measures.
Therefore, we now propose to characterize the NFL condition using new \lq{}\lq{}num\'{e}raires\rq\rq{}. 
% (which may be based on the tenor specific num\'{e}raires as linear combination or minimum or maximum of them). This allows to generalize  classical results on FTAP  for the interest rate markets.
Proposition  \ref{propnum} will be used to prove our second main result on pricing issues.
\begin{definition}
\label{defQZ}
%Let $L_{>0}^{\infty}$ be the set of $Z\in L^{\infty}$ such that $\P(Z>0)=1$.
Let $Z \in L_{>0}^{\infty}$,  $\Q \in \Qc^{\tau,Z}$ if and only if $\Q \sim \P$,
${\frac{d\Q}{d\P}}/{Z} \in L^1$ and
$$\left(
\E_{\Q}\left(
\frac{
\bar S^{\t,0}_T
}
{
Z
}
| \Fc_t
\right)
\tilde{S}_t^{\t}
\right)_{t \geq 0} \mbox{ is a  local martingale under $\Q$.} $$
We  set $\Qc^{Z}=\cap_{\tau \in \Tc} \Qc^{\tau,Z}.$
\begin{itemize}
\item Let $\hat{\t} \in \Tc.$ We denote by $\hat \Qc^{\hat{\t}}=\Qc^{\hat{\t},\bar S^{\hat \t,0}_T}$ the usual uni-market set of local  martingale measures for $ \tilde S^{\hat{\t}}$ and we denote by   $ \Qc^{\hat{\t}}=\Qc^{\bar S^{\hat \t,0}_T}.$ Then, $ \Qc^{\hat{\t}} \subset \hat  \Qc^{\hat{\t}}.$
%\item For $Z=\sum_{\t \in \Tc} {S^{\t,0}_T}/{S^{\t,0}_0}$,   we denote  $\Qc^{sum}=\Qc^{Z}$.
\item For $Z=\sum_{\t \in \Tc} \l_{\t}\bar S^{\t,0}_T$ where $\l=(\l_{\t})_{\t \in \Tc}  \in \Lambda^{\Tc},$ we denote  $\Qc^{lc, \l}=\Qc^{Z}$ where
$$\Lambda^{\Tc}= \left\{(\l_{\t})_{\t \in \Tc} \subset \R^+| \; \P\left(\sum_{\t \in \Tc} \l_{\t}\bar S^{\t,0}_T>0\right)=1 \right\}.$$
\item For $Z=\max_{\t \in \Tc} \bar S^{\t,0}_T$, we denote $\Qc^{max}=\Qc^{Z}$.
\item For $Z=\min_{\t \in \Tc} \bar S^{\t,0}_T$, we denote $\Qc^{min}=\Qc^{Z}$.
\end{itemize}
\end{definition}

%=\{\Q \sim \P, \,
%\frac{d\Q}{d\P} \in L^1, \, \left(
%\E_{\Q}\left(
%\frac{S^{\t,0}_T}{S^{\t,0}_0}
%| \Fc_t
%\right)
%\tilde{S}_t^{\t}
%\right)_{t \geq 0} \mbox{ is a local martingale under $\Q$\}$$
\begin{proposition}
\label{propnum}
Assume that Assumption \ref{locbound} holds true.\\
{\bf 1.} If $(S,S^0)$ satisfies the NFL condition, then $\Qc^{Z}$ is not empty for all $Z \in L_{>0}^{\infty}$.
Conversely, if $\Qc^{Z}$ is not empty for some $Z  \in L_{>0}^{\infty}$,
then $(S,S^0)$ satisfies the  NFL condition.\\
{\bf 2.} Fix $Z  \in L_{>0}^{\infty}.$ Then, we get that  
\begin{eqnarray}
\label{enfin}
\Qc^{Z} & = & \left\{\Q|\, \exists X^* \in  \Xc^*,\; \frac{d\Q}{d\P}=
\frac{
X^*Z
}
{
\E \left(
X^* Z
\right)
}\right\}.
\end{eqnarray}
\end{proposition}
\begin{proof} See Section \ref{preuve2}.  $\Box$\end{proof}
\begin{remark}
\label{remuntenor}
Let
$\hat{\t} \in \Tc$. Then, as in Proposition \ref{propnum} or using Kreps-Yan Theorem, $(S,S^0)$ satisfies the NFL condition in the submarket  $\hat{\t}$ if and only if $\hat \Qc^{\hat{\t}}\neq \emptyset$. Moreover, 
$\hat \Qc^{\hat{\t}} =\left\{\Q|\, \exists X^{\hat{\t}} \in  \Xc^{\hat{\t}},\; \frac{d\Q}{d\P}=
\frac{
X^{\hat{\t}}\bar S^{\hat \t,0}_T
}
{
\E \left(
X^{\hat{\t}} \bar S^{\hat \t,0}_T
\right)
}\right\}.$
%Assume that  $\Xc^{\hat{\t}}\neq \emptyset.$  Let $X^* \in \Xc^{\hat{\t}}$ and $Z \in L_{>0}^{\infty}.$ Let  $\Q_Z$ be defined by
%$
%{d\Q_Z}/{d\P}=
%{
%(X^*Z)
%}/
%{
%\E \left(
%X^* Z
%\right)
%}.$
%Then, $\Q_Z \in \Qc^{\hat \tau,Z}.$ Moreover, if $\Qc^{\hat \tau,Z}\neq \emptyset$ for some $Z  \in L_{>0}^{\infty}$, then
%$\E_{\Q}\left(\frac{W}{Z}\right)  \leq 0 \mbox{ for all } W \in \bar{C}^{\t} \cap L^{\infty}, \; \Q \in \Qc^{\hat \tau,Z}.$
%The proof is similar to the proof of Proposition \ref{propnum} and thus omitted.
\end{remark}

\section{Superhedging theorem}
\label{s:surepli}

\subsection{Definitions}

We now turn to pricing issues of some contingent claim $H$. In the classical case, the superreplication cost for $H$ is the minimum selling price needed to superreplicate it by trading in the (uni-market) market.
In complete markets, the superreplication cost is just the cash flow expectation computed under the unique (local) martingale measure. However, in incomplete markets, this is no longer the case, as the risk-neutral measure is no longer unique.

Here, as we have several submarkets that coexist, several choices of superhedging are possible.
The first definition of price assumes that all submarkets are used to cover $H$.
%is given for the price of $H$ is the minimum value of a strategy using all markets of tenor $\tau$ which allows to cover $H$.
Specifically, the initial wealth $x$ is divided among the different submarkets. Then, in each submarket  $\tau$ starting from initial wealth $x^{\t}$, one can find some hedging strategy (using only the instruments of submarket $\tau$) such that when liquidating all those portfolios, one is in a position to superreplicate $H$. Recalling \eqref{stratrich} and \eqref{defK}, this can written as follows
\begin{eqnarray}
\label{prixsurep1}
\pi(H) & = & \inf \left\{\sum_{\t \in \Tc} x^{\t}| \; x^{\t} \geq 0, \exists  W \in \bar{C}\cap L^{\infty}, \sum_{\t \in \Tc} x^{\t}\bar S^{\t,0}_T+W \geq  H \, \mbox{a.s.} \right\}
\end{eqnarray}
with the convention that $\pi(H)=+ \infty$ if the above set is empty. The same convention applies to the prices defined in \eqref{prixsurep2} to \eqref{prixsureptau}.
%, \eqref{prixsurep3} and \eqref{prixsureptau}.
%{\red Note that instead of the condition $x^{\t} \geq 0$ in \eqref{prixsurep1}, one may assume that $x^{\t} \geq -b$ for some fixed $b>0$. Non car alors on n a plus $K=\sum K^{\t}$.}

We can think of the minimum costs of other types of hedging strategies. First, we consider the case in which it is possible to find one specific submarket where $H$ can be superreplicated using only assets from this submarket
\begin{eqnarray}
\label{prixsurep2}
\underline{\pi}(H) & = & \inf \left \{ x| \; x \geq 0, \exists  \t \in \Tc, \, \exists W \in \bar{C}^{\t}\cap L^{\infty}, x\bar S^{\t,0}_T+W \geq   H \, \mbox{a.s.}\right\}.
\end{eqnarray}
Recall that  $C^{\t}=K^{\t} -L^{\infty}_+.$

We can also consider the minimal initial cost for which it is possible to superreplicate $H$ in all submarkets, i.e., that for each submarket, $H$ can be superreplicated using only assets from this submarket
\begin{eqnarray}
\label{prixsurep3}
\overline{\pi}(H) & = & \inf \left \{ x| \; x \geq 0, \forall  \t \in \Tc, \, \exists W \in \bar{C}^{\t}\cap L^{\infty}, x\bar S^{\t,0}_T+W \geq   H \, \mbox{a.s.} \right\}.
\end{eqnarray}
Finally, we introduce prices which are related to some specific submarket $\t,$ for any  $\t \in \Tc$.  
The superreplication price ${\pi}^{\t}(H)$ is obtained investing  the entire initial wealth  in the submarket $\t$ but  trading  in the other submarkets too
\begin{eqnarray}
\label{prixsureptautout}
{\pi}^{\t}(H) & = & \inf \left \{ x| \, x \geq 0, \, \exists W \in \bar{C}\cap L^{\infty}, x\bar S^{\t,0}_T+W \geq  H \, \mbox{a.s.} \right\}.
\end{eqnarray}
It is clear that ${\pi}^{\t}(H) \geq {\pi}(H).$ We introduce the classical superreplication price of $H$ in the submarket  $\tau$
\begin{eqnarray}
\label{prixsureptau}
\hat{\pi}^{\t}(H) & = & \inf \left \{ x| \, x \geq 0, \, \exists W \in \bar{C}^{\t}\cap L^{\infty}, x\bar S^{\t,0}_T+W \geq  H \, \mbox{a.s.} \right\}.
\end{eqnarray}
The superreplication price $\hat{\pi}^{\t}(H)$ in the submarket $\t$ alone is higher than $\pi^{\t}(H)$: even if  the entire initial wealth $\pi^{\t}(H)$ is invested  in the submarket $\t$, trading may also take place in the other submarkets. This is not the case for $\hat{\pi}^{\t}(H)$, where all the trading activity is made in the submarket $\t$. 

It is clear that the same claim may have various prices. This is because they are associated with different choices of initial endowments and/or hedging strategies  and that one cannot trade between submarkets.   
We prove in Theorem \ref{main2} below that $\overline{\pi}(H) \geq \underline{\pi}(H) \geq \min_{\t \in \Tc} {\pi}^{\t}(H)\geq \pi(H)$.
This shows that using all submarkets together to superreplicate $H$ yields the lowest price when requiring that superreplication holds true in all the submarkets separably yields the highest price. This is intuitive, as the last alternative ensures the most certitude and therefore costs more. Taking the example of the credit risk markets, the price $\pi(H)$ corresponds to a hedge through markets with different credit risks while $\overline{\pi}(H)$ ensures a hedge in all markets and in particular with ``sure" assets having a low credit risk. 
%This shows that the lowest way to superreplicate $H$ is to use all the tenors and  justifies the choice of $ \pi(H)$ for the price of $H$.

We show that $\overline{\pi}(H)$ corresponds to the maximum of all the superreplication prices $\hat{\pi}^{\t}(H)$, while $\underline{\pi}(H)$ corresponds to the minimum of $\hat{\pi}^{\t}(H)$.
Note also that the following classical dual representation holds true
\begin{eqnarray}
\label{pitau}\hat{\pi}^{\t}(H) = \sup_{\Q \in \hat \Qc^{\tau}} \E_{\Q} \left(\frac{H}{\bar S^{\t,0}_T}\right).
\end{eqnarray}
This well-known result may be proven as below (recall that the NFL condition implies the NFL condition in every submarket  $\tau$). 

With those definitions, we can measure the difference of features between  the submarket $\t$ stand alone and the submarket $\t$ as part of the global market: $\hat{\pi}^{\t}(H)-{\pi}(H)$ measures the gain to invest and to trade in the whole market and not only in $\t$, while $\hat{\pi}^{\t}(H)-{\pi}^{\t}(H)$ measures the gain to trade in the whole market and not only in $\t$, while investing all the initial wealth in $\t$.  
%We now focus on the cost of liquidity for risky assets. 
%Assume again that the market $\tau_2$ is more illiquid. One may want to hedge the claims of the market $\tau_2$ with payoff $S_T^{\tau_2}=(\tilde{S}_T^{\tau,j})_{j \in \{1,\ldots, d_{\tau}\}}$ using also the other markets, which are more liquid. The cheapest hedge is to use all the markets together and  the cost of illiquidity for $\tilde{S}_T^{\tau,j}$ is measured by $\hat{\pi}^{\t_2}(S_T^{\tau_2,j})-\pi(S_T^{\tau_2,j})=S_0^{\tau_2,j}-\pi(S_T^{\tau_2,j})$.  
%So, using \eqref{carctprixsurep1bishat}, we get the following inequality
%\begin{eqnarray*}
%S_0^{\tau_2}-\sup_{\Q \in \Qc^{max}} \E_{\Q} \left(\frac{S_T^{\tau_2}}{\max_{\t \in \Tc} \bar S^{\t,0}_T}\right)
%\leq S_0^{\tau_2}- \pi(S_T^{\tau_2}) \leq S_0^{\tau_2}-\sup_{\Q \in \Qc^{min}} \E_{\Q} \left(\frac{S_T^{\tau_2}}{\min_{\t \in \Tc} \bar S^{\t,0}_T}\right). 
%\end{eqnarray*}
Below and in Section \ref{ecoillus}, we will compute those differences of features for several types of markets. 
\begin{remark}
In the different definitions of the superreplication prices, we take the intersection of $\bar{C}^{\t}$ or $\bar C$ with $L^{\infty}$ so that the infima are minima.
\end{remark}
%; see Remark \ref{AOAtenor}).

\subsection{Dual Characterization}

We now state our second main result on the dual formulation of the superreplication prices, called the superhedging theorem.
\begin{theorem}
\label{main2}
Assume that Assumption \ref{locbound} and that the NFL condition hold true. \\
Assume that $\frac{H}{\sum_{\t \in \Tc} \l_{\t}\bar S^{\t,0}_T} \in L^{\infty}$ for some $(\l_{\t})_{\t \in \Tc}  \in \Lambda^{\Tc}$.
Then, the infimum in \eqref{prixsurep1} is attained, and there exists $(\hat{x}^{\t})_{\t \in \Tc}$ such that $\hat{x}^{\t} \geq 0$ for all $\t \in \Tc$ and
$$\pi(H) = \sum_{\t \in \Tc} \hat{x}^{\t}.$$  Moreover, the following holds true:
%for any $(\l_{\t})_{\t \in \Tc}  \in \Lambda^{\Tc}$
\begin{eqnarray}
\label{prixrepqsumhat}
\sup_{ \Q \in   \Qc^{lc, \l}}
\E_{\Q} \left( \frac{ H}{ \sum_{\t \in \Tc} \l_{\t}\bar S^{\t,0}_T}
-
\sum_{\t \in \Tc} \hat{x}^{\t} \frac{\bar S^{\t,0}_T}
{\sum_{\t \in \Tc} \l_{\t}\bar S^{\t,0}_T }
\right)  =0.
 \end{eqnarray}
Now, if $\frac{H}{\min_{\t \in \Tc} \bar S^{\t,0}_T}  \in L^{\infty}$, we have that
\begin{eqnarray}
\label{carctprixsurep1bishat}
%\pi_{inf}(H):=
\sup_{\Q \in \Qc^{max}} \E_{\Q} \left(\frac{H}{\max_{\t \in \Tc} \bar S^{\t,0}_T}\right)
\leq \pi(H) \leq \sup_{\Q \in \Qc^{min}} \E_{\Q} \left(\frac{H}{\min_{\t \in \Tc} \bar S^{\t,0}_T}\right).
%:=\pi_{sup}(H).
\end{eqnarray}
We also obtain that
\begin{eqnarray}
\label{carctprixsurep2hat}\underline{\pi}(H)  & = & \min_{\t \in \Tc} \hat{\pi}^{\t}(H) =\min_{\t \in \Tc}  \sup_{\Q \in \hat \Qc^{\tau}} \E_{\Q} \left(\frac{H}{\bar S^{\t,0}_T}\right)\\
\label{carctprixsurep3hat}\overline{\pi}(H)  & = & \max_{\t \in \Tc} \hat{\pi}^{\t}(H) =\max_{\t \in \Tc}  \sup_{\Q \in \hat \Qc^{\tau}} \E_{\Q} \left(\frac{H}{\bar S^{\t,0}_T}\right)\\
\label{defqui1}
{\pi}^{\t}(H) & = & \sup_{\Q \in \Qc^{\tau}} 
\E_{\Q} \left(\frac{H }{ \bar S^{\tau,0}_T}\right).
\end{eqnarray}
The infima in \eqref{prixsurep2},  \eqref{prixsurep3}, \eqref{prixsureptautout} and \eqref{prixsureptau} are attained. 
Finally, one can compare the initial values of the different hedging strategies
\begin{eqnarray}
\label{classement}{\pi}(H) \leq \min_{\t \in \Tc} {\pi}^{\t}(H)  \leq \underline{\pi}(H)  \leq \overline{\pi}(H).
\end{eqnarray}
\end{theorem}
\begin{proof} See Section \ref{preuve3}.  $\Box$\end{proof}
%\begin{remark}
%Fix $\hat{\t} \in \Tc$. Then \eqref{prixrepqsumhat} in Theorem \ref{main2} can be used with $\l_{\t}=0$ if ${\t \in \Tc} \setminus \{  \hat{\t}\}$ and $\l_{\hat{\t}}=1.$ Then $\Qc^{lc, \l}
%=\Qc^{\hat{\t}}$  and we get a characterization of the superreplication price using the num\'eraire related to the tenor $\hat{\t}$
%\begin{eqnarray*}
%\sup_{ \Q \in   \Qc^{\hat{\t}}}
%\E_{\Q} \left( \frac{ H}{ \frac{S^{\hat{\t},0}_T}{S^{\hat{\t},0}_0}}
%-
%\sum_{\t \in \Tc} \hat{x}^{\t} \frac{\frac{S^{\t,0}_T}{S^{\t,0}_0}}
%{\frac{S^{\hat{\t},0}_T}{S^{\hat{\t},0}_0} }
%\right)  =0.
% \end{eqnarray*}
%The theorem can also be used with $\l_{\t}=1$ for all ${\t \in \Tc}$ and we get a characterization of the superreplication price using the sum num\'eraire
%\begin{eqnarray*}
%\sup_{ \Q \in   \Qc^{sum}}
%\E_{\Q} \left( \frac{ H}{ \sum_{\t \in \Tc} \frac{S^{\t,0}_T}{S^{\t,0}_0}}
%-
%\sum_{\t \in \Tc} \hat{x}^{\t} \frac{\frac{S^{\t,0}_T}{S^{\t,0}_0}}
%{\sum_{\t \in \Tc} \frac{S^{\t,0}_T}{S^{\t,0}_0} }
%\right)  =0.
% \end{eqnarray*}
%
%
%\end{remark}
In the next proposition, we propose specific types of submarkets  where the superreplication price $\pi(H)$ has an exact dual representation. We will apply Proposition \ref{propmax} to concrete economic cases in Section \ref{Examples}. 
\begin{proposition}
\label{propmax}
1. Assume that there exist $(\l_{\t})_{\t \in \Tc}  \in \Lambda^{\Tc}$ and $(c_{\tau})_{\t \in \Tc} \in (0,\infty)^{\Tc}$ such that for all $\t \in \Tc$, for all $\Q \in \Qc^{lc, \l}$ 
\begin{eqnarray*}
\E_{\Q}
\left(\frac{\bar S^{\t,0}_T}{\sum_{\t \in \Tc} \l_{\t}\bar S^{\t,0}_T}\right)=c_{\t}.
\end{eqnarray*}
Assume that $\frac{H}{\sum_{\t \in \Tc} \l_{\t}\bar S^{\t,0}_T} \in L^{\infty}$  and let $\tau_{max}$ be the $\tau \in \Tc$ such that $c_{\t}$ is maximum. 
Then, $$\pi(H)= \frac{1}{c_{\tau_{max}}} \sup_{\Q \in \Qc^{lc, \l}} E_{\Q}
\left(\frac{H}{\sum_{\t \in \Tc} \l_{\t}\bar S^{\t,0}_T}\right).$$
2. Assume now that there exist some $\tau_{max}$ and some $(c_{\tau})_{\t \in \Tc} \in (0,1]^{\Tc}$ such that $\frac{H}{\bar S^{\tau_{max},0}_T} \in L^{\infty},$ $\E_{\Q}
\left(\frac{\bar S^{\t,0}_T}{\bar S^{\tau_{max},0}_T}\right)= c_{\t}$ for all $\t \in \Tc$ and for all $\Q \in  \Qc^{\tau_{max}}$. Then, 
%some $(c_{\tau})_{\t \in \Tc} \in \R^{\Tc}$ such that for all $\t \in \Tc$, for all $\Q \in  \Qc^{\tau_{max}}$ 
%\begin{eqnarray*}
%\E_{\Q}
%\left(\frac{\bar S^{\t,0}_T}{\bar S^{\tau_{max},0}_T}\right)=c_{\t} \leq 1.
%\end{eqnarray*} 
\begin{eqnarray}
\label{pluie}
\pi(H)=  \pi^{\tau_{max}}(H).
%\sup_{\Q \in  \Qc^{\tau_{max}}} E_{\Q}\left(\frac{H}{\bar S^{\tau_{max},0}_T}\right)
\end{eqnarray}
\end{proposition}
%Assume  there exists $\tau_{max}$ such that for all $\tau \in \Tc$, 
%$$\sup_{\Q \in \hat \Qc^{\tau_{max}}}\E_{\Q}
%\left(\frac{\bar S^{\tau,0}_T}{\bar S^{\tau_{max},0}_T}\right) \leq 1.$$ Then, if  $\frac{H}{\min_{\t \in \Tc} \frac{S^{\t,0}_T}{S^{\t,0}_0}}  \in L^{\infty}$
%%$\frac{H}{{S^{\t_{max},0}_T}/{S^{\t_{max},0}_0}}  \in L^{\infty}$
%$$\pi(H)=\underline{\pi}(H)= \pi^{\tau_{max}}(H).$$
\begin{proof}
%{\it of Proposition \ref{propmax}} 
%We choose first in Theorem \ref{main2} $\lambda_{{\t}_{max}}=1$  and $\lambda_{{\tau}}=0$ for all $\tau\neq {\t}_{max}$. Then, $\Qc^{lc, \l}
%=\Qc^{{\t}_{max}}$. Let $\Q \in \Qc^{{\t}_{max}}$, we compute  in \eqref{prixrepqsumhat}
%\begin{small}
%\begin{eqnarray*}
% \E_{\Q} \left( \frac{ H}{ \bar S^{\tau_{max},0}_T}\right)- 
%\sum_{\t \in \Tc} {x}^{\t} \E_{\Q} \left(
% \frac{\frac{S^{\t,0}_T}{S^{\t,0}_0}}
%{ \bar S^{\tau_{max},0}_T}
%\right) & \geq &   \E_{\Q} \left( \frac{ H}{ \bar S^{\tau_{max},0}_T}\right) - \sum_{\t \in \Tc} {x}^{{\t}} \sup_{\Q \in \Qc^{{\t}_{max}}}  \E_{\Q} \left(
% \frac{\frac{S^{\t,0}_T}{S^{\t,0}_0}}
%{\bar S^{\tau_{max},0}_T }\right)\\
%& \geq &   \E_{\Q} \left( \frac{ H}{ \bar S^{\tau_{max},0}_T}\right) - \sum_{\t \in \Tc} {x}^{{\t}} \\
%\end{eqnarray*}
%\end{small}
%So taking the supremum over all $\Q \in \Qc^{{\t}_{max}}$, we get that $$\sum_{\t \in \Tc} {x}^{{\t}} \geq \sup_{\Q \in \Qc^{{\t}_{max}}} \E_{\Q} \left( \frac{ H}{ \bar S^{\tau_{max},0}_T}\right)=\pi^{\t_{max}}(H)$$
%and $\pi(H) \geq \pi^{\t_{max}}(H)$. Thus, using \eqref{classement}
%\begin{eqnarray*}
%\pi^{\t_{max}}(H) \leq {\pi}(H) \leq \underline{\pi}(H)  =\pi^{\t_{max}}(H).
%\end{eqnarray*}
Let $\Q \in \Qc^{lc, \l}$ . We  compute  in \eqref{prixrepqsumhat}
\begin{small}
\begin{eqnarray*}
 \E_{\Q} \left( \frac{ H}{\sum_{\t \in \Tc} \l_{\t}\bar{S}_T^{\t,0}}\right)- 
\sum_{\t \in \Tc} \hat{x}^{\t} \E_{\Q} \left(
 \frac{\bar{S}_T^{\t,0}}
{\sum_{\t \in \Tc} \l_{\t}\bar{S}_T^{\t,0}}
\right) & = &   \E_{\Q} \left( \frac{ H}{\sum_{\t \in \Tc} \l_{\t}\bar{S}_T^{\t,0}}\right) - \sum_{\t \in \Tc} c_{\tau} \hat{x}^{{\t}}.
\end{eqnarray*}
\end{small}
So taking the supremum over all $\Q \in \Qc^{lc, \l}$, we get that $$\sum_{\t \in \Tc} c_{\tau} \hat{x}^{{\t}} =\sup_{\Q \in \Qc^{lc, \l}} \E_{\Q} \left( \frac{ H}{\sum_{\t \in \Tc} \l_{\t}\bar{S}_T^{\t,0}}\right).$$
Thus, \begin{eqnarray*}
\pi(H) & = & \inf\left\{ \sum_{\t \in \Tc}  {x}^{{\t}}|\; \sum_{\t \in \Tc} c_{\tau} {x}^{{\t}} =\sup_{\Q \in \Qc^{lc, \l}} \E_{\Q} \left( \frac{ H}{\sum_{\t \in \Tc} \l_{\t}\bar{S}_T^{\t,0}}\right)\right\} \\
 & = & \frac{1}{c_{\tau_{max}}} \sup_{\Q \in \Qc^{lc, \l}} E_{\Q}
\left(\frac{H}{\sum_{\t \in \Tc} \l_{\t}\bar{S}_T^{\t,0}}\right).
\end{eqnarray*}
The second part of the proposition is obtained choosing $\lambda_{{\t}_{max}}=1$  and $\lambda_{{\tau}}=0$ for all $\tau\neq {\t}_{max}$. Then, $\Qc^{lc, \l}=  \Qc^{\tau_{max}},$ $c_{\tau_{max}}=1$ and the result follows. 
$\Box$
\end{proof}
The proof above shows that all of the initial wealth is invested in the market $\tau_{max}$.  
If we go back to the example of the credit risk market,  $\pi(H),$ which is the cheapest superhedging price, is obtained by investing all the initial allocation in the market where the credit spread is the largest, which is intuitive because it is the submarket where the  risk is maximum. 

We now focus on the situation of 2. in Proposition \ref{propmax}. In this case, 
$$\hat{\pi}^{\t_{max}}(H) -{\pi}(H)=\sup_{\Q \in  \hat \Qc^{\tau_{max}}} E_{\Q}
\left(\frac{H}{\bar S^{\tau_{max},0}_T}\right)-\sup_{\Q \in  \Qc^{\tau_{max}}} E_{\Q}
\left(\frac{H}{\bar S^{\tau_{max},0}_T}\right) \geq 0,$$ 
since $  \Qc^{\tau_{max}}  \subset   \hat \Qc^{\tau_{max}}$ and the difference of features (for example liquidity) is related to the size difference between both sets $ \Qc^{\tau_{max}}$ and $\hat \Qc^{\tau_{max}}.$ We give a concrete illustration of this size difference just after Proposition \ref{propmatin}. 
%Note that the superreplication price in submarket $\t_{max}$ alone is higher than $\pi(H)$ since $  \Qc^{\tau_{max}}  \subset   \hat \Qc^{\tau_{max}}$ and $\hat{\pi}^{\t_{max}}(H) -{\pi}(H)$ measures the illiquidity of the market $\t_{max}$. 
%The superreplication price in submarket $\t_{max}$ alone is higher than $\pi(H)$: even if  the entire initial wealth $ {\pi}(H) $ is invested  in submarket $\t_{max}$, trading may also take place in the other submarkets. This is not the case for $\hat{\pi}^{\t_{max}}(H)$, where all the trading activity is made in submarket $\t_{max}$. 

\subsection{Focus on the prices of the risky assets}
In this section, we give with some elements on the prices of the risky assets. For that we assume that for all $\tau \in \Tc,$ the process $\widetilde S^{\tau}$ is a true martingale under any $\Q \in \hat \Qc^{\tau}.$ For example, we may assume that $\widetilde S^{\tau}$ is bounded or that $\E_{\Q} \sup_{0\leq s \leq t} |\widetilde S^{\tau}_t|<\infty$ for all $t>0$ or even that the quadratic variation of $\widetilde S^{\tau}$ is integrable under $\Q$. Then, for all $j \in \{1,\ldots, d_{\tau}\},$ using the dual formula given in \eqref{pitau}, we get that
%Then, for all $j \in \{1,\ldots, d_{\tau}\},$ $\widetilde S^{\tau,j}$ is a $\Q$-local martingale and  thus a $\Q$-supermartingale. Thus, 
%$$E_{\Q}
%\left(\frac{S^{\tau,j}_T}{\bar S^{\tau,0}_T}\right) \leq S^{\tau,j}_0.$$ 
%Using the dual formula given in \eqref{pitau}, $\hat{\pi}^{\t}(S^{\tau,j}_T)  \leq S_0^{\tau,j}$. Now, using the primal formula,\eqref{prixsureptau}, $S_0^{\tau,j} \geq \hat{\pi}^{\t}(S^{\tau,j}_T)$ and 
%$$S_0^{\tau,j} \in  \left \{ x| \, x \geq 0, \, \exists W \in \bar{C}^{\t}\cap L^{\infty}, x\bar S^{\t,0}_T+W \geq  S^{\tau,j}_T \, \mbox{a.s.} \right\}.$$
%and Remark  \ref{remuntenor}, 
%we see that for all $j \in \{1,\ldots, d_{\tau}\}$ 
\begin{eqnarray}
\label{eq1}
\hat{\pi}^{\t}(S^{\tau,j}_T) & = &  \sup_{\Q \in   \hat \Qc^{\tau}} \E_{\Q} \left(\frac{S^{\tau,j}_T}{\bar S^{\tau,0}_T}\right)
= S^{\tau,0}_0\sup_{\Q \in  \hat \Qc^{\tau}} \E_{\Q}
\left(\widetilde S^{\tau,j}_T\right)= S_0^{\tau,j}. 
\end{eqnarray} 
In the context of  the situation of 2. in Proposition \ref{propmax},  \eqref{pluie} implies that
\begin{eqnarray*}
\pi(S^{\tau_{max},j}_T) & =  & \sup_{\Q \in  \Qc^{\tau_{max}}} \E_{\Q}
\left(\frac{S^{\tau_{max},j}_T}{\bar S^{\tau_{max},0}_T}\right)
%= S^{\tau_{max},0}_0\sup_{\Q \in  \Qc^{\tau_{max}}} \E_{\Q}\left(\widetilde S^{\tau_{max},j}_T\right)
=S^{\tau_{max},j}_0.
\end{eqnarray*}
as $\widetilde S^{\tau_{max}}$ is a $\Q$-martingale for all $\Q \in  \Qc^{\tau_{max}} \subset \hat  \Qc^{\tau_{max}}$. 
Thus,  the illiquidity cost for $S^{\tau_{max}}_T$  is zero. 
%Indeed, in the case of 2. there exists some $\tau_{max}$ such that  $\E_{\Q}
%\left(\frac{\bar S^{\t,0}_T}{\bar S^{\tau_{max},0}_T}\right)= c_{\t} \leq 1$ for all $\t \in \Tc$ and for all $\Q \in  \Qc^{\tau_{max}}.$

The case of submarkets with only one risky asset is of one interest. Indeed, one may assume that each business unit only trades on one risky asset.  
We furthermore assume that there are only two submarkets $\tau_1$ and $\tau_2$.  We also introduce the contingent claim $S_T^{\tau_1}-S_T^{\tau_2}$ which allows to be short of the  asset $\t_2$ and long of the asset $\t_1$. This instrument is like a Basis swap in the interest rate market. 
%  and {\red allows to trade between both submarkets.} 
In this context, 
%if we assume that $\widetilde S^{\tau_i}$ is a true martingale under any $\Q \in \hat \Qc^{\tau_i}$  for $i \in \{1,2\}$, 
we are able to fully compute all the superreplication prices and also the illiquidity costs. We obtain that the first inequality in \eqref{classement} is in fact an equality.  
%Let \begin{eqnarray*}
%%\label{defqui1}
%{\pi}^{\t_2}(S^{\tau_1}_T) & = & \sup_{\Q \in \Qc^{\tau_2}} 
%\E_{\Q} \left(\frac{S^{\tau_1}_T }{ \bar S^{\tau_2,0}_T}\right).
%\end{eqnarray*}
\begin{proposition}
\label{propdeux}
Assume  that $\Tc=\{\tau_1,\tau_2\}$ and that for $i \in \{1,2\}$, $d_{\tau_i}=1$ and that  
$\widetilde S^{\tau_i}$ is a true martingale under any $\Q \in \hat \Qc^{\tau_i}.$ 
Then,  
\begin{eqnarray}
\label{qui0}
{\pi}^{\t_1}(S^{\tau_1}_T)  & =& \hat{\pi}^{\t_1}(S^{\tau_1}_T)=S^{\tau_1}_0\\
\label{qui}
{\pi}^{\t_2}(S^{\tau_1}_T) & = & 
S_0^{\tau_1}\sup_{\Q \in \Qc^{\tau_2}} 
\E_{\Q} \left(\frac{\bar S^{\tau_1,0}_T}{ \bar S^{\tau_2,0}_T}\right)
=   
\frac{S_0^{\tau_1}}{\inf_{\Q \in \Qc^{\tau_1}} 
\E_{\Q} \left(\frac{\bar S^{\tau_2,0}_T}{ \bar S^{\tau_1,0}_T}\right)}\\
\nonumber
\pi(S^{\tau_1}_T) & = &  {\pi}^{\t_1}(S_T^{\tau_1})  \wedge {\pi}^{\t_2}(S^{\tau_1}_T)\\
%=S_0^{\tau_1}\left(\sup_{\Q \in \Qc^{\tau_2}}\E_{\Q} \left(\frac{\bar S^{\tau_1,0}_T}{ \bar S^{\tau_2,0}_T}\right)\wedge 1 \right) \\& = &   
%\frac{S_0^{\tau_1}}{\inf_{\Q \in \Qc^{\tau_1}} 
%\E_{\Q} \left(\frac{\bar S^{\tau_2,0}_T}{ \bar S^{\tau_1,0}_T}\right) \vee 1}\\
\nonumber
\hat{\pi}^{\t_1}(S^{\tau_1}_T) - {\pi}(S^{\tau_1}_T) & =& \left({\pi}^{\t_1}(S^{\tau_1}_T) -{\pi}^{\t_2}(S^{\tau_1}_T) \right)_+, \mbox{ where $x_+=\max (x,0).$}
\end{eqnarray}
%$$
%\pi(S^{\tau_2}_T)={\pi}^{\t_1}(S^{\tau_2}_T)   \wedge \hat{\pi}^{\t_2}(S_T^{\tau_2}).$$
If $\pi(S^{\tau_1}_T)={\pi}^{\t_2}( S_T^{\tau_1}),$ then $\pi( S_T^{\tau_2})={\pi}^{\t_2}(S_T^{\tau_2}).$ \\
If ${\pi}^{\t_2}(S^{\tau_1}_T)   < {\pi}^{\t_2}(S_T^{\tau_2}),$ then $\pi(S^{\tau_1}_T-S^{\tau_2}_T)=+\infty$. Else, 
\begin{eqnarray}
\label{gilles}
\pi(S^{\tau_1}_T-S^{\tau_2}_T)=
\pi(S_T^{\tau_1}) -  \frac{S^{\tau_2}_0}{\max \left(\sup_{\Q \in \Qc^{\tau_2}} \E_{\Q} \left(\frac{\bar S^{\tau_1,0}_T}{ \bar S^{\tau_2,0}_T}\right) ,1 \right)}.
\end{eqnarray}
We now distinguish between three cases.\\
%if   $\pi (S^{\tau_1}_T) ={\pi}^{\t_2}( S_T^{\tau_1})$, then 
% $\pi(S^{\tau_1}_T-S^{\tau_2}_T)=\pi(S^{\tau_1}_T)- \pi(S^{\tau_2}_T).$\\
%If   $\pi (S^{\tau_1}_T) = \hat{\pi}^{\t_1}(S_T^{\tau_1}) $, then  $\pi(S^{\tau_1}_T-S^{\tau_2}_T)=\pi(S_T^{\tau_1}) - \pi(S_T^{\tau_2}) \frac{\inf_{\Q \in \Qc^{\tau_2}} 
%\E_{\Q} \left(\frac{\bar S^{\tau_1,0}_T}{ \bar S^{\tau_2,0}_T}\right) \vee 1}{\sup_{\Q \in \Qc^{\tau_2}} 
%\E_{\Q} \left(\frac{\bar S^{\tau_1,0}_T}{ \bar S^{\tau_2,0}_T}\right)}.$
%$$\pi(S^{\tau_1}_T-S^{\tau_2}_T)=
%\pi(S_T^{\tau_1}) - \pi(S_T^{\tau_2}) \frac{\inf_{\Q \in \Qc^{\tau_2}} 
%\E_{\Q} \left(\frac{\bar S^{\tau_1,0}_T}{ \bar S^{\tau_2,0}_T}\right) \vee 1}{\sup_{\Q \in \Qc^{\tau_2}} 
%\E_{\Q} \left(\frac{\bar S^{\tau_1,0}_T}{ \bar S^{\tau_2,0}_T}\right) \vee 1}.$$
Case 1 : $\pi(S^{\tau_1}_T)  =  {\pi}^{\t_2}(S^{\tau_1}_T)$ and $\pi(S^{\tau_2}_T)  =  {\pi}^{\t_2}(S^{\tau_2}_T).$ This is equivalent to $\sup_{\Q \in \Qc^{\tau_2}} 
\E_{\Q} \left(\frac{\bar S^{\tau_1,0}_T}{ \bar S^{\tau_2,0}_T}\right) <1$. Then, all the initial investments are made in the market $\tau_2$ and 
$$\pi(S^{\tau_1}_T-S^{\tau_2}_T) =\pi(S^{\tau_1}_T) -\pi(S^{\tau_2}_T) .$$
Case 2 :  $\pi(S^{\tau_1}_T)  =  {\pi}^{\t_1}(S^{\tau_1}_T)$ and $\pi(S^{\tau_2}_T)  =  {\pi}^{\t_1}(S^{\tau_2}_T).$ This is equivalent to 
 $\inf_{\Q \in \Qc^{\tau_2}} 
\E_{\Q} \left(\frac{\bar S^{\tau_1,0}_T}{ \bar S^{\tau_2,0}_T}\right) > 1.$ Then, all the initial investments are made in the market $\tau_1$ and
$$\pi(S^{\tau_1}_T-S^{\tau_2}_T) =\pi(S^{\tau_1}_T)  - \pi(S_T^{\tau_2}) \frac{\inf_{\Q \in \Qc^{\tau_2}} 
\E_{\Q} \left(\frac{\bar S^{\tau_1,0}_T}{ \bar S^{\tau_2,0}_T}\right) }{\sup_{\Q \in \Qc^{\tau_2}} 
\E_{\Q} \left(\frac{\bar S^{\tau_1,0}_T}{ \bar S^{\tau_2,0}_T}\right) } .$$
Case 3: $\pi(S^{\tau_1}_T)  =  {\pi}^{\t_1}(S^{\tau_1}_T)$ and $\pi(S^{\tau_2}_T)  =  {\pi}^{\t_2}(S^{\tau_2}_T).$ This is equivalent to $\sup_{\Q \in \Qc^{\tau_2}} 
\E_{\Q} \left(\frac{\bar S^{\tau_1,0}_T}{ \bar S^{\tau_2,0}_T}\right) \geq 1$ and 
 $\inf_{\Q \in \Qc^{\tau_2}} 
\E_{\Q} \left(\frac{\bar S^{\tau_1,0}_T}{ \bar S^{\tau_2,0}_T}\right) \leq 1.$ Then, the initial investments for the assets $S^{\tau_1}_T$ and $S^{\tau_1}_T-S^{\tau_2}_T$ (resp.  $S^{\tau_2}_T$) are made in the market $\tau_1$ (resp. $\tau_2$)  and 
$$\pi(S^{\tau_1}_T-S^{\tau_2}_T) =\pi(S^{\tau_1}_T) - \pi(S_T^{\tau_2}) \frac{1}{\sup_{\Q \in \Qc^{\tau_2}} 
\E_{\Q} \left(\frac{\bar S^{\tau_1,0}_T}{ \bar S^{\tau_2,0}_T}\right)} .$$
%\begin{eqnarray*}
%\pi(S^{\tau_1}_T-S^{\tau_2}_T) & = & 
%\left(\frac{\hat{\pi}^{\t_1}(S_T^{\tau_1})}{{\pi}^{\t_2}(S^{\tau_1}_T) } \wedge  1\right)
%\left({\pi}^{\t_2}(S^{\tau_1}_T) -  \hat{\pi}^{\t_2}(S_T^{\tau_2})
%\right) \\
%& = & \pi(S_T^{\tau_1}) - \pi(S_T^{\tau_2}) \frac{\inf_{\Q \in \Qc^{\tau_2}} 
%\E_{\Q} \left(\frac{\bar S^{\tau_1,0}_T}{ \bar S^{\tau_2,0}_T}\right) \vee 1}{\sup_{\Q \in \Qc^{\tau_2}} 
%\E_{\Q} \left(\frac{\bar S^{\tau_1,0}_T}{ \bar S^{\tau_2,0}_T}\right) \vee 1}.
%\end{eqnarray*}
\end{proposition}
\begin{proof} See Section \ref{preuve4}.  $\Box$\end{proof}
We  give some interpretations of Proposition \ref{propdeux}  for a difference of liquidity but it may be a difference of credit risk. For that, we assume that $\tilde S_T^{\tau_1}$ and $\tilde S_T^{\tau_2}$ are bounded.\\
We have found that the price of $S_T^{\tau_1},$ $\pi(S^{\tau_1}_T),$ is the minimum between its initial value, $\p^{\tau_1}(S^{\tau_1}_T)=S_0^{\tau_1},$ which corresponds to the case where the trading takes place in  $\tau_1,$  i.e.,
$$ 
\p^{\tau_1}(S^{\tau_1}_T) \bar S_T^{\tau_1,0} + S_T^{\tau_1,0}( \tilde S_T^{\tau_1}- \tilde S_0^{\tau_1}) =  S_T^{\tau_1}$$
and ${\pi}^{\t_2}(S^{\tau_1}_T),$ which corresponds to the case where the initial investment is made in  $\tau_2 $ and there exists 
$W \, \in \bar{C}\cap L^{\infty}$ such that 
\begin{eqnarray*}
{\pi}^{\t_2}(S^{\tau_1}_T)\bar S_T^{\tau_2,0} +W  & = & S_T^{\tau_1}. 
\end{eqnarray*}
When $\p(S^{\tau_1}_T)=\p^{\tau_1}(S^{\tau_1}_T),$  the illiquidity cost  $\hat{\pi}^{\t_1}(S^{\tau_1}_T) - {\pi}(S^{\tau_1}_T)=0$. So, the market $\tau_1$ is liquid. 
When $\p(S^{\tau_1}_T)=\p^{\tau_2}(S^{\tau_1}_T),$ we get that $\pi( S_T^{\tau_2})={\pi}^{\t_2}(S_T^{\tau_2}).$ So, the market $\tau_2$ is liquid, while the market $\tau_1$ is illiquid as $\hat{\pi}^{\t_1}(S^{\tau_1}_T) - {\pi}(S^{\tau_1}_T)>0$.
Note that in the first situation, the market $\tau_2$ may be illiquid or liquid as we will see below.  

Now, we comment on the price of the Basis swap and how the investments are made. 
Case 1 corresponds to the situation where the asset $S^{\tau_1}$ is illiquid and the asset $S^{\tau_2}$ is liquid. Proposition \ref{propdeux} shows that all the initial investments are made in the liquid market $\tau_2$. 
Recalling the definition of the different superreplication prices, there exist $W, \, W_1 \in \bar{C}\cap L^{\infty}$ such that 
\begin{eqnarray*}
\pi (S^{\tau_1}_T)\bar S_T^{\tau_2,0} +W_1  & = & S_T^{\tau_1}\\
\pi (S^{\tau_2}_T) \bar S_T^{\tau_2,0} + S_T^{\tau_2,0}( \tilde S_T^{\tau_2}- \tilde S_0^{\tau_2})& = & S_T^{\tau_2}\\
\left( \pi(S^{\tau_1}_T)-\pi(S^{\tau_2}_T)\right) \bar S_T^{\tau_2,0} +W  & = & S_T^{\tau_1}-S_T^{\tau_2}.
\end{eqnarray*}
So, replicating the illiquid asset  $S_T^{\tau_1}$ by buying the assets $S_T^{\tau_2}$ and $S^{\tau_1}_T-S^{\tau_2}_T$ does not provide a cheaper price. As expected, there is no free lunch. \\ 
The case 2   corresponds to the situation where the asset $S^{\tau_1}$ is liquid and the asset $S^{\tau_2}$ is illiquid. Proposition \ref{propdeux} proves  that all the initial investments are made in the liquid market $\tau_1$ and 
there exists 
$W \, \in \bar{C}\cap L^{\infty}$ such that 
\begin{eqnarray*}
\pi (S^{\tau_1}_T) \bar S_T^{\tau_1,0} + S_T^{\tau_1,0}( \tilde S_T^{\tau_1}- \tilde S_0^{\tau_1})& = & S_T^{\tau_1}\\
\pi (S^{\tau_2}_T)\bar S_T^{\tau_1,0} +W_2  & = & S_T^{\tau_2}\\
\left(\pi(S_T^{\tau_1}) - \frac{S_0^{\tau_2}}{\sup_{\Q \in \Qc^{\tau_2}} 
\E_{\Q} \left(\frac{\bar S^{\tau_1,0}_T}{ \bar S^{\tau_2,0}_T}\right)}
\right) \bar S_T^{\tau_1,0} +W  & = & S_T^{\tau_1}-S_T^{\tau_2}. 
\end{eqnarray*}
So, setting $\bar W=W_2+W - S_T^{\tau_1,0}( \tilde S_T^{\tau_1}- \tilde S_0^{\tau_1}) \, \in \bar{C}\cap L^{\infty},$ we obtain that 
\begin{eqnarray*}
\left(\frac{S_0^{\tau_2}}{\inf_{\Q \in \Qc^{\tau_2}} 
\E_{\Q} \left(\frac{\bar S^{\tau_1,0}_T}{ \bar S^{\tau_2,0}_T}\right)}-
\frac{S_0^{\tau_2}}{\sup_{\Q \in \Qc^{\tau_2}} 
\E_{\Q} \left(\frac{\bar S^{\tau_1,0}_T}{ \bar S^{\tau_2,0}_T}\right)}
\right) \bar S_T^{\tau_1,0} + \bar W = 0.
\end{eqnarray*}
This is not a free lunch since $$\frac{S_0^{\tau_2}}{\inf_{\Q \in \Qc^{\tau_2}} 
\E_{\Q} \left(\frac{\bar S^{\tau_1,0}_T}{ \bar S^{\tau_2,0}_T}\right)}-
\frac{S_0^{\tau_2}}{\sup_{\Q \in \Qc^{\tau_2}} 
\E_{\Q} \left(\frac{\bar S^{\tau_1,0}_T}{ \bar S^{\tau_2,0}_T}\right)}\geq \pi(0)=0.$$
Finally, in case 3 both assets are liquid. In this case, the initial investements are made in both markets. 
There exists 
$W \in \bar{C}\cap L^{\infty}$ such that 
\begin{eqnarray*}
\pi (S^{\tau_1}_T) \bar S_T^{\tau_1,0} + S_T^{\tau_1,0}( \tilde S_T^{\tau_1}- \tilde S_0^{\tau_1})& = & S_T^{\tau_1}\\
\pi (S^{\tau_2}_T) \bar S_T^{\tau_2,0} + S_T^{\tau_2,0}( \tilde S_T^{\tau_2}- \tilde S_0^{\tau_2})& = & S_T^{\tau_2}\\
\left(\pi(S_T^{\tau_1}) - \frac{S_0^{\tau_2}}{\sup_{\Q \in \Qc^{\tau_2}} 
\E_{\Q} \left(\frac{\bar S^{\tau_1,0}_T}{ \bar S^{\tau_2,0}_T}\right)}
\right) \bar S_T^{\tau_1,0} +W  & = & S_T^{\tau_1}-S_T^{\tau_2}.
\end{eqnarray*}
%\begin{eqnarray*}
%\left(\pi(S_T^{\tau_1}) - \frac{S_0^{\tau_2}}{\sup_{\Q \in \Qc^{\tau_2}} 
%\E_{\Q} \left(\frac{\bar S^{\tau_1,0}_T}{ \bar S^{\tau_2,0}_T}\right)}
%\right) \bar S_T^{\tau_1,0}+\pi (S^{\tau_2}_T) \bar S_T^{\tau_2,0} + W + S_T^{\tau_2,0}( \tilde S_T^{\tau_2}- \tilde S_0^{\tau_2}) & = & S_T^{\tau_1}. 
%\end{eqnarray*}
%Thus, $\pi(S_T^{\tau_1})\leq \pi(S_T^{\tau_1}) - \frac{S_0^{\tau_2}}{\sup_{\Q \in \Qc^{\tau_2}} 
%\E_{\Q} \left(\frac{\bar S^{\tau_1,0}_T}{ \bar S^{\tau_2,0}_T}\right)} + \pi (S^{\tau_2}_T)$ 
So, setting $\bar W=W+ S_T^{\tau_2,0}( \tilde S_T^{\tau_2}- \tilde S_0^{\tau_2}) - S_T^{\tau_1,0}( \tilde S_T^{\tau_1}- \tilde S_0^{\tau_1}) \, \in \bar{C}\cap L^{\infty},$ we obtain that 
\begin{eqnarray*}
\left( - \frac{S_0^{\tau_2}}{\sup_{\Q \in \Qc^{\tau_2}} 
\E_{\Q} \left(\frac{\bar S^{\tau_1,0}_T}{ \bar S^{\tau_2,0}_T}\right)}
\right) \bar S_T^{\tau_1,0}+\pi (S^{\tau_2}_T) \bar S_T^{\tau_2,0} +\bar W = 0. 
\end{eqnarray*}
Again, as $\sup_{\Q \in \Qc^{\tau_2}} 
\E_{\Q} \left(\frac{\bar S^{\tau_1,0}_T}{ \bar S^{\tau_2,0}_T}\right)\geq 1,$  this is not a free lunch because
$$S_0^{\tau_2}- \frac{S_0^{\tau_2}}{\sup_{\Q \in \Qc^{\tau_2}} 
\E_{\Q} \left(\frac{\bar S^{\tau_1,0}_T}{ \bar S^{\tau_2,0}_T}\right)} \geq \pi (0)=0. $$

An illustration of these results in the Brownian framework is given  in  Section \ref{concret} below. 
\section{Economic illustrations}
\label{ecoillus}
We now propose some economic illustrations. First, we provide some further evidence for our choice of model (self-financing  and no free lunch conditions, definition of the set $K$) coming from the multicurve financial market. Then, we give some explicit cases where there exists a common martingale measure. We also discuss completeness issues. Finally, we give a complete illustration of our result in a Brownian setting. 
\subsection{Multicurve financial market}
\label{remtaux}
In this section, we focus on the case of the multicurve financial market, which gives a good justification for our modeling choices. 
 Before the financial crisis, interest rate swaps (denominated in the same currency) were priced using the same zero-coupon (ZC) curve used as a num\'{e}raire, regardless of the priced swaps' payment structure frequency. %%EDITOR'S NOTE: Abbreviations and acronyms are typically defined the first time the term is used within the main text and then used throughout the remainder of the manuscript. Please consider adhering to this convention. The target journal may have a list of abbreviations that are considered common enough that they do not need to be defined.
In Europe, typical ZC curves were constructed using instruments with a 6-month payment frequency. In 2007, important distortions began to appear in the swap rates indexed by different frequencies (or tenors): 1 month, 3 month, 6 month and 1 year. 
Therefore, the market practice for interest rate swap valuation has evolved and considers several ZC curves that are tenor based while retaining the assumption of the existence of a common martingale measure. For example the 3-month ZC curve is
%: one curve for discounting and another one for estimating future cash flows. The curve associated to future cash flows is
built using only instruments with a tenor of 3 months such as the forward rate agreements (FRA)  3 months in 3 months or 3 months in 6 months or such as 
swaps rates against Euribor 3M for several maturities. %%EDITOR'S NOTE: Please ensure that the intended meaning has been maintained in this edit.
%, while the curve used for discounting is the standard ZC curve based on the overnight rate (EONIA in the Euro zone).
%{\red At the end of 2009, considerable bias in the evaluation of swaps appeared when using pre-crisis methods instead of the new market practice, which revealed that there is arbitrage if the 1-, 3-, 6- and 12-month markets are considered to be a single market without specific tenor risks. 
%The different submarkets with frequencies of 1, 3, 6 and 12 months are arbitrage free, but there is arbitrage in the global swap rates market when using a common num\'{e}raire. }

Usually, in the fixed income  theory, in order to state the no-arbitrage condition and then obtain FTAP, one takes the ZC bonds as basic tradable underlying assets. In the presence of multiple tenors, the  existence of such tradable ZC bonds is more than questionable. Firstly, if such bonds exist and are tradable, why do several ZC curves co-exist? There should be arbitrage opportunities between, say, the ZC bond in the 3-month tenor with maturity 6 months and the one in the 6-month tenor with the same maturity. In particular, if their prices are not equal at time 0, there is an arbitrage opportunity, because they both pay 1 euro at time 6 months. 
So, it appears clearly that in the multi-curve market the basic tradable underlying assets cannot be the ZC bonds of several tenors. So, one may use FRA of different tenors as basic risky assets, see \cite{FGGS}, \cite{GrbacRunggaldier16} and the references therein. We argue below that if 
one  assumes the existence of ZC bonds traded in the global market for all maturities (for example, associating a tradable zero-coupon bond to  the overnight indexed swap (OIS) rate  as in \cite{FGGS}), then there should not co-exist FRA of different tenors  by a text-book no-arbitrage argument as before. Indeed,  let $T>0$ denote a finite time horizon for the whole market and fix $I\leq M\leq T$. We recall that an FRA is an over-the-counter derivative that allows the holder to lock in at any date $0 \leq t\leq I$ 
the interest rate between the
inception date $I$ and the maturity $M$ at a fixed value $R$. At the maturity $M$, a payment based on $R$ is exchanged for a cash flow indexed on the underlying floating rate (generally the spot Libor rate
$L(I;I,M)$),
so giving the payoff 
$$
(M-I)(L(I;I,M) - R).
$$
The FRA rate $R(t;I,M)$ will be the rate for which the FRA is at pair. We assume that the notional amount is equal to one for simplicity. 
Assuming that the ZC bonds are denoted by $(B(s,t))_{0\leq s \leq t \leq T}$ and the FRA are tradable, we can implement the following strategy. 
A time $t$ buy one bond maturing $I$, short $\frac{B(t,I)}{B(t,M)}$ bonds maturing $M$  and enter an FRA with inception date $I$ and maturity $M$. At time $I$, place one euro on Libor. 
The net cash at $t$ is  $-B(t,I) + \frac{B(t,I)}{B(t,M)}B(t,M)=0$, at time $I$ it is $1-1=0$ and at time $M,$ it is 
$-(M-I)(L(I;I,M)-R(t;I,M))+(1+ (M-I)L(I;I,M) )-\frac{B(t,I)}{B(t,M)}= 1+ (M-I)R(t;I,M)-\frac{B(t,I)}{B(t,M)}$. So, by no-arbitrage the FRA rate is 
$$R(t;I,M)=\frac{1}{M-I}\left(\frac{B(t,I)}{B(t,M)} -1\right).$$
%A time $t$ buy one bond maturing $M$, short $\frac{B(t,M)}{B(t,I)}$ bonds maturing $I$ , borrow $\frac{B(t,M)}{B(t,I)}$  cash at the forward rate $L(t;I,M).$ 
%The net cash at $t$ is  $-B(t,M) + \frac{B(t,M)}{B(t,I)}B(t,I)=0$, at time $I$ it is $-\frac{B(t,M)}{B(t,I)}+\frac{B(t,M)}{B(t,I)}=0$ and at time $M$ 
%$1- \frac{B(t,M)}{B(t,I)} (1+ (M-I)L(t;I,M)$. So, by no-arbitrage
%$$L(t;I,M)=\frac{1}{M-I}\left(\frac{B(t,I)}{B(t,M)} -1\right).$$
The formula above shows that as the ZC bonds are not tenor dependent so are the FRA rates. So, we see that the existence of a tradable ZC bonds for all maturities is not consistent with a multicurve financial market, where FRA of different tenors are the basic assets. 

Here, we have addressed this issue by assuming that there is a  tradable num\'eraire for each tenor $\tau$ (the  ZC of tenor $\tau$) which allows to deflate only the instruments of tenor $\tau$.  The submarket of tenor $\tau$ consists only in the instruments having tenor $\tau$ and we model the case where 
there is no one num\'{e}raire available for trading to every submarkets. For example, the 3-month num\'eraire will be constructed using the OIS rate, but taking into account some roll-over risk, which will be of course different from, say, the  6-month tenor.  The fact that both num\'eraires are traded in two different submarkets and that  it is not allowed to borrow on one submarket again others  avoid the arbitrage opportunity described above.  

We have seen that in general, it is not possible to identify a common risk-neutral measure, as usually assumed in practice and in the literature (see \cite{GrbacRunggaldier16} and \cite{CuchieroFontanaGnoatto16} and the references therein). Nevertheless, we will see in Section \ref{Examples} below that this is true when the  spreads between the different submarkets are deterministic (and in particular when there is a common num\'eraire) or when the num\'eraires are constant.  
Note that the existence of a common martingale measure in the post-crisis market is  justified by some  FTAP in \cite{FGGS} assuming that the overnight indexed swap ZC bonds are freely tradable for all maturities. The authors also show the existence of martingale measures for a given  num\'{e}raire  assuming  that this  num\'{e}raire is traded.

An alternative approach to ours would be to incorporate the risks implicit in interbank transactions like for example  liquidity risk. This can be done introducing  illiquidity cost in the self-financing condition \eqref{autofo}. To the best of our knowledge this has not been done in this context. 
%, so our multi-num\'eraire approach is the first one addressing 
%
\subsection{Examples of submarkets with a common martingale measure}
\label{Examples}
We present two situations where for a specific choice of num\'eraires $(S^{\t,0})_{\t \in \Tc},$ there exists a (local)  martingale measure common to all submarkets. Using Proposition \ref{propmax}, we provide an explicit formula for the price $\pi(H).$ Assume that  $(S,S^0)$ satisfies the no free lunch condition. Then, Theorem \ref{main1} shows that   $\Xc^* \neq \emptyset$. Let $X^* \in \Xc^*. $
\begin{example}
\label{expleIndep}
%Recall from \eqref{Qtau}  that we have
%$$\frac{d\Q^{\t}}{d\P}\Big|_{\Fc}= \frac{X^* S_{T}^{\t, 0}}{\E [X^* S_{T}^{\t, 0} ]}$$
The first situation is the one where the  num\'eraires $S^{\t,0}$ are deterministic for any $\tau \in \Tc$. Then, \eqref{Qtau} implies that 
\begin{eqnarray*}
\frac{d\Q^{\t}}{d\P}\Big|_{\Fc}=\frac{X^*}{\E(X^*)}
\end{eqnarray*}
and the (local) martingale measures clearly do not depend on the submarket.

Let $\t_{max}$ be the $\tau$ such that $ \bar{S}_T^{\t,0}$ is  maximum.  Then, $ \Qc^{\tau_{max}}=\Qc^{max}$, 
$
c_{\t}= \frac{\bar S^{\tau,0}_T}{\bar S^{\tau_{max},0}_T} \leq 1
$
and \eqref{pluie} implies that for $H \in L^{\infty}$ 
\begin{eqnarray*}
\pi(H) & = & \pi^{\t_{max}}(H)=\sup_{\Q \in  \Qc^{\tau_{max}}} \E_{\Q} \left( \frac{ H}{ \bar S^{\tau_{max},0}_T}\right)=\sup_{\Q \in \Qc^{max}} \E_{\Q} \left(\frac{H}{\max_{\t \in \Tc} \bar S^{\t,0}_T}\right)
\end{eqnarray*}
and $\pi(H) $ and the lower bound in \eqref{carctprixsurep1bishat} are equal.
%then \eqref{prixrepqsumhat} implies that
%\begin{eqnarray*}
%\sum_{\t \in \Tc} \hat{x}^{\t} \bar{S}_T^{\t,0} =\sup_{ \Q \in   \Qc^{lc, \l}}
%\E_{\Q} ({ H})=\sup_{ \Q \in   \Qc^{one}}
%\E_{\Q} ({ H}),
% \end{eqnarray*}
%as for any  deterministic $Z>0$,
%$\Qc^{Z}=\Qc^{one}$ where $\Qc^{one}$ is the set $\Qc^{Z}$ obtained with $Z=1$.
%Thus,
%\begin{eqnarray*}
%\pi(H) & = & 
%\inf \left\{\sum_{\t \in \Tc} x^{\t}, \, x^{\t} \geq 0, \sum_{\t \in \Tc} \hat{x}^{\t} \bar{S}_T^{\t,0} =  \sup_{ \Q \in   \Qc^{one}}
%\E_{\Q} ({ H}) \right\}\\
%& = &   \frac{S^{\t_{max},0}_0}{S^{\t_{max},0}_T}
%\sup_{ \Q \in   \Qc^{one}}\E_{\Q} ({ H}) \\
%& = & \sup_{{\Q\in \Qc^{max}} \E_{\Q} \left(\frac{H}{
%\max_{\t \in \Tc} 
%\bar{S}_T^{\t,0}
%}\right), 
%\end{eqnarray*}
%where $\t_{max}$ is the $\tau$ such that $ \bar{S}_T^{\t,0}$ is  maximum.
\end{example}
\begin{example}
\label{exple}
In this example, we consider the case where the spreads are deterministic. 
First, we rewrite the num\'eraires $S^{\t,0}_t$ for each  $\t \in \mathcal T$ as follows:
$$
S^{\t,0}_t =S^{\t,0}_0 \exp \left( \int_0^t (r_u + s^\t_u ) du \right),
$$
where $r$ is some short-term submarket-free rate\footnote{For the multi-curve interest rates model, $r$ may be the Overnight Indexed Swap (OIS) rate, which is a reference basis rate (see, e.g., \cite{GrbacRunggaldier16}).} and $s^\t$ denotes the short-term spread specific\footnote{In the example of credit risk markets, $s^{\tau}$ is the default intensity of the class of risk $\tau$.} to the submarket $\t$.
Let
\begin{eqnarray*}
\frac{d \P^*}{d\P}\Big|_{\Fc}=\frac{X^* \exp\left(\int_0^T r_u du\right)}{\E\left(X^*\exp\left(\int_0^T r_u du\right)\right)}. 
\end{eqnarray*}
%\begin{eqnarray*}
%\frac{d\Q^{\t}}{d\P}\Big|_{\Fc}=\frac{d \P^*}{d\P}\Big|_{\Fc}  \exp\left(\int_0^T s^\t_u du\right)\frac{\E\left(X^*\exp\left(\int_0^T r_u du\right)\right)}{\E\left(X^*\exp\left(\int_0^T r_u du\right)\exp\left(\int_0^Ts^\t_u du\right)\right)}. 
%\end{eqnarray*}
%Thus, if for all $\t \in \mathcal T$ $\exp\left(\int_0^T s^\t_u du\right)$ is independent of $X^*$$\exp\left(\int_0^T r_u du\right)$, we obtain that
%\begin{eqnarray*}
%\frac{d\Q^{\t}}{d\P}\Big|_{\Fc}=\frac{d \P^*}{d\P}\Big|_{\Fc}  \frac{\exp\left(\int_0^T s^\t_u du\right)}{\E\left(\exp\left(\int_0^Ts^\t_u du\right)\right)}. 
%\end{eqnarray*}
%In the remaining of this example, 
As the spreads are assumed to be deterministic,  \eqref{Qtau} shows that 
$\Q^{\t}=\P^*$ for all $\t \in \mathcal T$ . This is in particular the case if we assume that there exists a common num\'eraire for all  $\t \in \mathcal T$ (i.e., $s^\t=0$). 

Let $\t_{max}$ be the $\tau$ such that $ \exp ( \int_0^T s^\t_u  du )$ is maximum. Then, 
$\max_{\t \in \Tc} \bar{S}_T^{\t,0}=\bar S^{\tau_{max},0}_T.$
So, $ \Qc^{\tau_{max}}=\Qc^{max},$ 
$c_{\tau}= 
\exp \left( \int_0^T ( s^\t_u -s^{\t_{max}}_u) du \right) \leq 1,$ 
and  \eqref{pluie} implies that 
\begin{eqnarray*}
\pi(H) & = & \pi^{\t_{max}}(H)= \sup_{\Q \in  \Qc^{\tau_{max}}} \E_{\Q} \left(\frac{H}{ \bar S^{\tau_{max},0}_T}\right) =\sup_{\Q \in \Qc^{max}} \E_{\Q} \left(\frac{H}{\max_{\t \in \Tc} \bar S^{\t,0}_T}\right)\\
& = &  
\exp \left( -\int_0^T s^{\t_{max}}_u  du \right)
\sup_{ \Q \in   \Pc^{R}}
\E_{\Q} \left( { H}\exp \left( -\int_0^T r_u  du \right) \right),
\end{eqnarray*}
where  $\Q \in \Pc^{R}$ if and only if $\Q \sim \P$,
${\frac{d\Q}{d\P}}e^{-\int_0^T r_u  du } \in L^1$ and for all $\tau \in \Tc$, 
$\left(
{S}_t^{\t}\exp \left( -\int_0^T r_u  du \right)
\right)_{t \geq 0}$ is a  local martingale under $\Q$.  It is clear that
$  \Qc^{\tau_{max}}=\Pc^{R}$ and again $\pi(H) $ and the lower bound in \eqref{carctprixsurep1bishat} are equal.

\end{example}
\subsection{Completeness}
\label{complet}

As mentioned after Theorem \ref{main1}, we now discuss completeness. To do so, we provide simple examples that allow a better understanding of the link between the completeness of all  submarkets $\tau$ considered as separate markets and that of the global market. 
The completeness of the submarket $\tau$ is well known. It amounts to saying that the set $\hat{\Qc}^{\tau}$ of local martingale measures for the discounted asset $\tau$ is a singleton.  Now, we define the completeness of the global market as follows:  each set ${\Qc}^{\tau}$ is a singleton. 

For simplicity, we consider the case of Example \ref{expleIndep}, where  the  num\'eraires $S^{\t,0}$ are deterministic for all $\tau \in \Tc$. In this case, the (local) martingale measures  for the global market clearly do not depend on the submarket. Proposition \ref{propnum} shows that for all $\tau \in \Tc$
\begin{eqnarray*}
\Qc^{\tau} & = & \left\{\P^*|\, \exists X^* \in  \Xc^*,\; \frac{d\P^*}{d\P}=
\frac{
X^*
}
{
\E \left(
X^* 
\right)
}\right\}.
\end{eqnarray*}
Moreover, Remark \ref{remuntenor} shows that for all  
${\t} \in \Tc,$
$$\hat \Qc^{{\t}} =\left\{\Q^{\t}|\, \exists X^{{\t}} \in  \Xc^{{\t}},\; \frac{d\Q^{\t}}{d\P}=
\frac{
X^{{\t}}
}
{
\E \left(
X^{{\t}} 
\right)
}\right\}.$$
Now, Definition \ref{defQZ} implies that for deterministic num\'eraires 
 \begin{eqnarray}
 \label{det}
 {\Qc}^{\tau}=\cap_{\tau\rq{} \in \Tc}{\Qc}^{\tau\rq{},\bar S^{ \t,0}_T}=\cap_{\tau\rq{} \in \Tc} \hat{\Qc}^{\tau\rq{}}.
 \end{eqnarray}
 So, the only case where the completeness of all  submarkets $\tau$ implies the completeness of the global market is when the sets $\hat{\Qc}^{\tau}$ reduce to the same singleton. Otherwise, even if every submarket $\tau$ is complete, the global market will not satisfy the NFL as  ${\Qc}^{\tau}$ is empty. 
 We provide a concrete example of that in Example \ref{bino}. 
 
We present also in Example \ref{trino} a situation where the global market is complete but each submarket $\tau$ is incomplete. The intuition beyond is that there are more constraints in the definition of  ${\Qc}^{\tau}$ than in the one of  $\hat{\Qc}^{\tau}.$

\begin{example}
\label{bino}
We consider two Binomial models, i.e., for $i \in\{1,2\}$, $S_1^{\tau_i}=S_0^{\tau_i}(1+U^i)$, where $U^i$ can take only two values: $u_i$  with probability $p>0$ in the  \lq\lq{}up\rq\rq{} state and $d_i$ with probability $1-p>0$ in the  \lq\lq{}down\rq\rq{} state . We suppose that $d_1=1/2$, $u_1=2$, $d_2=5/2$ and $u_2=4$. Moreover, $S_1^{\tau_i,0}=S_0^{\tau_i,0}(1+r_i),$ where $r_1=1$ and $r_2=3$. Then, it is easy to see that the markets $\tau_1$ and $\tau_2$ are arbitrage free and complete and that the global market is also arbitrage free. Indeed, we have  that
$$\hat \Qc^{{\t_1}}=\hat \Qc^{{\t_2}}= \Qc^{{\t_1}}= \Qc^{\t_2}=\{\P^*\},$$ 
where $\P^*$ is the probability that gives the weight $1/3$ to the   \lq\lq{}up\rq\rq{} state and $2/3$ to the  \lq\lq{}down\rq\rq{} state.

Making a choice of coefficients such that $d_i<r_i<u_i$ for $i \in \{1,2\}$ but 
$\frac{u_1-r_1}{u_1-d_1} \neq \frac{u_2-r_2}{u_2-d_2}$ illustrates a case where the markets $\tau_1$ and $\tau_2$ are arbitrage free and complete and the global market is not arbitrage free (the set $\Xc^*$ is empty). 
% Nevertheless, any $X^*$ which takes any value $x\in \R$ in the \lq\lq{}down\rq\rq{} state and the value $\frac12 \frac{1-p}px$ in the \lq\lq{}up\rq\rq{} state satisfies \eqref{condmart}. So, $\Xc^* $ is not a singleton. 
\end{example}
\begin{example}
\label{trino}
We consider two Trinomial models, i.e., for $i \in\{1,2\}$, $S_1^{\tau_i}=S_0^{\tau_i}(1+U^i)$, where $U^i$ can take only three values $\{u_i,0,d_i\}$  with probability $p$ in the  \lq\lq{}up\rq\rq{} state, $v$  in the  \lq\lq{}middle\rq\rq{} state and $q$ in the  \lq\lq{}down\rq\rq{} state. 
We assume that $-1<d_i<0<u_i$, $p,v,q \in (0,1)$ and $p+v+q=1$ 
Moreover, $S_1^{\tau_i,0}=S_0^{\tau_i,0}(1+r_i)$ and we assume that $d_i<r_i<u_i.$ 
The set $\hat \Qc^{{\t_i}}$ is characterized by the following equations
\begin{eqnarray*}
p \frac{1+u_i}{1+r_i} +v \frac{1}{1+r_i} + q \frac{1+d_i}{1+r_i} = 1\\
p + v+q=1. 
\end{eqnarray*}
The preceding system admits the following set of solutions
\begin{eqnarray*}
v \in \left(0, 1-\frac{r_i}{u_i} \right), \quad p =\frac{r_i -(1-v)d_i}{u_i-d_i} \quad   q =\frac{(1-v)u_i -r_i}{u_i-d_i} 
\end{eqnarray*}
and, in general, the submarkets $\tau_1$ and $\tau_2$ are arbitrage free but incomplete.  

Now, for the global market, recalling \eqref{det}, we have to solve the system 
\begin{eqnarray*}
p \frac{1+u_1}{1+r_1} +v \frac{1}{1+r_1} + q \frac{1+d_1}{1+r_1} = 1\\
p \frac{1+u_2}{1+r_2} +v \frac{1}{1+r_2} + q \frac{1+d_2}{1+r_2} = 1\\
p + v+q=1. 
\end{eqnarray*}
This system admits the following unique solution 
\begin{eqnarray*}
p & = & \frac{r_2d_1 -r_1d_2}{d_1u_2-d_2u_1},\\ 
v  & = & \frac{(u_2-d_2)(u_1-r_1) -(u_1-d_1)(u_2-r_2)}{d_1u_2-d_2u_1}, \\
 q  & = &\frac{r_1u_2 -r_2u_1}{d_1u_2-d_2u_1}.
\end{eqnarray*}
We still have to check that $0<v <1-\max(\frac{r_1}{u_1}, \frac{r_2}{u_2})$. We propose below an example where this condition holds true. We choose 
$d_1=\frac12$, $u_1=2$ and $r_1=1$ together with $d_2=3$, $u_2=15$ and $r_2=\frac{29}4.$ We find that $v=\frac14<1-\max(\frac{1}{2}, \frac{29}{60})$, $p=\frac5{12}$ and $q=\frac13.$  So, all the conditions are satisfied: the submarkets $\tau_1$ and $\tau_2$ are arbitrage free  but incomplete whereas the global market is complete. 
\end{example}
\subsection{Brownian model}
\subsubsection{General model}
Let $W$ be a $d$-dimensional Brownian motion on $(\Omega, \Fc, \P)$, adapted to the filtration $(\Fc_t)_{0\leq t \leq T}$, where we suppose that $d \geq \sum_{\t \in \mathcal T} d_\t$. We assume in the rest of the section (and of the next one) that all processes have c\`ad-l\`ag trajectories and are adapted  to this filtration. We also use the notation $\la \cdot, \cdot \ra$ for the scalar product. 
We postulate that the risky assets $S^{\tau}$ follow the following SDE.
\begin{assumption}
\label{a:riskyasset}
Let $\t \in \mathcal T,$ there exist an $\R^{d_{\tau}}$-valued process $(M_t^\t)_{t\geq 0}$ such that $\int_0^{T} \| M_t^\t \| dt < \infty$ a.s.,   
a $d_{\tau} \times d$ matrix-valued process $(H_t^\t)_{t\geq 0}$ such that \\$\sum_{i=1}^{d_{\tau}} \sum_{j=1}^{d}\int_0^{T}|(H_t^\t)_{i,j}|^2 dt < \infty$ a.s. and
\begin{equation}
\label{eq:actifrisque}
d S_t^{\tau}= \mbox{diag}(S_t^{\tau})\left(M_t^\t dt + \la H_t^\t, d W_t \ra \right),
\end{equation}
where $\mbox{diag}(S_t^{\tau})$ is the $d_{\tau} \times d_{\tau}$ diagonal matrix with the $d_{\tau}$-vector $S_t^{\tau}$ on the diagonal. 
\end{assumption}
We adopt a spread form for the num\'eraires $S^{\t,0}$ for each  $\t \in \mathcal T,$ i.e., 
\begin{equation}
\label{eq:actifsansrisque}
S^{\t,0}_t =S^{\t,0}_0 \exp \left( \int_0^t (r_u + s^\t_u ) du \right),
\end{equation}
where the process $r$ is the reference risk-free rate and the process $s^\t$ denotes the short-term spread for the tenor $\t$. We suppose that 
$\int_0^{T} | r_t| dt < \infty$ a.s. and $\int_0^{T} | s_t^\t | dt < \infty$ a.s. 
%and assume also that the $\tilde S^{\tau}$ are locally bounded processus (for example, we may assume that $r$ and $s$ have continuous trajectories). A VOIR.
\\
We assume that the density process ${d\Q^{\t}}/{d\P}$ related to the measure change has the following property.
\begin{assumption}
\label{a:historical-density}
Let $\t \in \mathcal T.$ For every $\Q^{\t} \in \Qc^{\tau},$  there exists an $\R^d$-valued process $(\Lambda_t^\t)_{t\geq 0}$ such that $\int_0^{T} \| \Lambda_t^\t \|^2 dt < \infty$ a.s. and
\begin{equation}
\label{eq:lambda}
\frac{d\Q^{\t}}{d\P}\Big|_{\Fc} = \exp \left( \int_0^T  \la \Lambda^\tau_t, d W_t \ra - \frac12  \int_0^T \| \Lambda^\tau_t \|^2 dt \right).
\end{equation}
\end{assumption}
Then, the process
\begin{eqnarray}
\label{eq:brownien}
d W^{\tau}_t= d W_t -\Lambda^\tau_t dt
\end{eqnarray}
 is a $\Q^{\t}$-Brownian motion by Girsanov's theorem. 

Let $\t\in \Tc$ and $\Q^{\t} \in \Qc^{\tau}$ as in \eqref{eq:lambda}. Definition \ref{defQZ} implies that $\tilde{S}^{\tau}$ is a $\Q^{\t}$ local martingale. Using It\^o\rq{}s formula, we find that for all $\t \in \mathcal T,$ 
\begin{eqnarray*}
d \tilde{S}_t^{\tau}  & = & \mbox{diag}(\tilde{S}_t^{\tau})\left( \left(M_t^\t -(r_t +s^\t_t)1_{d_{\tau}}\right)  dt + \la H_t^\t, d W_t \ra \right)\\
& = & \mbox{diag}(\tilde{S}_t^{\tau}) \left(\left(M_t^\t -(r_t +s^\t_t)1_{d_{\tau}} + \la H_t^\t, \Lambda^{\tau}_t \ra\right)  dt + \la H_t^\t, d W^{\tau}_t \ra\right),
\end{eqnarray*}
where $1_{d_{\tau}}$ is the $d_{\tau}$-vector with all coordinates equal to one. 
As $W^{\tau}$ is a $\Q^{\t}$-Brownian motion,    $\Lambda^{\tau}_t$ must satisfy for all  $\t\in \Tc$  that 
\begin{eqnarray}
\label{drift}
 \la H_t^\t, \Lambda^{\tau}_t \ra=(r_t +s^\t_t)1_{d_{\tau}} - M_t^\t\; a.s.
\end{eqnarray}  
Then,
\begin{eqnarray}
\label{driftbis}
d \tilde{S}_t^{\tau}  & = &  \mbox{diag}(\tilde{S}_t^{\tau})  \la H_t^\t, d W^{\tau}_t \ra. 
\end{eqnarray}
We want to find the other relations that the $\Lambda^{\tau}_t $ should satisfy. 
Let $\t^* \in \ \mathcal T.$  Recalling Definition \ref{defQZ},
$\Q^{\t^*} \in \Qc^{\tau^*}$  if and only if $\Q^{\t^*} \sim \P$,
${\frac{d\Q^{\t^*}}{d\P}}/{\bar{S}^{{\t^*},0}_T} \in L^1$ and for all $\t \in \ \mathcal T$ 
$$\left(
\E_{\Q^{\tau^*}}\left(
\frac{
\bar S^{\t,0}_T
}
{
\bar{S}^{\t^*,0}_T
}
| \Fc_t
\right)
\tilde{S}_t^{\t}
\right)_{t \geq 0} \mbox{ is a local  martingale under $\Q^{\tau^*}$.} $$
So, we need to compute $X_t=\E_{\Q^{\t^*}} \left(\frac{\bar S^{\t,0}_T}{ \bar S^{\t^*,0}_T}| \Fc_t\right)=
\E_{\Q^{\t^*}} \left(\exp \left( \int_0^T s_u  du\right) | \Fc_t\right), $ 
%for all  $\t \in T$ such that $\t \neq \t^*,$ 
where $s= s^{\t} -  s^{\t^*}$ denotes the  difference between the spreads $ s^{\t}$ and $ s^{\t^*}$. 
For that,  we assume that the spreads  follow an affine Hull-White (time dependent Vasicek) dynamic under their respective measures $\Q^{\t} \in \Qc^{\tau},$ where the coefficient $b_t$ does not depend from $\tau$.
\begin{assumption}
\label{a:spread-1}
Let $\t \in \mathcal T,$ let $\Q^{\t} \in \Qc^{\tau}$ satisfying Assumption \ref{a:historical-density} and let $W^{\tau}$ be 
the $\Q^{\t}$-Brownian motion defined in \eqref{eq:brownien}.  The spread $s^{\tau}$  follows
\begin{eqnarray*}
d s^{\tau}_t = (a_t^{\tau} - b_t s^{\tau}_t) dt + \la \Sigma^{\tau}_t, d W^{\tau}_t \ra,
\end{eqnarray*}
where $t \mapsto a_t^{\t}, \, b_t $ are nonrandom, real-valued,  positive functions of time and $t \mapsto \Sigma_t^{\t}$ is a nonrandom,  $\R^d_+$-valued function of time.  
\end{assumption}
%Using It\^o formula, we get that 
%\begin{eqnarray*}
%s^{\tau}_t  & = &  s^{\tau}_0 e^{-\int_0^t b_u du }+ \int_0^t e^{-\int_s^t b_u du }a_s^{\tau}  ds + \int_0^t e^{-\int_s^t b_u du }\la \Sigma^{\tau}_s , d W^{\t}_s \ra\\
%s_t & = &  s_0 e^{-\int_0^t b_u du }+ \int_0^t e^{-\int_s^t b_u du }a_s  ds + \int_0^t e^{-\int_s^t b_u du }\left(\la \Sigma^{\tau}_s , d W^{\t}_s \ra 
%-\la\Sigma^{\tau^*}_s , d W^{\t^*}_s \ra \right)\\
%\end{eqnarray*}
Let $\tau, \tau^* \in \Tc$ and let $\Q^{\t} \in \Qc^{\t}$ and $\Q^{\t^*} \in \Qc^{\t^*},$ both satisfying Assumption \ref{a:historical-density}. Let $\Lambda^{\tau}$ (resp. $\Lambda^{\tau^*}$) be the measure change associated to $\Q^{\t}$  (resp. $\Q^{\t^*}$) in Assumption \ref{a:historical-density}. We  set 
$\Lambda= \Lambda^{\t} - \Lambda^{\t^*}.$ Using that $dW^{\t}_t =  dW^{\t^*}_t - \Lambda_t dt$, which follows from \eqref{eq:brownien}, we obtain that 
$$
d s^{\t}_t = (a^{\t}_t - \la \Sigma^{\t}_t, \Lambda_t \ra - b_t s^{\t}_t) dt + \la \Sigma^{\t}_t, d W^{\t^*}_t \ra
$$
and hence, setting $\Sigma_t=  \Sigma^{\t}_t - \Sigma^{\t^*}_t$ and $a_t=a_t^{\t} - a_t^{\t^*},$ 
we get that 
\begin{eqnarray*}
%\label{eq:spread-2}
d s_t = \left(a_t -\la \Sigma^{\t}_t, \Lambda_t \ra- b_t s_t\right) dt + \la \Sigma_t , d W^{\t^*}_t \ra.
\end{eqnarray*}
At this stage, in order to obtain explicit results and apply the standard affine machinery, we have to assume that $T$ is deterministic  and that  the spread $s$ has an affine Vasicek-type dynamic with time-dependent coefficients. 
\begin{assumption}
\label{a:deterministic}
The stopping time $T$ is deterministic and for all $\t \in \mathcal T,$ the function  
$
 t \mapsto \la \Sigma^{\t}_t, \Lambda_t \ra
$
is  a deterministic function of time. 
\end{assumption}
%Under this assumption, using It\^o formula, we get that 
%\begin{eqnarray*}
%%\label{eq:spread-2}
%s_t = s_0 e^{-\int_0^t b_u du }+ \int_0^t e^{-\int_s^t b_u du }\left(a_s -\la \Sigma^{\t}_s, \Lambda_s \ra \right) ds + \int_0^t e^{-\int_s^t b_u du }\la \Sigma_s , d W^{\t^*}_s \ra.
%\end{eqnarray*}
Assuming that $T$ is deterministic is indeed an approximation that allows to obtain explicit results. The computation of the law of $T$ and the adaptation of the affine machinery would be very technical and would not add much to the article message.   Then,  using the standard affine machinery (see for example \cite[Theorem 10.4]{Filipovic09} or formula (6.5.8)-(6.5.11) in \cite{shreve2004}), we can calculate explicitly that 
\begin{equation}
\label{eq:spread-int}
\E_{\Q^{\t^*}} \left(\exp  \left( \int_t^T s_u  du \right)  \Big| \Fc_t \right) = \exp \left(m^{\Lambda}(t, T) + \widetilde{b}(t, T) s_t\right),
\end{equation}
where $m^{\Lambda}(t, T)$ and $\widetilde{b}(t, T)$ are deterministic functions obtained by solving the associated system of Riccati equations~:
\begin{eqnarray*} 
\frac{\partial  \widetilde{b}(t, T)}{\partial t} & = & b_t \widetilde{b}(t, T) - 1, \quad 
\widetilde{b}(t, T)  =  \int_t^T e^{- \int_t^s b_udu} ds \\
%\frac{1}{b} \left( 1 - e^{-b (T-t)} \right)\\
%dm(t,T) & = & \frac12  \|\Sigma_t \|^2 \widetilde{b}^2(t, T)  + \left( a_t - \la \Sigma^{\t}_t, \Lambda_t \ra \right)\widetilde{b}(t, T) \\
m^{\Lambda}(t,T) & = & \frac12 \int_t^T \|\Sigma_s \|^2 \widetilde{b}^2(s, T) ds + \int_t^T a_s\widetilde{b}(s, T) ds -\int_t^T \la \Sigma^{\t}_s, \Lambda_s \ra\widetilde{b}(s, T) ds.
\end{eqnarray*}
Note that $\widetilde{b}(t, T) \geq 0$. 

%Thus,
%\begin{eqnarray*}
%\E_{\Q^{\t^*}} \left(\exp  \left( \int_0^T s_u  du \right)   \right) & = & \exp \left(\frac12 \|\Sigma \|^2 \int_0^T \widetilde{b}^2(s, T) ds +a\int_0^T\widetilde{b}(s, T) ds  + \widetilde{b}(0, T) s_0 \right)  \\
%& &   \times \exp\left(-  \int_0^T  \la \Sigma^{\t}_t, \Lambda_s^{\t}  \ra \widetilde{b}(s, T) ds \right) 
% \exp\left(  \int_0^T  \la \Sigma^{\t}_t, \Lambda_s^{\t^*} \ra \widetilde{b}(s, T) ds \right). 
%\end{eqnarray*}
So, \eqref{eq:spread-int} implies that  $X_t=\exp  \left( \int_0^t s_u  du  \right)  \exp (m^{\Lambda}(t, T) + \widetilde{b}(t, T) s_t)$ and 
 It\^o\rq{}s formula yields that 
\begin{eqnarray*}
dX_t & = & X_t \left(
\left(s_t -
\frac12  \|\Sigma_t \|^2 \widetilde{b}^2(t, T)  - \left( a_t - \la \Sigma^{\t}_t, \Lambda_t \ra \right)\widetilde{b}(t, T)
+ b_t \widetilde{b}(t, T)s_t - s_t   \right)dt
\right.\\
%- \frac12 \|\Sigma \|^2 \widetilde{b}^2(t, T)  -  \left(a- \la \Sigma^{\t}_t, \Lambda_t \ra \right)\widetilde{b}(t, T) -e^{-b (T-t)}s_t \right)dt+\right.\\
& &\left. + \widetilde{b}(t, T) ds_t+\frac12\widetilde{b}^2(t, T)d<s>_t \right) \\
& = & X_t 
\widetilde{b}(t, T) \la \Sigma_t , d W^{\t^*}_t \ra.
\end{eqnarray*}
Moreover, recalling \eqref{driftbis}, for all $ \t\in \Tc \setminus\{\t^*\}$
$$d \tilde{S}_t^{\t} =   \mbox{diag}(\tilde{S}_t^{\tau}) \left(-\la H_t^{\t}, \Lambda_t \ra  dt + \la H_t^{\t}, d W^{\t^*}_t \ra \right).$$
So, using the integration by part formula, we get that 
\begin{eqnarray*}
d(X_t \tilde{S}^{\t}_t)& = & X_t  \mbox{diag}(\tilde{S}_t^{\tau})\left(\left( -\la H_t^{\t}, \Lambda_t \ra + \widetilde{b}(t, T) \la H_t^{\t}, \Sigma_t \ra \right)dt+ 
\widetilde{b}(t, T) \la  \Sigma_t, d W^{\t^*}_t \ra  1_{d^{\tau}} \right. \\
& & \left. + \la   H_t^{\t}, d W^{\t^*}_t \ra \right) .
\end{eqnarray*}
As $X \tilde{S}^{\t}$ is a $\Q^{\t^*}$ local martingale, we get that for all $ \t\in \Tc \setminus\{\t^*\}$
\begin{eqnarray}
\label{vol}
\la H_t^{\t}, \Lambda_t \ra = \widetilde{b}(t, T) \la H_t^{\t}, \Sigma_t \ra\; a.s.
\end{eqnarray}
%As $ \la H_t^{\t}, \Lambda^{\t}_t \ra  =  (r_t +s^{\t}_t)1_{d_{{\t}}} - M_t^{\t},$  
%$$\la H_t^{\t}, \Lambda^{\t^*}_t \ra = -\widetilde{b}(t, T) \la H_t^{\t}, \Sigma \ra +(r_t +s^{\t}_t)1_{d_{{\t}}} - M_t^{\t} \; a.s.$$
%
%\begin{proposition}
%\label{propwhat2}
%Assume that Assumptions \ref{a:riskyasset}-\ref{a:deterministic} hold. Then, 
%$\Q^{\t^*} \in \Qc^{\tau^*}$  if and only if 
%for all $t\geq 0$. 
%\begin{eqnarray*}
% \la H_t^{\t^*}, \Lambda^{\t^*}_t \ra & = & (r_t +s^{\t^*}_t)1_{d_{{\t^*}}} - M_t^{\t^*}\; a.s.\\
% \la H_t^{\t}, \Lambda^{\t^*}_t \ra & = & -\widetilde{b}(t, T) \la H_t^{\t}, \Sigma \ra +(r_t +s^{\t}_t)1_{d_{{\t}}} - M_t^{\t} \; a.s.\; \mbox{for all } \t\in \Tc \setminus\{\t^*\},
%%\la H_t^{\t}, \Lambda^{\t} - \Lambda^{\t^*} \ra &=& \widetilde{b}(t, T) \la H_t^{\t}, \Sigma \ra\; a.s. \; \mbox{for all } \t\in \Tc \Sigmaetminus\{\t^*\}, \; \Lambda^{\t} \mbox{ as in } \eqref{eq:lambda}.  
%\end{eqnarray*}
%\end{proposition}
%\begin{proof}
%\end{proof}
So, we have proved the following proposition. 
%${\frac{d\Q^{\t^*}}{d\P}}/{\bar{S}^{{\t^*},0}_T} =\exp \left( \int_0^T  \la \Lambda^{\t^*}_t, d W_t \ra - \frac12  \int_0^T \| \Lambda^{\t^*}_t \|^2 dt - \int_0^T (r_t + s^{\t^*}_t ) dt\right)$
\begin{proposition}
\label{propmatin}
Assume that Assumptions \ref{a:riskyasset}, \ref{a:historical-density}, \ref{a:spread-1} and \ref{a:deterministic} hold. 
Then, $\Q^{\t^*} \in \Qc^{\tau^*}$  if and only if 
${\frac{d\Q^{\t^*}}{d\P}}/{\bar{S}^{{\t^*},0}_T} \in L^1$ and
\begin{eqnarray*}
 \la H_t^{\t^*}, \Lambda^{\t^*}_t \ra & = & (r_t +s^{\t^*}_t)1_{d_{\t^*}} - M_t^{\t^*}\; a.s.\\
\la H_t^{\t}, \Lambda^{\t^*}_t \ra & = & \la H_t^{\t}, \Lambda^{\t}_t \ra-\widetilde{b}(t, T) \la H_t^{\t}, \Sigma^{\t}_t - \Sigma^{\t^*}_t \ra\; a.s. \mbox{ for all } \t\in \Tc \setminus\{\t^*\},
\end{eqnarray*}
where  for all $\tau \in \Tc,$ $\Lambda^{\tau}$ is the measure change associated to $\Q^{\t} \in \Qc^{\t}$ by Assumption \ref{a:historical-density}. 
\end{proposition}
Note that if $\hat \Q^{\t^*}$ satisfies Assumption \ref{a:historical-density} (denoting by $\hat\Lambda^{\t^*}$ the measure change associated to $\hat \Q^{\t^*}$), then under Assumption \ref{a:riskyasset},  $\hat \Q^{\t^*} \in \hat \Qc^{\tau^*}$  if and only if 
${\frac{d\hat\Q^{\t^*}}{d\P}}/{\bar{S}^{{\t^*},0}_T} \in L^1$ and
\begin{eqnarray*}
 \la H_t^{\t^*}, \hat \Lambda^{\t^*}_t \ra & = & (r_t +s^{\t^*}_t)1_{d_{\t^*}} - M_t^{\t^*}\; a.s.
\end{eqnarray*}
We see that $\hat \Qc^{\tau^*}$ may be much larger than $ \Qc^{\tau^*}.$ 

Now, we give the link between the spread $s$ and the risk premium process $\Lambda.$
\begin{proposition}
\label{propwhat}
Assume that Assumptions \ref{a:historical-density}, \ref{a:spread-1} and \ref{a:deterministic} hold. Let $\tau, \tau^* \in \Tc$ and let $\Q^{\t} \in \Qc^{\t}$ and $\Q^{\t^*} \in \Qc^{\t^*}.$ Let $\Lambda^{\tau}$ (resp. $\Lambda^{\tau^*}$) be the measure change associated to $\Q^{\t}$  (resp. $\Q^{\t^*}$) in Assumption \ref{a:historical-density}. 
Then, the process $\Lambda =\Lambda^{\tau}- 
\Lambda^{\tau^*}$ satisfies the following equalities 
for all $t\geq 0$. 
\begin{eqnarray}
\nonumber
\la  \Lambda_t, d W_t^{\t^*} \ra &  = & \widetilde{b}(t, T)\la \Sigma_t, d W_t^{\t^*} \ra \\
\label{norme}
\|\Lambda_t \| &  = &  \widetilde{b}(t, T) \|\Sigma_t \|. 
\end{eqnarray}
\end{proposition}
We see from Proposition \ref{propwhat}, that $t \mapsto \Lambda_t$ is a deterministic function of time. 
We also see that \eqref{vol} and \eqref{norme} provide $\sum_{\t \in \Tc} d_{\tau}+1$ equations in order to get the $d$ coordinates of $\Lambda$. 
Thus, if we assume that $d= \sum_{\t \in \Tc} d_{\tau}+1$, we may be able to characterize $\Lambda$. This will be done in the next section. 
%Morever, $ \eqref{drift}$ allows to obtain the 
%$\Lambda^{\tau}$.
%As $\la H_t^{\t}, \Lambda^{\t}_t \ra  =  (r_t +s^{\t}_t)1_{d_{{\t}}} - M_t^{\t}\; a.s.$ If $t \mapsto H_t^{\t}, \, M_t^{\t},\, r_t $  are deterministic function of time, then so should be $t \mapsto s^{\t}_t$, a contradiction!
\begin{proof}
From Assumption \ref{a:historical-density}, we have
\begin{eqnarray*}
\frac{d\Q^{\t}}{d\Q^{\t^*}}\Big|_{\Fc} & = & \exp \left( \int_0^T  \la \Lambda^{\t}_s - \Lambda^{\t^*}_s ,d W_s  \ra- \frac12 \int_0^T (\| \Lambda^{\t}_s \|^2 -  \| \Lambda^{\t^*}_s \|^2) ds \right) \\
 & = &  \exp \left( \int_0^T  \la \Lambda_s , d W^{\t^*}_s \ra -\frac12 \int_0^T \| \Lambda_s \|^2  ds \right).
\end{eqnarray*}
As $W^{\t^*}$ is a $\Q^{\t^*}$ Brownian motion, we get that 
\begin{eqnarray}
\label{calculsto}
\E_{\Q^{\t^*}} \left(\frac{d\Q^{\t}}{d\Q^{\t^*}}\Big|_{\Fc}\Big|{\Fc_t} \right)& = &   \exp \left( \int_0^t  \la \Lambda_s , d W^{\t^*}_s \ra -\frac12 \int_0^t \| \Lambda_s \|^2  ds \right).
\end{eqnarray}
Using \eqref{Qtau}, we obtain
\begin{eqnarray}\label{density-Qs-1}
\frac{d\Q^{\t}}{d\Q^{\t^*}} \Big|_{\Fc} & = & 
\frac{\E(X^*S^{\t^*,0}_T)}{\E(X^*S^{\t,0}_T)} 
\frac{S^{\t,0}_T}{S^{\t^*,0}_T} = C
\exp  \left( \int_0^T s_u  du \right),
\end{eqnarray}
where $C= \frac{\E(X^*S^{\t^*,0}_T)}{\E(X^*S^{\t,0}_T)}$ is a constant. Hence, combining  \eqref{eq:spread-int},  \eqref{calculsto} and \eqref{density-Qs-1}, we obtain that
$$ \int_0^t  \la \Lambda_s , d W^{\t^*}_s \ra -\frac12 \int_0^t \| \Lambda_s \|^2  ds = \ln C +  \int_0^t s_u  du
+ m^{\Lambda}(t, T) + \widetilde{b}(t, T) s_t.$$
Thus,
\begin{eqnarray*}  \la \Lambda_t , d W^{\t^*}_t \ra -\frac12  \| \Lambda_t \|^2  dt  &=&   s_t  dt 
+ dm^{\Lambda}(t, T) +  s_t d\widetilde{b}(t, T) +\\ 
& & \widetilde{b}(t, T) \left( (a_t- \la \Sigma^{\t}_t, \Lambda_t \ra  - b_t s_t) dt + \la \Sigma_t , d W^{\t^*}_t \ra\right).
\end{eqnarray*}
So, we see that $\la \Lambda_t , d W^{\t^*}_t \ra  =   \widetilde{b}(t, T) \la \Sigma_t , d W^{\t^*}_t \ra$ and that
\begin{eqnarray*}
-\frac12  \| \Lambda_t \|^2  &=& 
s_t -
\frac12  \|\Sigma_t \|^2 \widetilde{b}^2(t, T)  - \left( a_t - \la \Sigma^{\t}_t, \Lambda_t \ra \right)\widetilde{b}(t, T) + 
 \\
 & &
 \widetilde{b}(t, T)  (a_t- \la \Sigma^{\t}_t, \Lambda_t \ra  - b_t s_t) + b_t \widetilde{b}(t, T)s_t - s_t  \\
& = &  - \frac12 \|\Sigma_t \|^2 \widetilde{b}^2(t, T),
\end{eqnarray*}
which concludes the proof.  $\Box$
\end{proof}

\subsubsection{Application}
\label{concret}
In order to illustrate Proposition \ref{propdeux}, we assume now that there are only two submarkets with one risky asset in each. Let $d=3$ and $W=(W_0,W_1,W_2)$ be a $3$-dimensional Brownian motion on $(\Omega, \Fc, \P)$. Again, in order to obtain explicit results, we make an approximation and assume that $T$ is deterministic. 

We propose below an incomplete model where each submarket is driven by its own Brownian motion and by the common noise $W_0$. We are in the situation mentioned after Proposition \ref{propwhat} and we will be able to calculate $\Lambda$  as a solution of \eqref{eqsyst}. This will allow us to calculate the different prices of Proposition \ref{propdeux}.  
%For that we assume some Novikov condition so that, for $i \in \{1,2\},$ $\widetilde S^{\tau_i}$ is a true martingale under any $\Q \in \hat \Qc^{\tau_i},$ see \eqref{driftbis}. 
\begin{assumption}
\label{a:apply}
For $i \in \{1,2\}$, we consider  that 
%$\R$-valued processes $(\mu_{i,t} )_{t\geq 0}$ and $(\eta_{i,t} )_{t\geq 0}$ such that $\int_0^{T} | \mu_{i,t}  | dt < \infty$ a.s., $\eta_{i,t}>0$ a.s. and 
%$\E e^{\frac12 \int_0^{T} \eta_{i,t}^2 dt} < \infty.$ We assume that 
 $t \mapsto  \mu_{i,t}, \; \eta_{i,t}, \; \rho_{i,t}, \; \kappa_{i,t}, \; \sigma_{i,t}$  are deterministic functions of time such that $0<\rho_{i,t}< 1, \;|\kappa_{i,t}| \leq 1,\; \sigma_{i,t}>0, \; \eta_{i,t}>0,$ 
 $\int_0^{T} | \mu_{i,t}  | dt < \infty$, $\int_0^{T} \eta_{i,t}^2 dt < \infty$ 
 and 
$\int_0^{T} \sigma_{i,t}^2 dt < \infty.$ 
 % such that   
%$\int_0^{T} |\rho_{i,t} \eta_{i,t}|^2 dt < \infty,$ $\int_0^{T} |\sqrt{1- \rho_{i,t}} \eta_{i,t}|^2 dt < \infty.$ 
%Let also $\sigma_i>0 $ and $\kappa_i \in (-1,1).$  
Then, we set 
\begin{eqnarray*}
M_t^{\t_1} & = & \mu_{1,t}  \mbox{ and } M_t^{\t_2}  =  \mu_{2,t}\\
H^{\t_1}_t & = & \left(\eta_{1,t}\sqrt{1- \rho_{1,t}^2},\eta_{1,t}\rho_{1,t} , 0 \right)  \mbox{ and }
H^{\t_2}_t   =  \left(\eta_{2,t}\sqrt{1- \rho_{2,t}^2} ,0,\eta_{2,t}\rho_{2,t}\right) \\ 
\Sigma^{\t_1}_t & = & \left(\sigma_{1,t}\sqrt{1- \kappa_{1,t}^2}, \sigma_{1,t} \kappa_{1,t}, 0 \right)  \mbox{ and }
\Sigma^{\t_2}_t  =  \left(\sigma_{2,t}\sqrt{1- \kappa_{2,t}^2},0,\sigma_{2,t} \kappa_{2,t}\right). 
\end{eqnarray*}
\end{assumption}
Under, these conditions, for $i \in \{1,2\},$ $\widetilde S^{\tau_i}$ is a true martingale under any $\Q \in \hat \Qc^{\tau_i},$ see \eqref{driftbis}. We may take more general assumptions as some Novikov condition 
$\E_{\Q} e^{\frac12 \int_0^{T} \eta_{i,t}^2 dt} < \infty,$ but we wanted  to keep it simple. 
Then,  we set 
\begin{eqnarray*}\Sigma_t & = &  \Sigma^{\t_1}_t -\Sigma^{\t_2}_t=(\Sigma_{1,t},\Sigma_{2,t}, \Sigma_{3,t}) \\& = & \left(\sigma_{1,t}\sqrt{1- \kappa_{1,t}^2}-\sigma_{2,t}\sqrt{1- \kappa_{2,t}^2}, \sigma_{1,t} \kappa_{1,t}, -\sigma_{2,t} \kappa_{2,t} \right).
\end{eqnarray*}
Let $\Q^{\t_1} \in \Qc^{\t_1}$ and $\Q^{\t_2} \in \Qc^{\t_2}.$ Let $\Lambda^{\tau_1}$ (resp. $\Lambda^{\tau_2}$) be the measure change associated to $\Q^{\t_1}$  (resp. $\Q^{\t_2}$) in Assumption \ref{a:historical-density}. We set 
$\Lambda=\Lambda^{\t_1}-\Lambda^{\t_2}$ and $s=s^{\t_1}- s^{\t_2} $.

We first want to compute ${\pi}^{\t_2}(S^{\tau_1}_T) $ and recalling \eqref{qui}, \eqref{eq:actifsansrisque} and \eqref{eq:spread-int} 
\begin{eqnarray}
\label{cestclair}
%\sup_{\Q^{\t_2} \in \Qc^{\tau_2}} 
\E_{\Q^{\t_2}} \left(\frac{\bar S^{\tau_1,0}_T}{ \bar S^{\tau_2,0}_T}\right)
& = & 
%\sup_{\Q^{\t_2} \in \Qc^{\tau_2}} 
\E_{\Q^{\t_2}} \left(\exp \left( \int_0^T s_u  du\right) \right)  =
%\\ & = &   \sup_{\Lambda} 
\exp (m^{\Lambda}(0, T) + \widetilde{b}(0, T) s_0).
\end{eqnarray}
So, we need to determine the possible values for $\Lambda$. Using Proposition \ref{propmatin} for $\tau=\tau_1$ and $\tau^*=\tau_2$ and for $\tau=\tau_2$ and $\tau^*=\tau_1$ and Proposition \ref{propwhat} for $\tau=\tau_1$ and $\tau^*=\tau_2,$ we get that  
\begin{equation}\label{eqsyst}
\begin{array}{rcl}  \displaystyle 
\la H_t^{\t_1}, \Lambda_t \ra  & =  & \widetilde{b}(t, T) \la H_t^{\t_1}, \Sigma_t\ra\; \\
\la H_t^{\t_2}, \Lambda_t \ra &  = &  \widetilde{b}(t, T) \la H_t^{\t_2}, \Sigma_t\ra\; \\
\|\Lambda_t \|^2 &  = &  \widetilde{b}^2(t, T) \|\Sigma_t\|^2, 
\end{array}
 \end{equation}
where $\Lambda_t=\Lambda^{\t_1}_t-\Lambda^{\t_2}_t=(\Lambda_{1,t},\Lambda_{2,t}, \Lambda_{3,t}).$ We now solve \eqref{eqsyst}. 
Setting $\sqrt{\frac1{ \rho_{i,t}^2}-1}=r_{i,t}$ for $i\in \{1,2\}$, after some computations, we get that 
\begin{eqnarray*}
\Lambda_{1,t }^2 \left(r^2_{1,t} +r^2_{2,t} +1\right)  
  -2\widetilde{b}(t, T)\Lambda_{1,t }\left(\Sigma_{1,t} \left(r^2_{1,t} +r^2_{2,t} \right)  +
   \Sigma_{2,t}r_{1,t} +  \Sigma_{3,t} r_{2,t}\right) 
  \\
 + \widetilde{b}^2(t, T) \left(
  \Sigma^2_{1,t} \left(r^2_{1,t} +r^2_{2,t} -1\right)  +  2 \Sigma_{1,t}\left(\Sigma_{2,t}  r_{1,t}
    + \Sigma_{3,t} r_{2,t} \right) \right)  =0. 
\end{eqnarray*}
We compute the (reduced) discriminant of this quadratic equation. 
\begin{eqnarray*}
\Delta & = & \widetilde{b}^2(t, T)\left( \Sigma_{1,t} \left(r^2_{1,t} +r^2_{2,t} \right)  +
   \Sigma_{2,t}r_{1,t} +  \Sigma_{3,t} r_{2,t} \right)^2 - \\ 
   & & \widetilde{b}^2(t, T)\left(r^2_{1,t} +r^2_{2,t} +1\right)   \left(
  \Sigma^2_{1,t} \left(r^2_{1,t} +r^2_{2,t} -1\right)  +  2 \Sigma_{1,t}\left(\Sigma_{2,t}  r_{1,t}
   + \Sigma_{3,t} r_{2,t} \right) \right) \\
 & = &   \left(\widetilde{b}(t, T) \left( \Sigma_{1,t}-\Sigma_{2,t} r_{1,t} -\Sigma_{3,t} r_{2,t} \right)\right)^2.
\end{eqnarray*}
The two solutions of the quadratic equation are 
\begin{eqnarray*}
\Lambda_{t }^1 & = & \widetilde{b}(t, T)  \Sigma_t\\
\Lambda_{t }^2& = &\frac{\widetilde{b}(t, T)}{r^2_{1,t} +r^2_{2,t} +1}\left(
\Sigma_{1,t} \left(r^2_{1,t} +r^2_{2,t}-1 \right)  +
  2 \Sigma_{2,t}r_{1,t} +  2\Sigma_{3,t} r_{2,t} \; ,  \;
  2 \Sigma_{1,t}   r_{1,t} + \right.  \\
  & & \left.
\Sigma_{2,t}  \left(  
-r^2_{1,t} +r^2_{2,t} +1
 \right) 
-  2\Sigma_{3,t} r_{1,t} r_{2,t}\; ,  
2 \Sigma_{1,t}   r_{2,t}   -  2\Sigma_{2,t} r_{1,t} r_{2,t} 
+\Sigma_{3,t}  \left(  
r^2_{1,t} -r^2_{2,t} +1
 \right)  \right).
\end{eqnarray*}
So, solving \eqref{eqsyst} provides two possible values for $\Lambda$. Recalling \eqref{cestclair}, we now compute the associated values of $m^{\Lambda}(t,T)$. 
%Note that using \eqref{drift}, we can obtain for $i\in \{1,2\},$ $\Lambda^{\tau_1,i}_t$ and $\Lambda^{\tau_2,i}_t$ as a function of $\Lambda^{\tau_2,i}_{1,t}=x$ 
%\begin{scriptsize}
%\begin{eqnarray*}
% \Lambda^{\tau_1,i}_{t}  & = &  \left(\Lambda^i_{1,t}  +x\; ,  \;
% \frac{r_t +s^{\t_1}_t - \mu_{1,t}}{\rho_{1,t} \eta_{1,t} } - \sqrt{\frac1{\rho_{1,t}^2}-1} \left( \Lambda^i_{1,t}  +x\right)\; ,  \; \Lambda^i_{3,t}  +\frac{r_t +s^{\t_2}_t - \mu_{2,t}}{\rho_{2,t} \eta_{2,t} } -   x\sqrt{\frac{1}{ \rho_{2,t}^2}-1}\right)\; \\
% \Lambda^{\tau_2,i}_{t} & = & \left(x \; ,  \;-\Lambda^i_{2,t}+\frac{r_t +s^{\t_1}_t - \mu_{1,t}}{\rho_{1,t} \eta_{1,t} } - \sqrt{\frac1{\rho_{1,t}^2}-1} \left( \Lambda^i_{1,t}  +x\right)  \; ,  \; \frac{r_t +s^{\t_2}_t - \mu_{2,t}}{\rho_{2,t} \eta_{2,t} } -  x \sqrt{\frac{1}{ \rho_{2,t}^2}-1} \right). 
%\end{eqnarray*} 
%\end{scriptsize}
First, we find that 
\begin{eqnarray*}
\la \Sigma^{\t_1}_t, \Lambda_t^1 \ra
& = & \widetilde{b}(t, T) \left(\sigma^2_{1,t}-\sigma_{1,t}\sigma_{2,t}\sqrt{1- \kappa_{1,t}^2}\sqrt{1- \kappa_{2,t}^2}\right).
\\ 
 \la \Sigma^{\t_1}_t, \Lambda_t^2 \ra &  = & 
\widetilde{b}(t, T) \left(\sigma^2_{1,t}-\sigma_{1,t}\sigma_{2,t}\sqrt{1- \kappa_{1,t}^2}\sqrt{1- \kappa_{2,t}^2}\right) - \frac{2\widetilde{b}(t, T) }{\frac1{\r^2_{1,t}} +\frac1{\r^2_{2,t}} -1}  
%\times \\
%& & 
\left(\hat \sigma^2_{1,t} -\hat \sigma_{1,t} \hat \sigma_{2,t} \right), 
%\left(\sigma^2_{1,t} \left(\sqrt{1- \kappa_{1,t}^2} -  \kappa_{1,t}\sqrt{\frac{1}{ \rho_{1,t}^2}-1}   \right)^2
%-\sigma_{1,t}\sigma_{2,t} \left(\sqrt{1- \kappa_{1,t}^2} -  \kappa_{1,t}\sqrt{\frac{1}{ \rho_{1,t}^2}-1}   \right)\left( 
%\sqrt{1- \kappa_{2,t}^2}- \kappa_{2,t}\sqrt{\frac{1}{ \rho_{2,t}^2}-1} \right)  
%\right),
\end{eqnarray*}
where  $\hat \sigma_{i,t}= \sigma_{i,t} \left(\sqrt{1- \kappa_{i,t}^2} -  \kappa_{i,t}\sqrt{\frac{1}{ \rho_{i,t}^2}-1}   \right)$, for $i \in \{1,2\}$. So, we get that 
\begin{eqnarray*}
m^{\Lambda^1}(t,T) & = & \frac12\int_t^T  \|\Sigma_s\|^2 \widetilde{b}^2(s, T) ds +    \int_t^T (a_s^{\t_1}- a_s^{\t_2} )\widetilde{b}(s, T) ds -\int_t^T \la \Sigma^{\t_1}_t, \Lambda^1_s \ra\widetilde{b}(s, T) ds\\
 & = & -   \int_t^T \frac{\sigma^2_{1,s}-\sigma^2_{2,s}}2\widetilde{b}^2(s, T)ds +   \int_t^T (a_s^{\t_1}- a_s^{\t_2} )\widetilde{b}(s, T) ds\\
m^{\Lambda^2}(t,T) & = & \frac12 \int_t^T  \|\Sigma_s\|^2\widetilde{b}^2(s, T) ds +  \int_t^T(a_s^{\t_1}- a_s^{\t_2} )  \widetilde{b}(s, T) ds -\int_t^T \la \Sigma^{\t_1}_t, \Lambda^2_s \ra\widetilde{b}(s, T) ds
\\
 & = & m^{\Lambda^1}(t,T) +
2 \int_t^T
\frac{\widetilde{b}(s, T) }{\frac1{\r^2_{1,s}} +\frac1{\r^2_{2,s}} -1}
\left(\hat \sigma_{1,s}^2 - \hat \sigma_{1,s} \hat \sigma_{2,s} \right)ds. 
%2 \int_t^T
%\frac{\widetilde{b}(s, T) }{\frac1{\r^2_{1,s}} +\frac1{\r^2_{2,s}} -1}
%\left(\sigma_{1,t}^2 \left(\sqrt{1- \kappa_{1,t}^2} -  \kappa_{1,t}\sqrt{\frac{1}{ \rho_{1,s}^2}-1}   \right)^2  \right.\\
%&&  \left.
%-\sigma_{1,t}\sigma_{2,t} \left( 
%\sqrt{1- \kappa_{2,t}^2}- \kappa_{2,t}\sqrt{\frac{1}{ \rho_{2,s}^2}-1} \right)   \left(\sqrt{1- \kappa_{1,t}^2} -  \kappa_{1,t}\sqrt{\frac{1}{ \rho_{1,s}^2}-1}   \right)
%\right)ds.
\end{eqnarray*}
 Using \eqref{cestclair}, we get that 
\begin{eqnarray*}
\sup_{\Q^{\tau_2} \in \Qc^{\tau_2}} 
\E_{\Q^{\tau_2}} \left(\frac{\bar S^{\tau_1,0}_T}{ \bar S^{\tau_2,0}_T}\right)
& = & \max \left(\exp (m^{\Lambda_1}(0, T) + \widetilde{b}(0, T) s_0), \exp (m^{\Lambda_2}(0, T) + \widetilde{b}(0, T) s_0) \right)\\
%\sup_{\Q \in \Qc^{\tau_2}} \E_{\Q} \left(\exp \left( \int_0^T s_u  du\right) \right) \\
%& = &  S_0^{\tau_1} \sup_{\Q \in \Qc^{\tau_2}} \exp (m^{\Lambda}(t, T) + \widetilde{b}(t, T) s_0)\\
& = &  \exp \left( \widetilde{b}(0, T) s_0 +\max \left(m^{\Lambda_1}(0, T), m^{\Lambda_2}(0, T)\right)\right)= \exp Y^M \\ 
\inf_{\Q^{\tau_2} \in \Qc^{\tau_2}} 
\E_{\Q^{\tau_2}} \left(\frac{\bar S^{\tau_1,0}_T}{ \bar S^{\tau_2,0}_T}\right)
%& = & \inf_{\Q \in \Qc^{\tau_2}} \E_{\Q} \left(\exp \left( \int_0^T s_u  du\right) \right) \\
%& = &  S_0^{\tau_1} \sup_{\Q \in \Qc^{\tau_2}} \exp (m^{\Lambda}(t, T) + \widetilde{b}(t, T) s_0)\\
& = &  \exp \left( \widetilde{b}(0, T) s_0 +\min \left(m^{\Lambda_1}(0, T), m^{\Lambda_2}(0, T)\right)\right)= \exp Y^m, \\ 
% &&  \exp\left( 2 \left[\int_0^T
%\frac{\widetilde{b}(s, T) }{\frac1{\r^2_{1,s}} +\frac1{\r^2_{2,s}} -1}
%\left(\sigma_{1,t}^2 \left(\sqrt{1- \kappa_{1,t}^2} -  \kappa_{1,t}\sqrt{\frac{1}{ \rho_{1,s}^2}-1}   \right)^2 \right. \right. \right.\\
%&&  \left.\left.\left.
%-\sigma_{1,t}\sigma_{2,t} \left( 
%\sqrt{1- \kappa_{2,t}^2}- \kappa_{2,t}\sqrt{\frac{1}{ \rho_{2,s}^2}-1} \right)   \left(\sqrt{1- \kappa_{1,t}^2} -  \kappa_{1,t}\sqrt{\frac{1}{ \rho_{1,s}^2}-1}   \right)
%\right)ds \right]_+\right)
\end{eqnarray*}
where, with the notations  $x_+=\max (x,0)$ and $x_-=\max (-x,0),$ we have set 
\begin{eqnarray*}
 Y^M 
& = &  \widetilde{b}(0, T) s_0 +m^{\Lambda_1}(0, T)+2 \left[\int_0^T
\frac{\widetilde{b}(s, T) }{\frac1{\r^2_{1,s}} +\frac1{\r^2_{2,s}} -1}
\left(\hat\sigma_{1,s}^2-\hat\sigma_{1,s}\hat\sigma_{2,s} 
\right)ds \right]_+ \\
Y^m
& = &  \widetilde{b}(0, T) s_0 +m^{\Lambda_1}(0, T)-2 \left[\int_0^T
\frac{\widetilde{b}(s, T) }{\frac1{\r^2_{1,s}} +\frac1{\r^2_{2,s}} -1}
\left(\hat\sigma_{1,s}^2-\hat\sigma_{1,s}\hat\sigma_{2,s} 
\right)ds \right]_-. 
\end{eqnarray*}
Using Proposition \ref{propdeux}, we obtain that 
\begin{eqnarray*} 
{\pi}^{\t_2}(S^{\tau_1}_T) & = & S^{\tau_1}_0 \exp Y^M \mbox{ and }
{\pi}^{\t_1}(S^{\tau_2}_T) =  S^{\tau_2}_0 \exp \left(-Y^m\right) \\
\pi(S^{\tau_1}_T) & = &{\pi}^{\t_1}(S_T^{\tau_1})  \wedge  {\pi}^{\t_2}(S^{\tau_1}_T) =
S_0^{\tau_1} \exp \left(- Y^M_-\right)\\
\pi(S^{\tau_2}_T) & = & 
S_0^{\tau_2} \exp \left(-Y^m_+\right). 
\end{eqnarray*}
If $S_0^{\tau_1} \exp Y^M   < S_0^{\tau_2},$ then $\pi(S^{\tau_1}_T-S^{\tau_2}_T)=+\infty$. Else, recalling \eqref{gilles}
\begin{eqnarray*} 
\pi(S^{\tau_1}_T-S^{\tau_2}_T) 
%& = &   \exp \left(-Y^M_+\right) \left(S_0^{\tau_1} \exp Y^M-  S_0^{\tau_2} \right)\\
 & = & S_0^{\tau_1}\exp \left(- Y^M_-\right)-  S_0^{\tau_2} \exp \left(-Y^M_+\right).
\end{eqnarray*}
We now illustrate the discussion after Proposition \ref{propdeux}. Case 3 corresponds to  $Y^m\leq 0$ and $Y^M\geq 0.$ The cost of illiquidity is equal to zero in both markets and 
\begin{eqnarray*}
\pi(S_T^{\tau_2}) \exp(-Y^M)+ \pi(S^{\tau_1}_T-S^{\tau_2}_T) 
& = & \pi(S_T^{\tau_1}).
\end{eqnarray*}
Case 2 corresponds to   $Y^m > 0.$  The cost of illiquidity for $S^{\tau_2}_T$ is $S_0^{\tau_2}(1-\exp \left(-Y^m\right))>0$, while the cost of illiquidity for $S^{\tau_1}_T$ is 0 and 
\begin{eqnarray*}
%\pi(S_T^{\tau_2}) \frac{\exp (Y^m}{\exp Y_M }+ \pi(S^{\tau_1}_T-S^{\tau_2}_T) & = & \pi(S_T^{\tau_1}).\\
\pi(S_T^{\tau_2}) \exp {\left\{-2 \left|\int_0^T
\frac{\widetilde{b}(s, T) }{\frac1{\r^2_{1,s}} +\frac1{\r^2_{2,s}} -1}
\left(\hat\sigma_{1,s}^2-\hat\sigma_{1,s}\hat\sigma_{2,s} 
\right)ds \right|\right\}} + \pi(S^{\tau_1}_T-S^{\tau_2}_T) 
& = & \pi(S_T^{\tau_1}).
\end{eqnarray*}
Finally, Case 1 occurs when $Y^M\ <0.$ Then, $\pi (S^{\tau_1}_T) = S_0^{\tau_1} \exp Y^M$ and $\pi (S^{\tau_2}_T) =S^{\tau_2}_0.$ The cost of illiquidity 
of $S^{\tau_2}_T$ is equal to zero, while the cost of illiquidity 
of $S^{\tau_1}_T$ is $S_0^{\tau_1}(1- \exp Y^M)>0$.   
Moreover, $\pi (S^{\tau_1}_T-S^{\tau_2}_T)+\pi(S^{\tau_2}_T)=\pi (S^{\tau_1}_T).$ Thus, we don\rq{}t get a cheaper price for the illiquid asset being  long of  the liquid asset and of the Basis swap $S^{\tau_1}_T-S^{\tau_2}_T.$ 
%where we short the liquid asset. 
% it is equivalent in term of prices to be long of  the liquid asset and of the Basis swap where we short the liquid asset  or of the illiquid asset. 

%
%So, one may me long that $ <1$ share of the liquid asset and of an share on the Basis swap.  
%Now, the cost of being long of  the liquid asset and of the Basis swap $S^{\tau_1}_T-S^{\tau_2}_T$ is 
%$$\pi(S^{\tau_2}_T)+\pi (S^{\tau_1}_T-S^{\tau_2}_T)=S_0^{\tau_2} (\exp \left(-Y^m\right) -\exp \left(-Y^M\right))+ 
%S_0^{\tau_1}> \pi(S_T^{\tau_1})$$
%ARBITRAGE?

%If we consider the Brownian application, we have see that if $ Y^M \geq 0$, then $\pi(S^{\tau_1}_T)= S_0^{\tau_1}$. 
%If $ Y^M \leq 0$, then $\pi(S^{\tau_1}_T)={\pi}^{\t_2}( S_T^{\tau_1})= S_0^{\tau_1} \exp \left(Y^M\right)$ and  $\pi( S_T^{\tau_2})=S_0^{\tau_2}.$ \\
%If $S_0^{\tau_1} \exp \left(Y^M\right)   < S_0^{\tau_2},$ then $\pi(S^{\tau_1}_T-S^{\tau_2}_T)=+\infty$. Else, 
%\begin{eqnarray*}
%\pi(S^{\tau_1}_T-S^{\tau_2}_T) & = &  
%\left(S_0^{\tau_1} \exp \left(Y^M\right)-  S_0^{\tau_2} \right) \mbox{ if } Y^M \leq 0\\
%& = & 
%\left(S^{\tau_1}_0 - S_0^{\tau_2}\exp \left(-Y^M\right)\right)
%\mbox{ if } Y^M \geq 0.
%\end{eqnarray*}

We now give  conditions on the parameters that allow us to obtain the signs of $Y^M$ and $Y^m$. 
We have that $Y^M \geq 0$ if  $s^{\tau_1}_0\geq s^{\tau_2}_0,$ $ \sigma_{1,s}\leq \sigma_{2,s}$   and  $a_s^{\t_1} \geq  a_s^{\t_2}$ for all $s$. 
This is an example where the spread of the market $\tau_1$ dominates the one of the market $\tau_2$. Indeed, $s^{\tau_1}_t$ starts with a bigger initial value, has a higher drift and a smaller standard deviation. If we add the condition $\hat\sigma_{1,s}^2-\hat\sigma_{1,s}\hat\sigma_{2,s} \geq 0,$ for all $s$, then we also have that  $Y^m \geq 0.$ 

Symmetrically, we have that $Y^m \leq 0$ if  $s^{\tau_1}_0\leq s^{\tau_2}_0,$  $a_s^{\t_1}\leq a_s^{\t_2}$ and $\sigma_{1,s}\geq \sigma_{2,s}$ for all $s$. If we add the condition  $\hat\sigma_{1,s}^2-\hat\sigma_{1,s}\hat\sigma_{2,s} \leq 0$ for all $s$, then  $Y^M \leq 0.$ 
%Remark that $\sqrt{1- \kappa_{1,s}^2} -  \kappa_{1,s}\sqrt{\frac{1}{ \rho_{1,s}^2}-1} \geq 0 \Leftrightarrow \rho_{1,s}^2 \geq \kappa^2_{1,s}. $ 
Let 
$$\bar \kappa_s= \frac{
\sqrt{1- \kappa_{2,s}^2}- \kappa_{2,s}\sqrt{\frac{1}{ \rho_{2,s}^2}-1} }{\sqrt{1- \kappa_{1,s}^2} -  \kappa_{1,s}\sqrt{\frac{1}{ \rho_{1,s}^2}-1}  }. $$ 
Then, $\hat\sigma_{1,s}^2-\hat\sigma_{1,s}\hat\sigma_{2,s}\leq 0$ if and only if 
%$\rho_{1,s}^2 \geq \kappa^2_{1,s}$ and 
$
\frac{\sigma_{1,s}}{\sigma_{2,s}} \leq  \bar \kappa_s.$ 

%or   $\rho_{1,s}^2 \leq  \kappa^2_{1,s}$ and   $\frac{\sigma_{1,s}}{\sigma_{2,s}} \geq \bar \kappa_s. $ 
%\begin{eqnarray*}
%% \Leftrightarrow  & & \\ 
%%\sigma_{1,s} \left(\sqrt{1- \kappa_{1,s}^2} -  \kappa_{1,s}\sqrt{\frac{1}{ \rho_{1,s}^2}-1}   \right)^2
%%-\sigma_{2,s} \left( 
%%\sqrt{1- \kappa_{2,s}^2}- \kappa_{2,s}\sqrt{\frac{1}{ \rho_{2,s}^2}-1} \right)  \left(\sqrt{1- \kappa_{1,s}^2} -  \kappa_{1}\sqrt{\frac{1}{ \rho_{1,s}^2}-1}   \right) \leq 0 \Leftrightarrow & & \\ 
%\rho_{1,s}^2 \geq \kappa^2_{1,s} \mbox{ and } 
%\frac{\sigma_{1,s}}{\sigma_{2,s}} \leq  \bar \kappa   
%%=\rho_{1,s}^2\frac{\left(\sqrt{1- \kappa_{2,s}^2}- \kappa_{2}\sqrt{\frac{1}{ \rho_{2,s}^2}-1} \right)\left(\sqrt{1- \kappa_{1,s}^2} + \kappa_{1}\sqrt{\frac{1}{ \rho_{1,s}^2}-1} \right)}{\rho_{1,s}^2- \kappa_{1,s}^2 }
%\mbox { or } \rho_{1,s}^2 \leq  \kappa^2_{1,s} \mbox{ and } 
%\frac{\sigma_{1,s}}{\sigma_{2,s}} \geq 
%\bar \kappa
%\end{eqnarray*}
%$$\sqrt{1- \kappa_{1,s}^2} -  \kappa_{1,s}\sqrt{\frac{1}{ \rho_{1,s}^2}-1} \geq 0 \Leftrightarrow 1- \kappa_{1,s}^2 \geq  \frac{\kappa^2_{1,s}}{ \rho_{1,s}^2}-\kappa^2_{1,s} \Leftrightarrow \rho_{1,s}^2 \geq \kappa^2_{1,s}. $$
Thus, to get that 
 $Y^M \leq 0,$  we can assume that  
$s^{\tau_1}_0\leq s^{\tau_2}_0,$ and for all $s,$ $a_s^{\t_1}\leq a_s^{\t_2}$    and 
%$\rho_{1,s} \geq |\kappa_{1,s}| $ and 
$1 \leq \frac{\sigma_{1,s}}{\sigma_{2,s}} \leq \bar \kappa_s. $ 
Sufficient conditions for 
 $Y^m \geq 0$  are 
$s^{\tau_1}_0\geq s^{\tau_2}_0$ and for all $s,$ $a_s^{\t_1}\geq a_s^{\t_2},$   
%$\rho_{1,s} \geq |\kappa_{1,s}| $ 
and $\bar \kappa_s \leq \frac{\sigma_{1,s}}{\sigma_{2,s}} \leq 1. $ 
%$$1 \leq \frac{\sigma_{1,s}}{\sigma_{2,s}} \leq 
%\frac{
%\sqrt{1- \kappa_{2,s}^2}- \kappa_{2,s}\sqrt{\frac{1}{ \rho_{2,s}^2}-1} }{\sqrt{1- \kappa_{1,s}^2} -  \kappa_{1,s}\sqrt{\frac{1}{ \rho_{1,s}^2}-1}  } .$$
%One may also assume that $s^{\tau_1}_0\geq s^{\tau_2}_0,$  $a_s^{\t_1}\geq a_s^{\t_2},$ $\rho_{1,s}\leq |\kappa_{1,s}| $  and $  \frac{\sigma_{1,s}}{\sigma_{2,s}} \leq \bar \kappa_s \wedge 1. $ 

We now consider the case where the coefficients are not time dependent i.e., $b_t=b$, $a^{\tau_i}_{t}=a_{i},$ $\sigma_{i,t}=\sigma_{i} $, $\rho_{i,t}=\rho_{i},$ $\eta_{i,t}=\eta_{i} $ and $\kappa_{i,t}=\kappa_{i} $
for all $t$ and $i \in \{1,2\}$. This allows to obtain closed formula for which we provide numerical examples where $Y^M\geq 0$ and $Y^m \leq 0.$  
%First,
%\begin{eqnarray*}
%%\label{eq:spread-2}
%s_t  &  = &  s_0 e^{-b t  }+ \int_0^t e^{-b (t -s) }\left(a -\la \Sigma^{\t}_s, \Lambda_s \ra \right) ds + \int_0^t e^{-b (t -s) }\la \Sigma, d W^{\t^*}_s \ra \\
% & \sim & \Nc\left(s_0 e^{-b t  }+ \int_0^t e^{-b (t -s) }\left(a -\la \Sigma^{\t}_s, \Lambda_s \ra \right) ds, \| \Sigma\|^2\int_0^t e^{-b (t -s) } d s \right)
%\end{eqnarray*}
If $ b>0$, we obtain that $\widetilde{b}(t, T)   
 =  \frac{1}{b} ( 1 - e^{-b (T-t)} )$ and 
\begin{eqnarray*}
%\widetilde{b}(t, T)   
%& = & \frac{1}{b} \left( 1 - e^{-b (T-t)} \right) \\
%\int_t^T \widetilde{b}(s, T) ds  &= & \frac{T-t}{b}-\frac1{b^2} \left( 1-e^{-b(T-t)}\right) \\
%\int_t^T \widetilde{b}^2(s, T) ds  &= & \frac{T-t}{b^2}-\frac2{b^3} \left( 1-e^{-b(T-t)}\right) +\frac1{2b^3} \left( 1-e^{-2b(T-t)}\right) \\
Y^M & = &  \frac{s_0}{b} \left( 1 - e^{-b T} \right) -  \frac{\sigma^2_{1}-\sigma^2_{2}}2
\left(\frac{T}{b^2}-\frac2{b^3} \left( 1-e^{-bT}\right) +\frac1{2b^3} \left( 1-e^{-2bT}\right)  \right)
+ \\
& &   \left(\frac{T}{b}-\frac1{b^2} \left( 1-e^{-bT}\right) \right) \left( a^{\t_1}- a^{\t_2} + \frac{2\left[\hat\sigma_{1}^2-\hat\sigma_{1}\hat\sigma_{2}\right]_+ }{\frac1{\r^2_{1}} +\frac1{\r^2_{2}} -1} \right).
\end{eqnarray*}
If $ b=0$, we get that $\widetilde{b}(t, T)    =  T-t$ and 
\begin{eqnarray*}
%\widetilde{b}(t, T)   & =  T-t  \\
%\int_t^T \widetilde{b}(s, T) ds  &= &  \frac{(T-t)^2}2\\
%\int_t^T \widetilde{b}^2(s, T) ds  &= &  \frac{(T-t)^3}3\\
Y^M & = &   s_0T-  \frac{\sigma^2_{1}-\sigma^2_{2}}2 \frac{T^3}3 +   \frac{T^2}2  \left (a^{\t_1}- a^{\t_2}  
+ \frac{2\left[\hat\sigma_{1}^2-\hat\sigma_{1}\hat\sigma_{2}\right]_+ }{\frac1{\r^2_{1}} +\frac1{\r^2_{2}} -1}\right).
\end{eqnarray*}
We obtain $Y^m$ changing in the equations above $\left[\hat\sigma_{1}^2-\hat\sigma_{1}\hat\sigma_{2}\right]_+$ by $-\left[\hat\sigma_{1}^2-\hat\sigma_{1}\hat\sigma_{2}\right]_-$. 

Now, we illustrate the three situations of liquidity/illiquidity with a numerical example. Assume that $T=2$, $s_0^{\tau_1}=3.4$, $s_0^{\tau_2}=3,$ $a_1=0.4,$ $a_2=0.8$, $b=0.1,$  $\s_1=0.3,$ $\s_2=0.4,$ 
 $\rho_1= 0.9,$ $\rho_2= 0.8,$ $\kappa_1= 0.5$ and $\kappa_2= -0.7.$ We find that 
$Y^M	=0.0564,$ $Y^m=-0.0639$  and
\begin{eqnarray*} 
{\pi}^{\t_2}(S^{\tau_1}_T) & = & S^{\tau_1}_0 \exp Y^M = 3.5972
\mbox{ and }
{\pi}^{\t_1}(S^{\tau_2}_T) =  S^{\tau_2}_0 \exp \left(-Y^m\right)=3.1981
 \\
\pi(S^{\tau_1}_T) & = &3.4  \wedge  3.5972= 3.4 \mbox{ and }
\pi(S^{\tau_2}_T)  =  3 \wedge 3.1981 =3\\
\pi(S^{\tau_1}_T-S^{\tau_2}_T)   &= & 0.5645. 
\end{eqnarray*}
Now, if we change $\kappa_2$ into $\kappa_2= 0.7,$ without changing the values of the other coefficients, we find that 
$Y^M	=0.0999,$ $Y^m=0.0564$  and
\begin{eqnarray*} 
{\pi}^{\t_2}(S^{\tau_1}_T) & = & S^{\tau_1}_0 \exp Y^M = 3.7572
\mbox{ and }
{\pi}^{\t_1}(S^{\tau_2}_T) =  S^{\tau_2}_0 \exp \left(-Y^m\right)=2.8355
 \\
\pi(S^{\tau_1}_T) & = &3.4  \wedge  3.7572= 3.4 \mbox{ and }
\pi(S^{\tau_2}_T)  =  3 \wedge 2.8355 =2.8355\\
\pi(S^{\tau_1}_T-S^{\tau_2}_T)   &= & 0.6852. 
\end{eqnarray*}
Finally, if furthermore we set $\s_1=0.4$ and $\s_2=0.3,$ and keep the same values for the other coefficients, we find that 
we find that 
$Y^M	=-0.0044,$ $Y^m=-0.1047$  and
\begin{eqnarray*} 
{\pi}^{\t_2}(S^{\tau_1}_T) & = & S^{\tau_1}_0 \exp Y^M = 3.3850
\mbox{ and }
{\pi}^{\t_1}(S^{\tau_2}_T) =  S^{\tau_2}_0 \exp \left(-Y^m\right)=3.3311
 \\
\pi(S^{\tau_1}_T) & = &3.4  \wedge  3.3850= 3.3850 \mbox{ and }
\pi(S^{\tau_2}_T)  =  3 \wedge 3.3311=3\\
\pi(S^{\tau_1}_T-S^{\tau_2}_T)   &= & 0.3850. 
\end{eqnarray*}

\subsection{Extensions}
The definition of NFL is tailored to easily obtain the existence of equivalent (local) martingale measures. However, the economic interpretation of this condition is not very natural. A free lunch is some non-negative and non-zero random variable for which one can find a convergent net in $C$. If one wants to return to the notion of sequence and take into account some budget constraints, one has to introduce the notion of no free lunch with vanishing risk (NFLVR) and consider general (and not simple) strategies; see \cite{DS94}. The version of Theorem \ref{main1} with the NFLVR condition instead of the NFL condition is left for further research.
Note that the definition of NFLVR (and in particular the admissibility of the trading strategies) has to be carefully generalized since there are several num\'eraires that are used together.
Finally, recall that it is shown in \cite{DS94} that for continuous and locally bounded processes with finite time horizons, it is sufficient to consider no free lunch with bounded risk for simple integrands (together with the NA condition) to obtain equivalent local martingale measures.
One may also study the no unbounded profit with bounded risk (NUPBR) condition of \cite{KK}, which is equivalent to the no arbitrage of the first kind (NA1) condition, 
%see \cite{KKS}, 
but again those definitions have to be adapted carefully because there are several num\'eraires.
Recall that the NFLVR condition is equivalent to the NA condition plus the NA1 condition.
\section{Conclusion}
\label{Conclu}
We propose a model, where several submarkets coexist, each one with its own num\'eraire, but where no tradeable num\'eraire exists for the whole market. Nevertheless, we consider the No Free Lunch condition at the level of the global market. This modelling is meaningful for credit or liquidity risk and for the multicurve markets. 

We prove that there is a common factor from which equivalent submarket-dependent local martingale measures are constructed. Next, we introduce several superhedging prices depending on the chosen type of initial wealth allocation as well as the selected submarket(s) for hedging. Those prices allow to measure the difference of features between submarkets. We provide dual relationships for the super-replication prices, as well as a comparison between them. 

In the special case of two submarkets each with a single risky asset, we also calculate the price of a Basis swap. We find that our measure of the difference of features  is indeed relevant: the initial investment is made in the submarket where this measure is equal to zero. 

Then, we provide several economic illustrations.  First, we show that in the case where the num\'eraire spreads are constant, a common local martingale measure exists for the whole market. Then, we show with several examples and counter-examples, that there is no implication between completeness of each submarket taken separately and completeness of the whole market.  We also propose a Brownian illustration, where under the assumption of a time dependent Vasicek model for the num\'eraires spread, we provide a characterisation of the sets $\Qc^{{\t}}$ of (local)  martingale measures. In the special case mentioned above, we fully compute the different superreplication prices and give conditions on the parameters to obtain one of the three possible configurations: market 1 is illiquid and market 2 is liquid, market 1 is liquid and market 2 is illiquid or finally both markets are liquid. 

\appendix
\section{Appendix}
\label{appendix}
%{\red POUR REPONSE : Note that we can not apply directly the Kreps-Yan theorem because  add}. \\
We now present the proofs of Theorems \ref{main1} and \ref{main2} and of Propositions \ref{propnum} and \ref{propdeux}.

\subsection{Proof of Theorem \ref{main1}} 
\label{preuve1}
%\begin{proof}{\it of Theorem \ref{main1}} 
First, we prove that 
\begin{eqnarray}
\label{laref1}
NFL \Longleftrightarrow \exists X^* \in L^{1}_{>0}, \; 
\E(X^* V) = 0, \forall V \in K. 
\end{eqnarray}
On the top of that, we show that 
\begin{eqnarray}
\label{laref2}
\Xc^* =\{ X^* \in L^{1}_{>0} | \; 
\E(X^* V) = 0, \forall V \in K\}. 
\end{eqnarray}
The proof of \eqref{laref1} is very similar to the proof of the Kreps-Yan theorem (see \cite{K81}, \cite{Y80}) given by \cite{S94} and \cite{DelSch05} and is given for the convenience of the reader. 
We first prove $\Leftarrow$ in \eqref{laref1}. For every $W \in C$, $\E(X^*W) \leq 0$. Let $W \in \bar{C} \cap L^{\infty}_+$. As
$W \mapsto \E(X^*W)$ is a weak-star continuous functional, $\E(X^*W) \leq 0$ follows. Thus, $W=0$ $\P$-a.s. since $\P(X^*>0)=1$.

We  prove $\Rightarrow$ in \eqref{laref1} in two steps. First, we prove that for any fixed $Y \in L^{\infty}_+$, $Y \neq 0$, there exists some
$X \in L^{1}$ such that $\E(X Y)>0$ and $\E(X W) \leq 0$ for all $W \in \bar{C}\cap L^{\infty}$.
As $\{Y\} \cap \bar{C}\cap L^{\infty} =\emptyset$, we can use the Hahn-Banach theorem (see Theorem II.9.2 in \cite{Sc99}) and find some $X \in L^{1}$ and $a <b$ such that $\E(XY) \geq b$ and $\E(X W) \leq a$ for all $W \in \bar{C}\cap L^{\infty}$.
As $0 \in \bar{C}$ we have that $b>a\geq 0$ and thus $\E(XY) >0$. As $\bar{C}\cap L^{\infty}$ is a cone, $\E(X W) \leq 0$ for all $W \in \bar{C}\cap L^{\infty}$. Finally, as $-L^{\infty}_+ \subset \bar C\cap L^{\infty}$, $\E(X W) \geq 0$ for all $W \in L^{\infty}_+$, and thus $X  \in L^{1}_+$. 
The second step is the so-called exhaustion argument. Let
$$\Dc=\{X \in L^{1}_+| \;\E(X W) \leq 0, \; \forall W \in C\}.$$
Then, $\Dc$ is non-empty ($0 \in \Dc$), and there exists some $X^* \in \Dc$ such that
\begin{eqnarray}
\label{support}
\P(X^*>0)= \sup_{X \in \Dc} \P(X>0).
\end{eqnarray}
Indeed, let $(X_n)_{n\geq 1}$ be a maximizing sequence  for \eqref{support} and $X^*=\sum_{n\geq 1} \frac{1}{2^n(1+ \E(X_n))} X_n$.
Then, $X^* \in \Dc$ and \eqref{support} holds true.
%let $(X_n)_{n\geq 1} \in \Dc$ such that $(\P(X_n>0))_{n\geq 1}$ goes to $\sup_{X \in \Dc} \P(X>0).$ Let $X^*=\sum_{n\geq 1} \frac{1}{2^n(1+ \E(X_n))} X_n$.
%Then, $X^* \in \Dc$ and \eqref{support} holds true.

Note that, $\P(X^*>0)=1$. Else, $\P(X^*=0)>0$, and applying step 1 with $Y=\one_{\{X^*=0\}}$ we obtain some $X_0 \in \Dc$ such that $\E(Y X_0)>0$. This last inequality implies that $\P(\{X^*=0\} \cap \{X_0>0\})>0$, and thus
$\P(X^*+X_0>0)>\P(X_0>0)$, which contradicts \eqref{support}.

From the first two steps, we have found some $X^* \in L^{1}_{>0}$ such that
$\E(X^* W) \leq 0$ for all $W \in C$ and as for all $V \in K$, as $\pm V \in C,$ $\E(X^* V) = 0,$ which concludes the proof of \eqref{laref1}.

Now we show $\subset$ in \eqref{laref2}. Let  $X^* \in \Xc^*$ and 
$V =\sum_{\t \in \Tc} S^{\t,0}_T (\sum_{j=1}^{n_{\t}}
\f^{\t}_j (\tilde{S}^{\t}_{\d_j^{\t}}-\tilde{S}^{\t}_{\d_{j-1}^{\t}} ))\in K,$
\begin{eqnarray*}
\E(X^*V) & = & \sum_{\t \in \Tc} \sum_{j=1}^{n_{\t}}  \E \left(X^*S^{\t,0}_T\f^{\t}_j
 \left(\tilde{S}^{\t}_{\d_j^{\t}}-\tilde{S}^{\t}_{\d_{j-1}^{\t}} \right) \right) = 0
\end{eqnarray*}
using \eqref{condmart}. For the other inclusion, fix some
$\t \in \Tc$, some stopping times $0\leq \b_1 \leq \b_2 \leq T$ and some $\R^{d_{\t}}$-valued, $\Fc_{\b_1}$-measurable random variable $\f.$ Then, 
$\pm S^{\t,0}_T \f \left(\tilde{S}^{\t}_{\b_2}-\tilde{S}^{\t}_{\b_1} \right)\in K$ and we obtain that \eqref{condmart} holds true. $\Box$

\subsection{Proof of Proposition \ref{propnum}} 
\label{preuve2}
\begin{lemma}
\label{newlem}
Assume that Assumption \ref{locbound} holds true and that $\Xc^* \neq \emptyset$. If $X^* \in \Xc^*$, then for all $Z \in L_{>0}^{\infty}$, $\Q_Z \in \Qc^{Z}$ where $\Q_Z$ is defined by
\begin{eqnarray}
\label{pasbol}
\frac{d\Q_Z}{d\P}=
\frac{
X^*Z
}
{
\E \left(
X^* Z
\right)
}.
\end{eqnarray}
\end{lemma}
\begin{proof}
%{\it of Lemma \ref{newlem}}
Let $X^* \in \Xc^*$ and fix some $Z \in L_{>0}^{\infty}$. Let $\Q_Z$ be defined by \eqref{pasbol}. 
%$
%{d\Q_Z}/{d\P}=
%({
%X^*Z
%})/
%{
%\E \left(
%X^* Z
%\right)
%}$.
%We show that $\Q_Z \in   \Qc^{Z}$. 
As $\P(X^*>0)=1$ and $\P(Z>0)=1$, we obtain that $\Q_Z \sim \P$ and $({d\Q_Z}/{d\P})/{Z} \in L^1$, as $X^* \in L^1$. Fix some  $\t \in \Tc$. Then, for all stopping times $ \b_1 \leq \b_2 \leq T$ and all $\R^{d_{\t}}$-valued, $\Fc_{\b_1}$-measurable random variables $\f^{\t}$, we obtain from \eqref{condmart} that
\begin{eqnarray*}
0 &= & \E\left(X^*\bar{S}_T^{\t,0}\f^{\t} \left(\tilde{S}^{\t}_{\b_2}-\tilde{S}^{\t}_{\b_1} \right)\right)
= 
\E \left(
X^*
Z
\right) \E_{\Q_Z}\left(
\frac{
\bar{S}_T^{\t,0}
}{
Z
}
\f^{\t} \left(\tilde{S}^{\t}_{\b_2}-\tilde{S}^{\t}_{\b_1} \right)\right).
\end{eqnarray*}
Thus, $\E_{\Q_Z}\left(
\frac{
\bar S^{\tau,0}_T
}{
Z
}
\left(\tilde{S}^{\t}_{\b_2}-\tilde{S}^{\t}_{\b_1} \right)| \Fc_{\b_1}\right)  =   0$, which implies that
\begin{eqnarray}
\label{espoir}
%\E_{\Q_Z}\left(
%\frac{
%\bar S^{\tau,0}_T
%}{
%Z
%}
%\left(\tilde{S}^{\t}_{\b_2}-\tilde{S}^{\t}_{\b_1} \right)| \Fc_{\b_1}\right) & =  & 0\\
\E_{\Q_Z}\left(
\E_{\Q_Z}\left(
\frac{
\bar S^{\tau,0}_T
}{
Z
}
| \Fc_{\b_2}\right)
\tilde{S}^{\t}_{\b_2} | \Fc_{\b_1}\right) & = &
\E_{\Q_Z}\left(
\frac{
\bar S^{\tau,0}_T
}{
 Z
}
| \Fc_{\b_1}\right) \tilde{S}^{\t}_{\b_1}
\end{eqnarray}
and $\Q_Z \in   \Qc^{Z}$.$\Box$
\end{proof}

\begin{proof}{\it of Proposition \ref{propnum}}
First, we prove that if $\Qc^{Z}\neq \emptyset$ for some $Z \in L_{>0}^{\infty},$ then 
\begin{eqnarray}
\label{comedie}
\E_{\Q}\left(\frac{W}{Z}\right) \leq 0, \; \forall \Q \in   \Qc^{Z},\;  \forall W \in \bar{C} \cap L^{\infty}.
\end{eqnarray}
Let  $V =\sum_{\t \in \Tc} S^{\t,0}_T \left(\sum_{j=1}^{n_{\t}}
\f^{\t}_j \left(\tilde{S}^{\t}_{\d_j^{\t}}-\tilde{S}^{\t}_{\d_{j-1}^{\t}} \right) \right)\in K$; 
\begin{small}
\begin{eqnarray*}
\E_{\Q}\left(\frac{V}{ Z}\right)
& = &
% \sum_{\t \in \Tc}
% \sum_{j=1}^{n_{\t}}
% \left(
%\E_{\Q} \left(\f^{\t}_jS^{\t,0}_0
%\E_{\Q}  \left(
%\frac{ \bar S^{\tau,0}_T}{Z}\tilde{S}^{\t}_{\d_j^{\t}}
%-\frac{ \bar S^{\tau,0}_T}{Z}\tilde{S}^{\t}_{\d_{j-1}^{\t}}
%| \Fc_{\d_{j-1}^{\t}} \right) \right)
%\right) \\
%& = &
 \sum_{\t \in \Tc}
 \sum_{j=1}^{n_{\t}}
\E_{\Q} \left(\f^{\t}_jS^{\t,0}_0
\E_{\Q}  \left(
\E_{\Q}  \left(
\frac{ \bar S^{\tau,0}_T}{Z}
| \Fc_{\d_{j}^{\t}} \right)
\tilde{S}^{\t}_{\d_j^{\t}}
-
\E_{\Q}  \left(
\frac{ \bar{S}_T^{\t,0}}{Z}
| \Fc_{\d_{j-1}^{\t}} \right) \tilde{S}^{\t}_{\d_{j-1}^{\t}}
| \Fc_{\d_{j-1}^{\t}} \right) \right)  \\
 & = & 0.
\end{eqnarray*}
\end{small}
%This implies that for every $Y \in C$, $\E_{\Q}\left(\frac{Y}{Z}\right)  \leq 0$.
Thus, \eqref{comedie} follows as 
$W \mapsto \E\left(\frac{d\Q}{d\P}\frac{W}{ Z}\right) $ is a weak-star continuous functional (recall that ${\frac{d\Q}{d\P}}/{Z} \in L^1$).

{\bf 1.} Assume that the NFL condition holds true. Theorem \ref{main1} implies that there exists some $X^* \in \Xc^*$ and Lemma \ref{newlem} implies that $\Qc^{Z}$ is not empty  for  all   $Z \in L_{>0}^{\infty}$.\\
Now, assume that there exists some $Z \in L_{>0}^{\infty}$ such that $\Qc^{Z}\neq \emptyset.$  Let $W \in \bar{C} \cap L^{\infty}_+$ and let $\Q \in   \Qc^{Z}.$  Then, 
\eqref{comedie} implies that   $\frac{W}{Z}=0$ $\Q$-a.s. As $\Q \sim \P$ and $\P(Z>0)=1$,
$W=0$ $\P$-a.s., and the NFL condition follows.

{\bf 2.} Assume first that $\Xc^* =\emptyset.$ Then, for all $Z \in L_{>0}^{\infty},$  $\{\Q|\, \exists X^* \in  \Xc^*,\; \frac{d\Q}{d\P}=
\frac{
X^*Z
}
{
\E \left(
X^* Z
\right)
}\}=\emptyset.$ Moreover, Theorem \ref{main1} and 
the contraposed of the second part of {\bf 1.} imply that $\Qc^{Z}=  \emptyset$ and both sets in \eqref{enfin}
are empty. Assume now that  for some  $Z \in L_{>0}^{\infty},$ $\Qc^{Z}=  \emptyset.$ Then, the contraposed of the first part of {\bf 1.} and Theorem \ref{main1} imply that $\Xc^* =\emptyset$ and again both sets in \eqref{enfin} are empty. We now assume that both sets are non empty. 

We have proved the reverse inclusion in Lemma \ref{newlem}. 
Now, let $\Q \in   \Qc^{Z}.$  Then, for all stopping times $ \b_1 \leq \b_2  \leq T,$  Definition \ref{defQZ} 
shows that \eqref{espoir} is satisfied for all $\tau \in \Tc$ with $\Q$ instead of $\Q_Z$. 
This implies that for  all $\R^{d_{\t}}$-valued, $\Fc_{\b_1}$-measurable random variables $\f^{\t}$
%,  for every  $X^*\in L_{>0}^{1}$
\begin{eqnarray*}
0 & = & \E_{\Q}\left(
\frac{
\bar{S}_T^{\t,0}
}{
Z
}
\f^{\t} \left(\tilde{S}^{\t}_{\b_2}-\tilde{S}^{\t}_{\b_1} \right)\right)   
%=  \E\left(\frac{\frac{d\Q}{dP}}{ZS^{\t,0}_0}S^{\t,0}_T\f^{\t} \left(\tilde{S}^{\t}_{\b_2}-\tilde{S}^{\t}_{\b_1} \right)\right) \\
 = \frac{1}{S^{\t,0}_0 }\E\left(\frac{d\Q}{d\P}\frac{1}{Z}S^{\t,0}_T\f^{\t} \left(\tilde{S}^{\t}_{\b_2}-\tilde{S}^{\t}_{\b_1} \right)\right).
\end{eqnarray*}
Then,    $X^*=\frac{d\Q}{d\P}\frac{1}Z\in \Xc^*$ and $\frac{d\Q}{d\P}= \frac{X^*Z}{\E(X^*Z)}.$
$\Box$
\end{proof}

%We start this Appendix with several simple results.  
%that will be used later for the effective computation of superreplication costs. 

\subsection{Proof of Theorem \ref{main2}}
\label{preuve3}
%\begin{proof}{\it of Theorem \ref{main2}}
Let  $(\l_{\t})_{\t \in \Tc} \in \Lambda^{\Tc}$ be such that $\frac{H}{\sum_{\t \in \Tc} \l_{\t}\bar{S}_T^{\t,0}} \in L^{\infty}$. Note that this implies that
$H \in L^{\infty}$.\\
{\it The infima are attained in $\pi(H),$ $\pi^{\tau}(H),$ and $\hat \pi^{\tau}(H)$.} Let 
$$F=\left\{x \in \R| \, \exists (x^{\t})_{\tau \in \Tc}\subset (0,\infty) \mbox{ s.t. } x=\sum_{\t \in \Tc} x^{\t}, \left(H-\sum_{\t \in \Tc} x^{\t}\bar{S}_T^{\t,0} \right) \in \bar{C}\cap L^{\infty} \right\}.$$
It is clear that $F$ is nonempty and that $\pi(H)=\inf F$. We show now that the set $F$ is closed. 
Let $(\hat y_n)_{n \geq 1} \subset F$ be such that $\hat y_n$ goes to $\hat  y$. For all $n \geq 1$, there exist $(\hat{y}_n^{\t})_{\t \in \Tc}$ with $\hat{y}_n^{\t} \geq 0,$ $\hat y_n=\sum_{\t \in \Tc} \hat{y}_n^{\t}$ and
$$Y_n=\left(H-\sum_{\t \in \Tc} \hat{y}_n^{\t}\bar{S}_T^{\t,0} \right) \in \bar{C}\cap L^{\infty}.$$
Let $\e >0$. For $n$ large enough, $  0 \leq \hat{y}_n^{\t} \leq \hat{y}_n \leq\hat{y} + \e$ for all $\t \in \Tc$, and we can extract a sub-sequence that we still denote by $(\hat{y}_n^{\t})_{\t \in \Tc}$ that converges to some $(\hat{y}^{\t})_{\t \in \Tc}$ such that $\hat{y}^{\t} \geq 0$ for all $\t \in \Tc$ and $\hat y=\sum_{\t \in \Tc} \hat{y}^{\t}$.
It is clear that $(Y_n)_{n \geq 1}$ converges a.s. and also weak-star to $\hat Y= \left(H-\sum_{\t \in \Tc} \hat{y}^{\t}\bar{S}_T^{\t,0} \right)$.
Indeed, let $A\in L^1$,
$$\left|\E\left( A\left(Y_n- \hat Y\right) \right)\right| \leq
\sum_{\t \in \Tc}\E\left( |A|\bar{S}_T^{\t,0} |\hat{y}_n^{\t} -\hat{y}^{\t}|\right) \leq \e \sum_{\t \in \Tc}\E\left( |A|\bar{S}_T^{\t,0} \right).$$
Thus, $\hat Y \in \bar{C}\cap L^{\infty}$ and $\hat y \in F$. As $\pi(H)=\inf F$, $\pi(H)$ belongs to $F$. From now,  we denote $\pi(H) =\sum_{\t \in \Tc} \hat{x}^{\t} \in F.$\\
Using the same kind of arguments,  we obtain that the infima  are attained in \eqref{prixsureptautout} and \eqref{prixsureptau}.\\

\noindent {\it  Proof of \eqref{audessussuphat}} 
For all $x=\sum_{\t \in \Tc} x^{\t} \in F$ and all $Z \in L_{>0}^{\infty},$ 
\begin{eqnarray}
\label{audessussuphat}
\sup_{ \Q \in   \Qc^{Z}}
\E_{\Q} \left( \frac{ H}{Z}
-
\sum_{\t \in \Tc} x^{\t} \frac{\bar{S}_T^{\t,0}}
{Z}
\right)
\leq 0.
\end{eqnarray}
%Let $x \in F.$ 
%There exist $(x^{\t})_{\t \in \Tc}$ and $W \in \bar{C}\cap L^{\infty}$ such that $x^{\t} \geq 0$ for all $\t \in \Tc$, $x=\sum_{\t \in \Tc} x^{\t}$ and $\sum_{\t \in \Tc} x^{\t}\bar{S}_T^{\t,0}+W \geq   H \, \mbox{a.s.}$ 
%Let $Z \in L_{>0}^{\infty}.$ As the NFL condition holds true, Proposition \ref{propnum} shows that $\Qc^{Z}$ is nonempty and that for all
%$\Q \in   \Qc^{Z},$  
Indeed, there exist $W \in \bar{C}\cap L^{\infty}$ $\sum_{\t \in \Tc} x^{\t}\bar{S}_T^{\t,0}+W \geq   H \, \mbox{a.s.}$ 
So, \eqref{comedie} implies that 
for all $\Q \in   \Qc^{Z}$ (which is non-empty, see   Proposition \ref{propnum}), 
\begin{eqnarray*}
%\label{audessushat}
 \sum_{\t \in \Tc} x^{\t}\E_{\Q} \left(\frac{\bar{S}_T^{\t,0}}
{Z}
\right) \geq \E_{\Q} \left( \frac{ H}{Z} \right). 
\end{eqnarray*}
%and as it is true for all $\Q \in   \Qc^{Z}$,

\noindent{\it Proof of \eqref{prixrepqsumhat}} 
Set  for all $\tau \in \Tc,$ 
$\hat{z}^{\t}= \hat{x}^{\tau} - \e \frac{\l_{\t}}{\sum_{\t \in \Tc} \l_{\t}}.$
Note that one may  choose $\e$ such that $\hat{z}^{\t} \geq 0$. Then,
$\sum_{\t \in \Tc} \hat{z}^{\t}=\pi(H)-\e$, and thus
$$\left\{H-\sum_{\t \in \Tc} \hat{z}^{\t}\bar{S}_T^{\t,0}\right\} \cap \bar{C}\cap L^{\infty}  =\emptyset.$$
% We use again Hahn-Banach theorem (see Theorem II.9.2 in \cite{Sc99}) and find some $\hat X \in L^{1}_+$ and $a <b$ such that
% $$\E\left(\hat X \left(H-\sum_{\t \in \Tc} \hat{z}^{\t}\bar{S}_T^{\t,0}\right)\right) \geq b \mbox{ and } \E(\hat X W) \leq a \mbox{ for all } W \in \bar{C}\cap L^{\infty}.$$
As in the proof of Theorem \ref{main1} using the Hahn-Banach argument, we obtain the existence of $\hat X \in \Xc^*$ such that
\begin{eqnarray}
\label{eqinterhat}
\E\left(\hat X H \right) >\E\left(\hat X \sum_{\t \in \Tc} \hat{z}^{\t}\bar{S}_T^{\t,0}\right).
 \end{eqnarray}
%Let $Z=\sum_{\t \in \Tc} \l_{\t}\bar{S}_T^{\t,0}$.
Let
$
{d\hat{\Q}}/{d\P}=
\left({
\hat{X}\sum_{\t \in \Tc} \l_{\t}\bar{S}_T^{\t,0}
}\right)/
{
\E \left(
\hat{X} \sum_{\t \in \Tc} \l_{\t}\bar{S}_T^{\t,0}
\right)
}
$.  Proposition  \ref{propnum} implies that $\hat{\Q} \in \Qc^{lc, \l}$ and
\begin{eqnarray*}
\E_{\hat{\Q}}\left( \frac{ H}{\sum_{\t \in \Tc} \l_{\t}\bar{S}_T^{\t,0}}  \right) >\E_{\hat{\Q}}\left(
\frac{ \sum_{\t \in \Tc} \hat{z}^{\t}\bar{S}_T^{\t,0}}{\sum_{\t \in \Tc} \l_{\t}\bar{S}_T^{\t,0}} \right) & = &
% \sum_{\t \in \Tc} \hat{x}^{\t}\E_{\hat{\Q}}\left(\frac{ \bar{S}_T^{\t,0}}{Z} \right)-\frac{\e}{\sum_{\t \in \Tc} \l_{\t}} \E_{\hat{\Q}}\left(\sum_{\t \in \Tc} \frac{ \l_{\t} \bar S^{\tau,0}_T}{Z} \right)\\
%& =  &
\E_{\hat{\Q}}\left(
\frac{\sum_{\t \in \Tc} \hat{x}^{\t} \bar{S}_T^{\t,0}}{\sum_{\t \in \Tc} \l_{\t}\bar{S}_T^{\t,0}} \right)
-\frac{\e}{\sum_{\t \in \Tc} \l_{\t}}.
 \end{eqnarray*}
This implies that
\begin{eqnarray*}
%\label{eqinterprobahat}
\sup_{ \Q \in   \Qc^{lc, \l}}
\E_{ \Q } \left( \frac{ H}{\sum_{\t \in \Tc} \l_{\t}\bar{S}_T^{\t,0}}
-
 \frac{\sum_{\t \in \Tc} \hat{x}^{\t}\bar{S}_T^{\t,0}}
{\sum_{\t \in \Tc} \l_{\t}\bar{S}_T^{\t,0}}
\right)  & \geq &\E_{\hat{\Q}}\left( \frac{ H}{\sum_{\t \in \Tc} \l_{\t}\bar{S}_T^{\t,0}}   -
\frac{ \sum_{\t \in \Tc} \hat{x}^{\t}\bar{S}_T^{\t,0}}{\sum_{\t \in \Tc} \l_{\t}\bar{S}_T^{\t,0}} \right)\\
& > & -\frac{\e}{\sum_{\t \in \Tc} \l_{\t}}.
 \end{eqnarray*}
Therefore, letting $\e$ go to zero 
and using \eqref{audessussuphat} for $Z=\sum_{\t \in \Tc} \l_{\t}\bar{S}_T^{\t,0}$, we obtain that \eqref{prixrepqsumhat} holds true.\\

\noindent {\it Proof of  \eqref{carctprixsurep1bishat} }
We assume now that $\frac{H}{\min_{\t \in \Tc} \bar{S}_T^{\t,0}}  \in L^{\infty}$, which implies that $\frac{H}{ \bar{S}_T^{\t,0}}  \in L^{\infty}$ for all
$\t \in \Tc$ and also that  $\frac{H}{\sum_{\t \in \Tc} \l_{\t}\bar{S}_T^{\t,0}} \in L^{\infty}$ for all $(\l_{\t})_{\t \in \Tc} \in \Lambda^{\Tc}$.
Using \eqref{audessussuphat} for $Z=\max_{\t \in \Tc}\bar{S}_T^{\t,0}$,
%(recall that in this case $\Qc^{Z}=\Qc^{max}$),
we obtain that if $ x=\sum_{\t \in \Tc} x^{\t} \in F$,
$$
\sup_{ \Q \in   \Qc^{max}}
\E_{\Q} \left( \frac{ H}{\max_{\t \in \Tc}\bar{S}_T^{\t,0}}
-
\sum_{\t \in \Tc} x^{\t}
\right)
%\leq\sup_{ \Q \in   \Qc^{max}}\E_{\Q} \left( \frac{ H}{\max_{\t \in \Tc}\bar{S}_T^{\t,0}}-\sum_{\t \in \Tc} x^{\t} \frac{\bar{S}_T^{\t,0}}{\max_{\t \in \Tc}\bar{S}_T^{\t,0}}\right)
\leq 0,
$$
and thus $$\sum_{\t \in \Tc} x^{\t}  \geq \sup_{ \Q \in   \Qc^{max}}
\E_{\Q} \left( \frac{ H}{\max_{\t \in \Tc}\bar{S}_T^{\t,0}}\right)$$ and taking the infimum on all  $ x=\sum_{\t \in \Tc} x^{\t} \in F$
%$(x^{\t})_{\t \in \Tc}$ such that $x^{\t} \geq 0$ for all $\t \in \Tc$ and such that there exists  $W \in \bar{C}\cap L^{\infty}$ satisfying $\sum_{\t \in \Tc} x^{\t}\bar{S}_T^{\t,0}+W \geq   H \, \mbox{a.s.}$
on the left-hand side, we obtain that
$$\pi(H) \geq \sup_{\Q \in \Qc^{max}}\E_{\Q}\left(
\frac{H}{\max_{\t \in \Tc}\bar{S}_T^{\t,0}}
\right).$$
Now, we use \eqref{eqinterhat}
to obtain that
$$\E\left(\hat X H \right) >\E\left(\hat X \sum_{\t \in \Tc} \hat{z}^{\t}\bar{S}_T^{\t,0}\right) \geq \E\left(\hat X \min_{\t \in \Tc}\bar{S}_T^{\t,0}\right)\sum_{\t \in \Tc} \hat{z}^{\t}.$$
Let
$
{d\Q_m}/{d\P}=
\left({\hat{X}\min_{\t \in \Tc}\bar{S}_T^{\t,0}}\right)/{ \E \left( \hat{X} \min_{\t \in \Tc}\bar{S}_T^{\t,0}\right)}
$. Then, from Proposition \ref{propnum}, $\Q_m \in \Qc^{min}$ and
$$\sup_{\Q \in \Qc^{min}}
\E_{\Q}\left(
\frac{H}{\min_{\t \in \Tc}\bar{S}_T^{\t,0}}
\right)
\geq
\E_{\Q_m}\left(
\frac{H}{\min_{\t \in \Tc}\bar{S}_T^{\t,0}}
\right)
>\sum_{\t \in \Tc} \hat{z}^{\t}.$$
Thus, letting $\e$ go to $0$, we obtain that $$\sup_{\Q \in \Qc^{min}}
\E_{\Q}\left(
\frac{H}{\min_{\t \in \Tc}\bar{S}_T^{\t,0}}
\right)\geq \pi(H)$$ 	and this achieves the proof of \eqref{carctprixsurep1bishat} (note that here we do not need the infimum in $\pi(H)$ to be attained; it is enough to choose
$\hat{z}^{\t}$ such that $\sum_{\t \in \Tc} \hat{z}^{\t}<\pi(H)$ and to let $\sum_{\t \in \Tc} \hat{z}^{\t}$ go to $\pi(H)$).\\

\noindent {\it Proof of  \eqref{defqui1}}
Let $x \geq 0$ such that there exists $W \in \bar{C}\cap L^{\infty},$ satisfying $x\bar S^{\t,0}_T+W \geq  H \, \mbox{a.s.}$ 
Using \eqref{audessussuphat}, 
$
\sup_{ \Q \in   \Qc^{\t}}
\E_{\Q} \left( \frac{ H}{\bar S^{\t,0}_T} \right)\leq x
$
 and the first inequality in \eqref{defqui1} is proved. For the reverse one, let  $\e>0$. As $\{H-(\pi^{\t}(H) -\e)\bar S^{\t,0}_T\} \cap \bar{C}\cap L^{\infty} =\emptyset$,
the proof of 
$\Rightarrow$ in \eqref{laref1} shows that there exists some
$\hat X \in \Xc^*$ such that $\E \left(\hat X \left(  H-(\pi^{\t}(H) -\e)\bar S^{\t,0}_T\right)\right)>0.$ 
Let
$
{d\hat{\Q}}/{d\P}=
\left({
\hat{X}\bar{S}_T^{\t,0}
}\right)/
{
\E \left(
\hat{X} \bar{S}_T^{\t,0}
\right)
}
$.  Proposition  \ref{propnum} implies that $\hat{\Q} \in \Qc^{\t}$ and
\begin{eqnarray*}
\sup_{ \Q \in   \Qc^{\t}}
\E_{\Q} \left( \frac{ H}{\bar S^{\t,0}_T} \right) \geq \E_{\hat{\Q}}\left( \frac{ H}{\bar{S}_T^{\t,0}}  \right) >\pi^{\t}(H) -\e
 \end{eqnarray*}
and the proof is complete letting $\e$ go to zero. \\
%is proved exactly as \eqref{eqinterhat}. 

\noindent {\it Proof of  \eqref{carctprixsurep2hat}}
Let $x\geq 0$ such that there exist $\t_x \in \Tc$, $W \in \bar{C}^{\t_x}\cap L^{\infty}$ satisfying $x\bar S^{\tau_x,0}_T+W \geq   H \, \mbox{a.s.}$
As the NFL condition implies the NFL in the submarket $\tau_x$, 
Remark
%\ref{AOAtenor} 
 \ref{remuntenor}  implies that $\hat \Qc^{\tau_x}$ is nonempty and as in \eqref{comedie} that for all $\Q \in \hat \Qc^{\tau_x},$
$ 
x \geq \E_{\Q}\left(
\frac{H}{\bar S^{\tau_x,0}_T}\right).$
%Let $X^* \in \Xc^{\t_x}$. Then  $\E(WX^*)\leq 0$
%%(recall that $W \mapsto \E(X^*W)$ is a weak-star continuous functional)
%and
%$$
%x \geq \frac{\E \left( X^* H \right)}{\E \left( X^* \bar S^{\tau_x,0}_T\right)}=\E_{\Q_{\tau_x}}\left(
%\frac{H}{\bar S^{\tau_x,0}_T}\right), \mbox{ where
%}
%\frac{d\Q_{\tau_x}}{d\P}=
%\frac{{X}^*\bar S^{\tau_x,0}_T}{ \E \left( {X}^* \bar S^{\tau_x,0}_T\right)}
%\in \Qc^{\tau_x},$$ see Remark  \ref{remuntenor}.
As this is true for all $\Q \in \hat \Qc^{\tau_x}$, one obtains that
$$x \geq \sup_{\Q \in \hat \Qc^{\tau_x}} \E_{\Q}\left(
\frac{H}{\bar S^{\tau_x,0}_T}\right)
\geq \min_{\t \in \Tc} \sup_{\Q \in \hat  \Qc^{\tau}} \E_{\Q}\left(
\frac{H}{\bar{S}_T^{\t,0}}\right).$$
Thus,
\begin{eqnarray}
\label{audessusbishat}
\underline{\pi}(H) \geq \min_{\t \in \Tc} \sup_{\Q \in \hat \Qc^{\tau}} \E_{\Q}\left(
\frac{H}{\bar{S}_T^{\t,0}}\right).
\end{eqnarray}
Let $z < \underline{\pi}(H)$; then, for all $\t \in \Tc$, $\left\{H-z\bar{S}_T^{\t,0}\right\} \cap \bar{C}^{\t} \cap L^{\infty}=\emptyset$. Following the same arguments as above, one finds $X^{\t} \in \Xc^{\t}$ such that
\begin{small}
$$z < \frac{\E\left( X^{\t} H \right)}{\E\left(X^{\t}\bar{S}_T^{\t,0}\right)}
= \E_{\Q_{\tau}}\left(
\frac{H}{\bar{S}_T^{\t,0}}\right)
\leq \sup_{\Q \in \hat \Qc^{\tau}} \E_{\Q}\left(
\frac{H}{\bar{S}_T^{\t,0}}\right)
\mbox{
where }
\frac{d\Q_{\tau}}{d\P}=
\frac{{X}^{\t}\bar{S}_T^{\t,0}}{ \E \left( {X}^{\t} \bar{S}_T^{\t,0}\right)}
\in \hat \Qc^{\tau}$$
\end{small}
see Remark  \ref{remuntenor}.
Letting $z$ go to $\underline{\pi}(H)$, one obtains that for all $\t \in \Tc$,
\begin{eqnarray}
\label{endessousbishat}
\underline{\pi}(H) & \leq &
\sup_{\Q \in \hat \Qc^{\tau}} \E_{\Q}\left(
\frac{H}{\bar{S}_T^{\t,0}}\right). 
%\leq \min_{\t \in \Tc} \sup_{\Q \in \hat \Qc^{\tau}} \E_{\Q}\left(\frac{H}{\bar{S}_T^{\t,0}}\right).
\end{eqnarray}
As this is true for all $\t \in \Tc$, the other inequality in \eqref{audessusbishat} is proven, and \eqref{carctprixsurep2hat} holds true (recall \eqref{pitau}).\\

%Putting  \eqref{audessusbishat} and \eqref{endessousbishat} together achieves the proof of \eqref{carctprixsurep2hat}.
\noindent {\it Proof of  \eqref{carctprixsurep3hat} }
It is clear from \eqref{prixsurep3} and \eqref{prixsureptau} that $\overline{\pi}(H) \geq \max_{\t \in \Tc} \hat{\pi}^{\t}(H).$ For the reverse inequality, fix $\e>0$.
Let $\hat \tau \in \Tc.$ There exists $W \in \bar{C}^{\hat \t}\cap L^{\infty}$ such that
$$\left( \max_{\t \in \Tc} \hat{\pi}^{\t}(H) + \e\right) \bar S^{\hat\t,0}_T+W \geq
\left(\hat{\pi}^{\hat\t}(H) + \e\right) \bar S^{\hat\t,0}_T+W \geq H \, \mbox{a.s. }$$
Therefore,  $\max_{\t \in \Tc} \hat{\pi}^{\t}(H) + \e \geq \overline{\pi}(H)$, and the reverse inequality is proven by letting $\e$ go to zero.
Now, \eqref{carctprixsurep3hat} follows from \eqref{pitau}.\\

\noindent{\it Proof of the fact that the infima in the definitions of the different superreplication prices are minima}
We have already proved that the infima are attained for $\pi(H),$ $\pi^{\tau}(H),$ and $\hat \pi^{\tau}(H)$. As the set $\Tc$ is finite 
and \eqref{carctprixsurep2hat} holds true, there exists some $\underline{\tau}\in \Tc$ such that $\underline{\pi}(H)=\hat \pi^{\underline{\tau}}(H).$ Now, as the infimum is attained in $\hat \pi^{\underline{\tau}}(H),$ there exists $W \in \bar{C}^{\underline{\tau}}\cap L^{\infty}$ such that
$$
\underline{\pi}(H) \bar S^{\underline{\tau},0}_T+W \geq H \, \mbox{a.s. }$$
So, the infimum is attained in \eqref{prixsurep2}. 
For all $\t \in \Tc,$ the infimum is attained in $ \hat{\pi}^{\t}(H)$ and there exists $W^{\tau} \in \bar{C}^{{\tau}}\cap L^{\infty}$ such that
$$\left(\max_{\t \in \Tc} \hat{\pi}^{\t}(H) \right) \bar S^{{\tau},0}_T+W^{\tau}\geq  \hat{\pi}^{\t}(H)  \bar S^{{\tau},0}_T+W^{\tau} \geq H \, \mbox{a.s. }$$
Recalling \eqref{carctprixsurep3hat}, the infimum  is also attained in   \eqref{prixsurep3}. \\

%From the definitions given in  \eqref{prixsurep2}, \eqref{prixsurep3} and \eqref{prixsureptau}, all the equalities or inequalities in \eqref{classement} are clear except the first one.
\noindent {\it Proof of  \eqref{classement} } 
Fix $\t \in \Tc.$  It is clear that  ${\pi}^{\t}(H) \geq \pi(H)$ and as this true for all $\t \in \Tc$, the first inequality in  \eqref{classement} holds true. 
The second one follows as ${\pi}^{\t}(H) \leq \hat {\pi}^{\t}(H)$ for  all $\t \in \Tc$ and recalling \eqref{carctprixsurep2hat}. 
%Let $x \geq 0$ such that there exist and $W \in \bar{C}\cap L^{\infty}$ such that $x\bar{S}_T^{\t,0}+W \geq  H \, \mbox{a.s.}$ 
%%It is clear that $ \bar{C}^{\t} \subset \bar{C}$, and 
%We obtain that
%$x \geq \pi(H)$. This, ${\pi}^{\t}(H) \geq \pi(H)$ follows from taking the infimum over all such $x$. 
$\Box$
%\end{proof}
\subsection{Proof of  Proposition \ref{propdeux}}
In this section, recall that the local martingale are assumed to be true martingale. 
\label{preuve4}
\begin{lemma}
Let  $Z \in L_{>0}^{\infty}$ and   $\Q \in \Qc^{Z}$. Then,  for all $\t \in \Tc,$
\begin{eqnarray}
\label{numjuillet}
\E_{\Q} \left(\frac{S^{\tau}_T}{ Z}\right) & = &
%S^{\tau,0}_0\E_{\Q} \left(\frac{\bar S^{\t,0}_T}{Z} \widetilde{S}^{\tau}_T\right) 
%& = & S^{{\t},0}_0 \E_{\Q} \left( \frac{S^{\t}_T}{S^{{\t},0}_T} \E_{\Q}\left(\frac{S^{{\t},0}_T/S^{{\t},0}_0}{S^{\hat{\t},0}_T/S^{\hat{\t},0}_0}|\Fc_T\right)\right)\\
S_0^{\tau}
\E_{\Q}\left(\frac{\bar S^{\t,0}_T}{Z}\right).
\end{eqnarray}
Assume that $\frac{H}{Z} \in L^{\infty}$ and  $\frac{Z}{H} \in L^{\infty}.$ Then, 
\begin{eqnarray}
\label{fou}
\sup_{\Q \in \Qc^{Z}} 
\E_{\Q} \left(\frac{ H }{ Z}\right)  & = &\frac{1}{
\inf_{\Q \in \Qc^{H} }
\E_{\Q} \left(\frac{Z }{ H}\right) 
}.
\end{eqnarray}
\end{lemma}
\begin{proof}
As $\E_{\Q} \left(\frac{S^{\tau}_T}{ Z}\right)  = 
S^{\tau,0}_0\E_{\Q} \left(
\frac{\bar S^{\t,0}_T}{Z} \widetilde{S}^{\tau}_T\right)$, \eqref{numjuillet} follows from Definition \ref{defQZ} (with true martingale). Then, if $\frac{H}{Z} \in L^{\infty},$ \eqref{enfin} implies that 
\begin{eqnarray}
\label{dingue}
\sup_{X^* \in \Xc^*} \frac{\E(X^*H)}{\E(X^*Z)} = \sup_{\Q \in \Qc^Z}\E_{\Q}\left(\frac{H}{Z}\right).
\end{eqnarray}
Note that the same equality holds true by changing suprema by infima or by changing respectively $\Qc^Z$ by $\Qc^{\tau,Z}$ and $\Xc^*$ by $\Xc^{\tau}$. 
Assume furthermore that $\frac{Z}{H} \in L^{\infty},$ using \eqref{dingue}, we get that 
\begin{eqnarray*}
\sup_{\Q \in \Qc^{Z}} 
\E_{\Q} \left(\frac{ H }{ Z}\right)  & = & 
\sup_{X^* \in \Xc^*} \frac{\E\left(X^*H\right)}{\E\left(X^*Z\right)}  = \frac{1}{
\inf_{X^* \in \Xc^*} \frac{\E\left(X^*Z\right)}{\E\left(X^*H\right)}
}
 = \frac{1}{
\inf_{\Q \in \Qc^{H} }
\E_{\Q} \left(\frac{Z }{ H}\right) 
}.
\end{eqnarray*} $\Box$
\end{proof}

\begin{proof}{\it of Proposition \ref{propdeux}} 
Note that \eqref{qui0} holds true since recalling \eqref{defqui1} and \eqref{eq1}
$${\pi}^{\t_1}(S^{\tau_1}_T)  =  \sup_{\Q \in \Qc^{\tau_1}} 
\E_{\Q} \left(\frac{S^{\tau_1}_T }{ \bar S^{\tau_1,0}_T}\right)
=S_0^{\tau_1}=\hat{\pi}^{\t_1}(S^{\tau_1}_T).$$

We choose in Theorem \ref{main2} $\lambda_{{\tau}_2}=1$  and $\lambda_{{\tau}_1}=0$. Then, $\Qc^{lc, \l}
=\Qc^{{\t}_{2}}$.\\

\noindent {\it Computation of $\pi(S^{\tau_1}_T)$} 
Let $\Q \in \Qc^{{\t}_{2}}$.  
Using \eqref{numjuillet} we get that 
\begin{eqnarray}
\label{rainingagain}
\E_{\Q} \left(\frac{S^{\tau_1}_T}{ \bar S^{{\t}_{2},0}_T}\right) 
&=& S_0^{\tau_1}
\E_{\Q}\left(\frac{\bar S^{\tau_1,0}_T}{\bar S^{{\t}_{2},0}_T}\right).
\end{eqnarray}
So, \eqref{qui} follows from \eqref{fou} and \eqref{rainingagain}. 
Then, we compute  in \eqref{prixrepqsumhat}
\begin{eqnarray}
\label{quimper}
\E_{\Q} \left(\frac{S^{\tau_1}_T}{ \bar S^{\tau_2,0}_T}\right) -\sum_{\t \in \Tc} \hat{x}^{\t} \E_{\Q} \left(
 \frac{\bar{S}_T^{\t,0}}{\bar S^{\tau_2,0}_T}
\right) & = & 
\left(S_0^{\tau_1}-\hat{x}^{\tau_1} \right)\E_{\Q} \left(\frac{\bar S^{\tau_1,0}_T}{ \bar S^{\tau_2,0}_T}\right)-\hat{x}^{\tau_2}.
\end{eqnarray}
Thus, we get that
$$\hat{x}^{\tau_2} = \sup_{\Q \in \Qc^{\tau_2}} \left(\left(S_0^{\tau_1}-\hat{x}^{\tau_1} \right) 
\E_{\Q} \left(\frac{\bar S^{\tau_1,0}_T}{ \bar S^{\tau_2,0}_T}\right) 
\right).$$
If $S_0^{\tau_1}-\hat{x}^{\tau_1} <0,$ $\hat{x}^{\tau_2}< 0$. Thus, assume that $0 \leq \hat{x}^{\tau_1} \leq S_0^{\tau_1}$. 
Then, \\$\hat{x}^{\tau_2}=\left(S_0^{\tau_1}-\hat{x}^{\tau_1} \right)\sup_{\Q \in \Qc^{\tau_2}} 
\E_{\Q} \left(\frac{\bar S^{\tau_1,0}_T}{ \bar S^{\tau_2,0}_T}\right) $ and 
\begin{eqnarray*}
\pi(S^{\tau_1}_T) & = & \inf \left\{ S_0^{\tau_1}\sup_{\Q \in \Qc^{\tau_2}} 
\E_{\Q} \left(\frac{\bar S^{\tau_1,0}_T}{ \bar S^{\tau_2,0}_T}\right) + \hat{x}^{\tau_1}\left(1-  
\sup_{\Q \in \Qc^{\tau_2}} 
\E_{\Q} \left(\frac{\bar S^{\tau_1,0}_T}{ \bar S^{\tau_2,0}_T}\right)
\right)| \; 0\leq\hat{x}^{\tau_1} \leq S^{\tau_1}_0
\right\}\\
%& = & S_0^{\tau_1}\left( \sup_{\Q \in \Qc^{\tau_2}} \E_{\Q} \left(\frac{\bar S^{\tau_1,0}_T}{ \bar S^{\tau_2,0}_T}\right)  \wedge 1\right)\\
& = & \left(S_0^{\tau_1} \sup_{\Q \in \Qc^{\tau_2}} 
\E_{\Q} \left(\frac{\bar S^{\tau_1,0}_T}{ \bar S^{\tau_2,0}_T}\right) \right) \wedge S_0^{\tau_1},
\end{eqnarray*}
and we conclude using \eqref{qui0} and \eqref{qui}.  \\
%Note that the formulation of $\pi(S^{\tau_1}_T)$ in \eqref{qui} with an infimum in $\Qc^{{\t}_{1}}$  may appear directly choosing $\Qc^{lc, \l}=\Qc^{{\t}_{1}}$.\\

\noindent {\it $\pi(S^{\tau_1}_T)={\pi}^{\t_2}( S_T^{\tau_1}) \Rightarrow \pi( S_T^{\tau_2})={\pi}^{\t_2}(S_T^{\tau_2}).$ }\\
Assume that 
$\pi(S^{\tau_1}_T)={\pi}^{\t_2}( S_T^{\tau_1})$. This is equivalent to 
$\sup_{\Q \in \Qc^{\tau_2}} 
\E_{\Q} \left(\frac{\bar S^{\tau_1,0}_T}{ \bar S^{\tau_2,0}_T}\right)  \leq 1.$ 
If ${\pi}^{\t_1}( S_T^{\tau_2}) < {\pi}^{\t_2}( S_T^{\tau_2})$, i.e., 
$$1>\sup_{\Q \in \Qc^{\tau_1}} 
\E_{\Q} \left(\frac{\bar S^{\tau_2,0}_T}{ \bar S^{\tau_1,0}_T}\right) = 
\frac{1}{\inf_{\Q \in \Qc^{\tau_2}} 
\E_{\Q} \left(\frac{\bar S^{\tau_1,0}_T}{ \bar S^{\tau_2,0}_T}\right)},$$
see \eqref{fou} and  we get a contraction. Thus, ${\pi}^{\t_1}( S_T^{\tau_2}) \geq  {\pi}^{\t_2}( S_T^{\tau_2})$ and $\pi( S_T^{\tau_2})={\pi}^{\t_2}( S_T^{\tau_2}).$ 
Then,
\begin{eqnarray*}
\hat{\pi}^{\t_1}(S^{\tau_1}_T) - {\pi}(S^{\tau_1}_T) & =& {\pi}^{\t_1}(S^{\tau_1}_T) -{\pi}^{\t_1}(S^{\tau_1}_T) \wedge {\pi}^{\t_2}(S^{\tau_1}_T)
= \left({\pi}^{\t_1}(S^{\tau_1}_T) -{\pi}^{\t_2}(S^{\tau_1}_T) \right)_+.
\end{eqnarray*}
%{\red Assume that 
%$\pi(S^{\tau_1}_T)=\hat{\pi}^{\t_1}( S_T^{\tau_1})$. This is equivalent to 
%$\sup_{\Q \in \Qc^{\tau_2}} 
%\E_{\Q} \left(\frac{\bar S^{\tau_1,0}_T}{ \bar S^{\tau_2,0}_T}\right)  \geq 1.$ 
%If ${\pi}^{\t_1}( S_T^{\tau_2}) > \hat{\pi}^{\t_2}( S_T^{\tau_2})$, i.e., 
%$$1<\sup_{\Q \in \Qc^{\tau_1}} 
%\E_{\Q} \left(\frac{\bar S^{\tau_2,0}_T}{ \bar S^{\tau_1,0}_T}\right) = 
%\frac{1}{\inf_{\Q \in \Qc^{\tau_2}} 
%\E_{\Q} \left(\frac{\bar S^{\tau_1,0}_T}{ \bar S^{\tau_2,0}_T}\right)},$$
%see \eqref{fou} and  we get a non contraction. }

\noindent {\it Computation of $\pi(S^{\tau_1}_T-S^{\tau_2}_T)$} 
 Let $\Q \in \Qc^{{\t}_{2}}$, using \eqref{rainingagain}, we compute  in \eqref{prixrepqsumhat}
\begin{eqnarray*}
\E_{\Q} \left(\frac{S^{\tau_1}_T-S^{\tau_2}_T}{ \bar S^{\tau_2,0}_T}\right) -\sum_{\t \in \Tc} \hat{x}^{\t} \E_{\Q} \left(
 \frac{\bar{S}_T^{\t,0}}{\bar S^{\tau_2,0}_T}
\right) & = & 
\left(S_0^{\tau_1}-\hat{x}^{\tau_1} \right)\E_{\Q} \left(\frac{\bar S^{\tau_1,0}_T}{ \bar S^{\tau_2,0}_T}\right)-S_0^{\tau_2}-\hat{x}^{\tau_2}.
\end{eqnarray*}
Thus, we get that
$$\hat{x}^{\tau_2} = \sup_{\Q \in \Qc^{\tau_2}} \left(\left(S_0^{\tau_1}-\hat{x}^{\tau_1} \right) 
\E_{\Q} \left(\frac{\bar S^{\tau_1,0}_T}{ \bar S^{\tau_2,0}_T}\right) 
\right)-S_0^{\tau_2}.$$
If $S_0^{\tau_1}-\hat{x}^{\tau_1} < 0,$ then $\hat{x}^{\tau_2} <0$. Thus, assume that $0 \leq \hat{x}^{\tau_1} \leq S_0^{\tau_1}$. Then, $\hat{x}^{\tau_2}=\left(S^{\tau_1}_0-\hat{x}^{\tau_1} \right)\sup_{\Q \in \Qc^{\tau_2}} 
\E_{\Q} \left(\frac{\bar S^{\tau_1,0}_T}{ \bar S^{\tau_2,0}_T}\right)  - S^{\tau_2}_0.$ 
As $\hat{x}^{\tau_2}\geq 0$, we need to impose that (recall \eqref{qui})
$$0 \leq \hat{x}^{\tau_1} \leq S_0^{\tau_1} - \frac{S_0^{\tau_2}}{\sup_{\Q \in \Qc^{\tau_2}} 
\E_{\Q} \left(\frac{\bar S^{\tau_1,0}_T}{ \bar S^{\tau_2,0}_T}\right) }=\frac{S_0^{\tau_1}}{{\pi}^{\t_2}( S_T^{\tau_1})} \left( {\pi}^{\t_2}( S_T^{\tau_1}) -S_0^{\tau_2} \right).$$
For that we impose that $ {\pi}^{\t_2}( S_T^{\tau_1}) \geq S_0^{\tau_2}.$ Else, $\pi(S^{\tau_1}_T-S^{\tau_2}_T)=+\infty$. Then, we get that 
%$$S_0^{\tau_1}\sup_{\Q \in \Qc^{\tau_2}} \E_{\Q} \left(\frac{\bar S^{\tau_1,0}_T}{ \bar S^{\tau_2,0}_T}\right)  \geq S_0^{\tau_2}.$$
%\begin{small}
%\begin{eqnarray*}
%\pi(S^{\tau_1}_T-S^{\tau_2}_T) & = & \inf \left\{ 0\leq x \leq S_0^{\tau_1} - \frac{S_0^{\tau_2}}{\sup_{\Q \in \Qc^{\tau_2}} 
%\E_{\Q} \left(\frac{\bar S^{\tau_1,0}_T}{ \bar S^{\tau_2,0}_T}\right) }|\; 
%S^{\tau_1}_0 \sup_{\Q \in \Qc^{\tau_2}} 
%\E_{\Q} \left(\frac{\bar S^{\tau_1,0}_T}{ \bar S^{\tau_2,0}_T}\right) 
%-  S^{\tau_2}_0+ x\left(1-  \sup_{\Q \in \Qc^{\tau_2}} 
%\E_{\Q} \left(\frac{\bar S^{\tau_1,0}_T}{ \bar S^{\tau_2,0}_T}\right)  \right)\right\}\\
%& = & 1_{\sup_{\Q \in \Qc^{\tau_2}} 
%\E_{\Q} \left(\frac{\bar S^{\tau_1,0}_T}{ \bar S^{\tau_2,0}_T}\right) \leq 1} \left(
%S^{\tau_1}_0 \sup_{\Q \in \Qc^{\tau_2}} 
%\E_{\Q} \left(\frac{\bar S^{\tau_1,0}_T}{ \bar S^{\tau_2,0}_T}\right) 
%-  S^{\tau_2}_0
%\right) + \\
%& &1_{\sup_{\Q \in \Qc^{\tau_2}} 
%\E_{\Q} \left(\frac{\bar S^{\tau_1,0}_T}{ \bar S^{\tau_2,0}_T}\right) > 1}\left( S^{\tau_1}_0
%   - \frac{S_0^{\tau_2}}{\sup_{\Q \in \Qc^{\tau_2}} 
%\E_{\Q} \left(\frac{\bar S^{\tau_1,0}_T}{ \bar S^{\tau_2,0}_T}\right) } \right).
%%\\
%%& = & 1_{\pi( S^{\tau_1}_T)= {\pi}^{\t_2}( S^{\tau_1}_T)} \left(
%%{\pi}^{\t_2}( S^{\tau_1}_T)
%%-  \hat{\pi}^{\t_2}( S^{\tau_2}_T)
%%\right) +
%%1_{\pi( S^{\tau_1}_T)= \hat{\pi}^{\t_1}( S^{\tau_1}_T)}\left( \hat{\pi}^{\t_1}(S^{\tau_1}_T)
%%   - {\pi}^{\t_1}( S^{\tau_2}_T) \right).
%\end{eqnarray*}
%\end{small}
\begin{small}
\begin{eqnarray*}
\pi(S^{\tau_1}_T-S^{\tau_2}_T) & = & \inf \left\{ 
{\pi}^{\t_2}( S_T^{\tau_1})
-  S^{\tau_2}_0+ \frac{\hat x^{\tau_1}}{S_0^{\tau_1}}\left(S_0^{\tau_1}-  {\pi}^{\t_2}( S_T^{\tau_1}) \right)
|\; 0\leq \hat x^{\tau_1} \leq \frac{S_0^{\tau_1}}{{\pi}^{\t_2}( S_T^{\tau_1})} \left( {\pi}^{\t_2}( S_T^{\tau_1}) -S_0^{\tau_2} \right)
\right\}.
\end{eqnarray*}
\end{small}
{\it Case  $\sup_{\Q \in \Qc^{\tau_2}} 
\E_{\Q} \left(\frac{\bar S^{\tau_1,0}_T}{ \bar S^{\tau_2,0}_T}\right)<1.$}\\
This is equivalent to 
${\pi}^{\t_2}( S_T^{\tau_1})<  S_0^{\tau_1}.$ Then, $\pi (S^{\tau_1}_T) ={\pi}^{\t_2}( S_T^{\tau_1})$ and also $\pi( S_T^{\tau_2})=S^{\tau_2}_0.$  This is the case where the asset $\tau_2$ is liquid and the asset $\tau_1$ is illiquid. In this case, $\hat x^{\tau_1}=0,$ all the initial investments are made in the liquid market $\tau_2$ and 
$$\pi(S^{\tau_1}_T-S^{\tau_2}_T)={\pi}^{\t_2}( S_T^{\tau_1})
-  S^{\tau_2}_0=\pi (S^{\tau_1}_T) - \pi( S_T^{\tau_2}).$$
Note that in this case we need to have that ${S^{\tau_2}_0} \leq {S^{\tau_1}_0}$.  \\
%Then, $\pi(S^{\tau_1}_T-S^{\tau_2}_T)=\pi(S^{\tau_1}_T)- S^{\tau_2}_0.$
{\it Case  $\sup_{\Q \in \Qc^{\tau_2}} 
\E_{\Q} \left(\frac{\bar S^{\tau_1,0}_T}{ \bar S^{\tau_2,0}_T}\right)\geq 1.$}\\ This is  equivalent to  
  ${\pi}^{\t_2}( S_T^{\tau_1}) \geq  S_0^{\tau_1}.$ Then, $\pi (S^{\tau_1}_T) = S_0^{\tau_1},$ the asset $\tau_1$ is liquid. Moreover,  $\hat{x}^{\tau_1}=\frac{S_0^{\tau_1}}{{\pi}^{\t_2}( S_T^{\tau_1})} \left( {\pi}^{\t_2}( S_T^{\tau_1}) -S_0^{\tau_2} \right)$ and 
  $\hat{x}^{\tau_2} =  \left(1-\frac{\hat{x}^{\tau_1}}{S^{\tau_1}_0} \right){\pi}^{\t_2}( S_T^{\tau_1})  - S^{\tau_2}_0=0$. Thus, 
%\begin{eqnarray*}
%\hat{x}^{\tau_2}& = & \left(1-\frac{\hat{x}^{\tau_1}}{S^{\tau_1}_0} \right){\pi}^{\t_2}( S_T^{\tau_1})  - S^{\tau_2}_0=
%\left(1-\frac{1}{{\pi}^{\t_2}( S_T^{\tau_1})} \left( {\pi}^{\t_2}( S_T^{\tau_1}) -S_0^{\tau_2} \right) \right){\pi}^{\t_2}( S_T^{\tau_1})  - S^{\tau_2}_0=
%\left(\frac{S_0^{\tau_2}}{{\pi}^{\t_2}( S_T^{\tau_1})} \right){\pi}^{\t_2}( S_T^{\tau_1})  - S^{\tau_2}_0= 0
%\\
%\end{eqnarray*}
\begin{eqnarray*}
\pi(S^{\tau_1}_T-S^{\tau_2}_T) & = & \hat{x}^{\tau_1}=
%=\frac{S_0^{\tau_1}}{{\pi}^{\t_2}( S_T^{\tau_1})} \left( {\pi}^{\t_2}( S_T^{\tau_1}) -S_0^{\tau_2} \right) = 
%{\pi}^{\t_2}( S_T^{\tau_1})-  S^{\tau_2}_0+ \frac{1}{{\pi}^{\t_2}( S_T^{\tau_1})} \left( {\pi}^{\t_2}( S_T^{\tau_1}) -S_0^{\tau_2} \right)\left(S_0^{\tau_1}-  {\pi}^{\t_2}( S_T^{\tau_1}) \right) \\
%\left( {\pi}^{\t_2}( S_T^{\tau_1}) -S_0^{\tau_2} \right) \frac{S_0^{\tau_1}}{{\pi}^{\t_2}( S_T^{\tau_1})}  = 
  \pi(S_T^{\tau_1}) - \frac{S_0^{\tau_2}}{\sup_{\Q \in \Qc^{\tau_2}} 
\E_{\Q} \left(\frac{\bar S^{\tau_1,0}_T}{ \bar S^{\tau_2,0}_T}\right)}.  
%=\pi(S_T^{\tau_1}) - \pi(S_T^{\tau_2}) \frac{\inf_{\Q \in \Qc^{\tau_2}} \E_{\Q} \left(\frac{\bar S^{\tau_1,0}_T}{ \bar S^{\tau_2,0}_T}\right) \vee 1}{\sup_{\Q \in \Qc^{\tau_2}} \E_{\Q} \left(\frac{\bar S^{\tau_1,0}_T}{ \bar S^{\tau_2,0}_T}\right)}.
\end{eqnarray*}
%It follows that 
%$$\left( \frac{S_0^{\tau_2}}{\sup_{\Q \in \Qc^{\tau_2}} 
%\E_{\Q} \left(\frac{\bar S^{\tau_1,0}_T}{ \bar S^{\tau_2,0}_T}\right)}
%\right) S_T^{\tau_1,0} +W_1-W   =  S_T^{\tau_2}$$
%and $$\pi^{\tau_1}(S_T^{\tau_2}) \leq \frac{S_0^{\tau_2}}{\sup_{\Q \in \Qc^{\tau_2}} 
%\E_{\Q} \left(\frac{\bar S^{\tau_1,0}_T}{ \bar S^{\tau_2,0}_T}\right)}= \frac{S_0^{\tau_2}S_0^{\tau_1}}{{\pi}^{\t_2}( S_T^{\tau_1})}.$$
{\it Case  $\inf_{\Q \in \Qc^{\tau_2}} 
\E_{\Q} \left(\frac{\bar S^{\tau_1,0}_T}{ \bar S^{\tau_2,0}_T}\right)\leq 1.$} Then, $\pi( S_T^{\tau_2})=S^{\tau_2}_0,$ the asset $\tau_2$ is also liquid.  
\begin{eqnarray*}
\pi(S^{\tau_1}_T-S^{\tau_2}_T) & = &  
\pi(S_T^{\tau_1}) - \pi(S_T^{\tau_2}) \frac{ 1}{\sup_{\Q \in \Qc^{\tau_2}} \E_{\Q} \left(\frac{\bar S^{\tau_1,0}_T}{ \bar S^{\tau_2,0}_T}\right)}.
\end{eqnarray*}
{\it Case  $\inf_{\Q \in \Qc^{\tau_2}} 
\E_{\Q} \left(\frac{\bar S^{\tau_1,0}_T}{ \bar S^{\tau_2,0}_T}\right)>1.$} Then, $\pi( S_T^{\tau_2})=\pi^{\tau_1}( S_T^{\tau_2}),$ the asset $\tau_2$ is illiquid, the initial investments are made in  the market $\tau_1$ and we get that 
$$\pi(S^{\tau_1}_T-S^{\tau_2}_T)=
\pi(S_T^{\tau_1}) - \pi(S_T^{\tau_2}) \frac{\inf_{\Q \in \Qc^{\tau_2}} 
\E_{\Q} \left(\frac{\bar S^{\tau_1,0}_T}{ \bar S^{\tau_2,0}_T}\right) }{\sup_{\Q \in \Qc^{\tau_2}} 
\E_{\Q} \left(\frac{\bar S^{\tau_1,0}_T}{ \bar S^{\tau_2,0}_T}\right) }.$$
%Thus, all together we obtain the following formula 
%$$\pi(S^{\tau_1}_T-S^{\tau_2}_T)=
%\pi(S_T^{\tau_1}) - \pi(S_T^{\tau_2}) \frac{\inf_{\Q \in \Qc^{\tau_2}} 
%\E_{\Q} \left(\frac{\bar S^{\tau_1,0}_T}{ \bar S^{\tau_2,0}_T}\right) \vee 1}{\sup_{\Q \in \Qc^{\tau_2}} 
%\E_{\Q} \left(\frac{\bar S^{\tau_1,0}_T}{ \bar S^{\tau_2,0}_T}\right) \vee 1}.$$
$\Box$

\end{proof}

\begin{acknowledgements}
The author would like to thank Zorana Gbrac and Martino Grasselli for helpful discussions.
\end{acknowledgements}



\bibliographystyle{spmpsci}
\bibliography{bibmultinum}

\end{document}